\documentclass[a4paper,11pt]{article}
\pdfoutput=1 
\usepackage{verbatim}

\usepackage{jheppub}
\usepackage{amsmath,amsfonts,amssymb,amsthm,graphicx,color}
\usepackage[T1]{fontenc} 
\usepackage{enumerate}
\usepackage{tikz, tikz-3dplot, pgfplots}
\usepackage[utf8]{inputenc}
\usepackage{slashed}
\usepackage{subcaption}
\usepackage{booktabs}

\usepackage{changepage}
\usepackage{float}
\usepackage{tikz}
\usetikzlibrary{decorations.text}
\usepackage{enumitem}
\setlist{itemsep=0pt}
\usetikzlibrary{math}

\usepackage{tikz}
\usepackage{pgfplots}
\usetikzlibrary{decorations.text,intersections, pgfplots.fillbetween}
\usetikzlibrary{math}
 \usetikzlibrary{snakes,3d,shapes.geometric,shadows.blur}
\usepackage{tikz-3dplot}

\makeatletter
\newcommand\footnoteref[1]{\protected@xdef\@thefnmark{\ref{#1}}\@footnotemark}
\makeatother

\newcommand{\comm}[1]{} 

\def\({\left(}
\def\){\right)}
\def\[{\left[}
\def\]{\right]}

\def\One{{\hbox{ 1\kern-.8mm l}}}

\def\barray{\begin{array}}
\def\earray{\end{array}}
\def\be{\begin{equation}}
\def\ee{\end{equation}}
\def\bea{\begin{eqnarray}}
\def\eea{\end{eqnarray}}
\def\bal{\begin{align}}
\def\eal{\end{align}}
\def\nn{\nonumber}

\def\-{\,-\,}
\def\={\,=\,}
\def\+{\,+\,}
\def\equi{\,\equiv\,}

\numberwithin{equation}{section} 

\definecolor{cardinal}{rgb}{0.6,0,0}
\definecolor{darkgreen}{rgb}{0,0.4,0}
\definecolor{golden}{rgb}{0.92, 0.7, 0}
\definecolor{midnight}{rgb}{0, 0, 0.5}
\definecolor{darkblue}{rgb}{0, 0, 0.7}
\definecolor{purple}{rgb}{0.5, 0, 0.5}

\newcommand{\subf}[2]{%
  {\small\begin{tabular}[t]{@{}c@{}}
  #1\\#2
  \end{tabular}}%
}

\def\IR{\mathbb{R}}

\def\cA{{\cal A}}

\def\cJ{{\cal J}}

\def\cM{{\cal M}}
\def\cN{{\cal N}}

\def\cO{{\cal O}}
\def\cQ{{\cal Q}}
\def\cR{{\cal R}}

\def\cZ{{\cal Z}}

\usetikzlibrary{positioning,calc,fadings,decorations.pathreplacing}

\tikzset{
 diffuse color/.initial = black,                       
}

\tikzset{
 linear opacity/.initial=0.5,                          
 linear stroke/.style = {                              
   preaction={                                         
     draw=\pgfkeysvalueof{/tikz/diffuse color},        
     line width = (2.0-#1)*\pgflinewidth,              
     opacity=\pgfkeysvalueof{/tikz/linear opacity},white}},  
 diffuse gradient/.style={                             
   draw = none,                                        
   linear opacity=#1,                                  
   linear stroke/.list={0.0,#1,...,1.0}},              
 diffuse gradient/.default=1,                          
}

\tikzset{
 non-linear stroke/.style = {                          
   preaction={                                         
     draw=\pgfkeysvalueof{/tikz/diffuse color},        
     line width = (2.0-#1)*\pgflinewidth,              
     opacity=#1,white}},                                     
 diffuse falloff/.style={                              
   draw = none,                                        
   non-linear stroke/.list={0.0,#1,...,1.0}},          
 diffuse falloff/.default=1,                           
}
 
\tikzset{%
  >=latex, 
  inner sep=0pt,%
  outer sep=2pt,%
  mark coordinate/.style={inner sep=0pt,outer sep=0pt,minimum size=3pt,
    fill=black,circle}%
}

\title{\boldmath
Schwarzschild-like Topological Solitons
}

\author[a,b]{Ibrahima Bah,}
\author[a]{Pierre Heidmann} 
\author[a]{ and Peter Weck} 
\affiliation[a]{Department of Physics and Astronomy, Johns Hopkins University,\\
3400 North Charles Street, Baltimore, MD 21218, USA}
\affiliation[b]{Institute for Advanced Study,\\
1 Einstein Drive, Princeton, NJ 08540, USA}

\emailAdd{iboubah@jhu.edu}
\emailAdd{pheidma1@jhu.edu}
\emailAdd{pweck1@jhu.edu}

\abstract{ We construct the first class of topological solitons in gravity that are supported by internal electromagnetic flux with vanishing net charges.  The solutions are obtained in a six-dimensional Einstein-Maxwell theory with a three-form flux, and admit an uplift to type IIB supergravity on T$^4$.  They are asymptotic to a torus fibration over four-dimensional Minkowski spacetime.   An interesting class corresponds to solitons with a BPS particle and its anti-BPS partner held apart by a vacuum bubble.  In type IIB, they correspond to bound states of BPS and anti-BPS D1-D5 extremal black holes.  These metrics are a particular limit of a larger class of axially symmetric metrics that we construct and that describe smooth horizonless topological solitons.  They correspond to bound states of three non-BPS bubbles on a line.  An important achievement is that the outer bubbles can carry arbitrary D1-D5 charges that we can tune to vanishing net charges.  We discuss their properties and compare them to a four-dimensional Schwarzschild black hole of the same mass. We show that they have a long throat with a large redshift, and that they are ultra-compact with a characteristic size of 1.52 times the Schwarzschild radius.}

\preprint{}
\begin{document}

\maketitle
\flushbottom

\newpage


\section{Introduction}
\label{sec:Intro}

Smooth gravitational solitons with interesting topology can exist in a variety of supergravity theories that descend from string theories.  They are staples in the construction of explicit models for holography \cite{Lin:2004nb,Lunin:2001jy}, and for microstate geometries in black hole physics \cite{Bena:2022ldq,Warner:2019jll,Bena:2007kg}.  More recently, their existence in generic theories of gravity have been demonstrated beyond the framework of supergravity and supersymmetry \cite{Bah:2020ogh,Bah:2020pdz,Bah:2021owp,Bah:2021rki,Heidmann:2021cms}.  The novel methods employed for such systems are generalizations of the well-known Weyl constructions of black holes \cite{Weyl:book,Papapetrou:1953zz,Emparan:2001wk} to include various electromagnetic fields as well as bubbling geometries. Asymptotically, the solitons are ultra-compact geometries that have the same mass and electromagnetic charges as non-supersymmetric four-dimensional Reisner-Nordstr\"om black holes, and for which the UV origin as bound states of strings and branes in string theory is well-established \cite{Heidmann:2021cms}.  

A major conundrum has been whether topological solitons can be Schwarzschild-like, i.e. ultra-compact geometries that have finite mass and vanishing net electromagnetic charges.  This has been a significant challenge over the years because of the no-go theorems that stand in the way. Indeed, it is known that vacuum four-dimensional Einstein theory has no asymptotically flat, topologically trivial, and globally stationary solutions other than flat spacetime \cite{Serini,Einstein,Einstein:1943ixi,Lichnerowicz}.  These no-go theorems can be avoided by adding extra compact dimensions, electromagnetic fluxes, and Chern-Simons interactions \cite{Gibbons:2013tqa}.  Most of the solutions that exploit these loopholes have non-trivial asymptotic electromagnetic charges, and they are generically supersymmetric. However, neutral topological solitons cannot be supersymmetric since the BPS conditions fix certain charges in terms of their mass.  Aside from supersymmetry, one can hope to construct bound states of solitons where all the charges add up to zero asymptotically.  The main result of this paper is an explicit construction for such neutral Schwarzschild-like solitons.  

We consider Einstein-Maxwell theory in six dimensions with a two-form potential, $C^{(2)}$, and field strength $F_3 = dC^{(2)}$.  The solitons live in the background $\IR^{1,3} \times \text{T}^2$, a torus fibration over four-dimensional Minkowski spacetime, and can carry magnetic and electric line charges along one of the T$^2$ directions.  This is a consistent truncation of type IIB supergravity on a rigid T$^4$ with D1-D5 brane sources.  With an appropriate ansatz for static and axially-symmetric solutions, the Einstein-Maxwell equations decompose into decoupled sectors of PDEs \cite{Bah:2021owp,Bah:2021rki,Heidmann:2021cms}. The equations in each sector are related to Einstein equations for four-dimensional static and axially-symmetric spaces.  These, famously, have hidden integrability structure which we can exploit. This framework opened a new door in the construction of non-trivial topological solitons beyond supersymmetry.

Generically, the equations of interest admit smooth bolts as sources.  These are defined as segments on the axis of symmetry, called rods, where a circle from the torus collapses to zero size.  In the three-dimensional base of the external spacetime, a rod source is a locus where there is a blown-up two-cycle corresponding to a smooth topological soliton, or bubble, with no horizon.  Each rod has a mass parameter associated to its length.  Electromagnetic charges, or D1-D5 charges, can be attached to it as flux wrapped on the two-cycle. These fluxes are necessary for stability of the solitons \cite{Stotyn:2011tv,Bah:2021irr}.  If we turn them off, the rods  correspond to static ``bubbles of nothing'' which are known to be unstable \cite{Witten:1981gj}.  

There is an interesting extremal limit for a charged rod \cite{Bah:2020ogh,Bah:2020pdz,Heidmann:2021cms}.  This corresponds to taking the length of the rod to zero while holding the charges fixed.  The source saturates BPS conditions and corresponds to a BPS black string. This provides an interpretation of the rod length as a non-BPS deformation, from the extremal black string to a smooth non-BPS bubble.  These special rod sources will be central in our constructions. 

The existence of charged rod sources raises the question of whether they can be combined to obtain bound states that are smooth and horizonless.  This is indeed a hard task as it requires confronting the full non-linear Einstein equations. In the presence of axial symmetry, our novel framework allows us to construct explicit metrics corresponding to rod sources stacked in a line in the three-dimensional base.  

Indeed, in \cite{Bah:2021owp,Bah:2021rki,Heidmann:2021cms}, a large class of smooth and regular metrics are obtained with an arbitrary sequence of rod sources inducing a chain of smooth bubbles.  In these solutions, the charge-to-mass ratio of all rods must be the same.  This followed from an emergent linear structure of the equations.  The background metric still has a non-linear and non-perturbative dependence on the emergent linear system.  It is surprising that such structure can exist far from under the lamppost of supersymmetry and supergravity. However, in this family of solutions, the total charges cannot vanish since all the rod sources have charges with the same sign.  The success and the resulting insight of \cite{Bah:2021owp,Bah:2021rki,Heidmann:2021cms} suggested the existence of a larger basin of smooth and regular non-supersymmetric solutions.  

In this paper, we construct new classes of solutions induced by three rods, in which two can carry arbitrary electromagnetic or D1-D5 charges. In particular, we find configurations where the asymptotic net charges vanish by considering rods with opposite charges. Such solutions cannot fit within the paradigm of {\cite{Bah:2021owp,Bah:2021rki,Heidmann:2021cms}  and new methods must be developed.  We exploit the integrability structure of the Ricci-flat equations by using inverse scattering methods.  In particular, we consider the solutions used in \cite{Alekseev:2007re,Alekseev:2007gt,Manko:2007hi,Manko:2008gb,PhysRevD.51.4192}  to construct bound states of two Reisner-Nordstr\"om black holes in four dimensions.  

In the first class of solutions studied in this paper, both charged rods degenerate to points, corresponding to BPS sources. These solutions describe bound states of a BPS D1-D5 black hole and an anti-BPS $\overline{\text{D1}}$-$\overline{\text{D5}}$ black hole at the poles of a vacuum bubble. We discuss the neutral geometries when the charges of the extremal black holes cancel out asymptotically, which can be considered as ``D1-D5 black hole dipoles'', and compare them to Schwarzschild geometries. 

In the second class of solutions, the two charged rods have small but finite size, and induce two smooth bolts where one of the T$^2$ directions degenerates. These configurations correspond to two bubbles carrying D1-D5 charges and connected by a vacuum bubble. They can be seen as smooth non-BPS resolutions of the BPS/anti-BPS black hole bound states. We also discuss neutral geometries in which the two bubbles have opposite charges, and compare these neutral topological solitons to Schwarzschild. This new class of horizonless geometries constitutes the first examples of smooth topological solitons wrapped by flux which arise from string theory but describe ultra-compact objects with the same asymptotic charges as a Schwarzschild black hole.

Before proceeding, we provide a summary of results and a road map for the paper.   In section \ref{sec:Weyl6d}, we review the charged Weyl framework we work with, and discuss how to adapt the results of  \cite{Alekseev:2007re,Alekseev:2007gt,Manko:2007hi,Manko:2008gb,PhysRevD.51.4192}, developed for double Reisner-Nordstr\"om black holes in four dimensions, to our setup.  In section \ref{sec:BaB} and \ref{sec:arbitraryB}, we derive and analyze the specific classes of solutions that contain the desired Schwarzschild-like solutions in six dimensions or in type IIB. In Appendix \ref{App:Profile4d}, we provide the effective four-dimensional theory for our framework. In Appendix \ref{App:tworodansatz}, \ref{App:BaBConstruction} and \ref{App:BubbleConstruction}, we give more details for the interested reader on the technical steps used to generate the solutions. 

\subsubsection*{BPS/ Anti-BPS system}

In section \ref{sec:BaB}, we present a class of solutions obtained from a three-rod system in which the outer rods are collapsed to point particles with BPS and anti-BPS charges. The middle rod induces a neutral bolt where $y_2$, one of the T$^2$ directions, smoothly degenerates.  In six dimensions, this system describes an asymptotically $\IR^{1,3}\times$T$^2$ suspension of a two-charge BPS black hole and a two-charge anti-BPS black hole at the poles of a vacuum bubble, as shown in Fig.\ref{fig:B/aBboundstatesintro}. Physically, the two extremal black holes attract each other but are prevented from collapse by the inherent pressure of the vacuum bubble, in a manner similar to \cite{Elvang:2002br}.

\begin{figure}[h]\centering
   \includegraphics[width=0.65\textwidth]{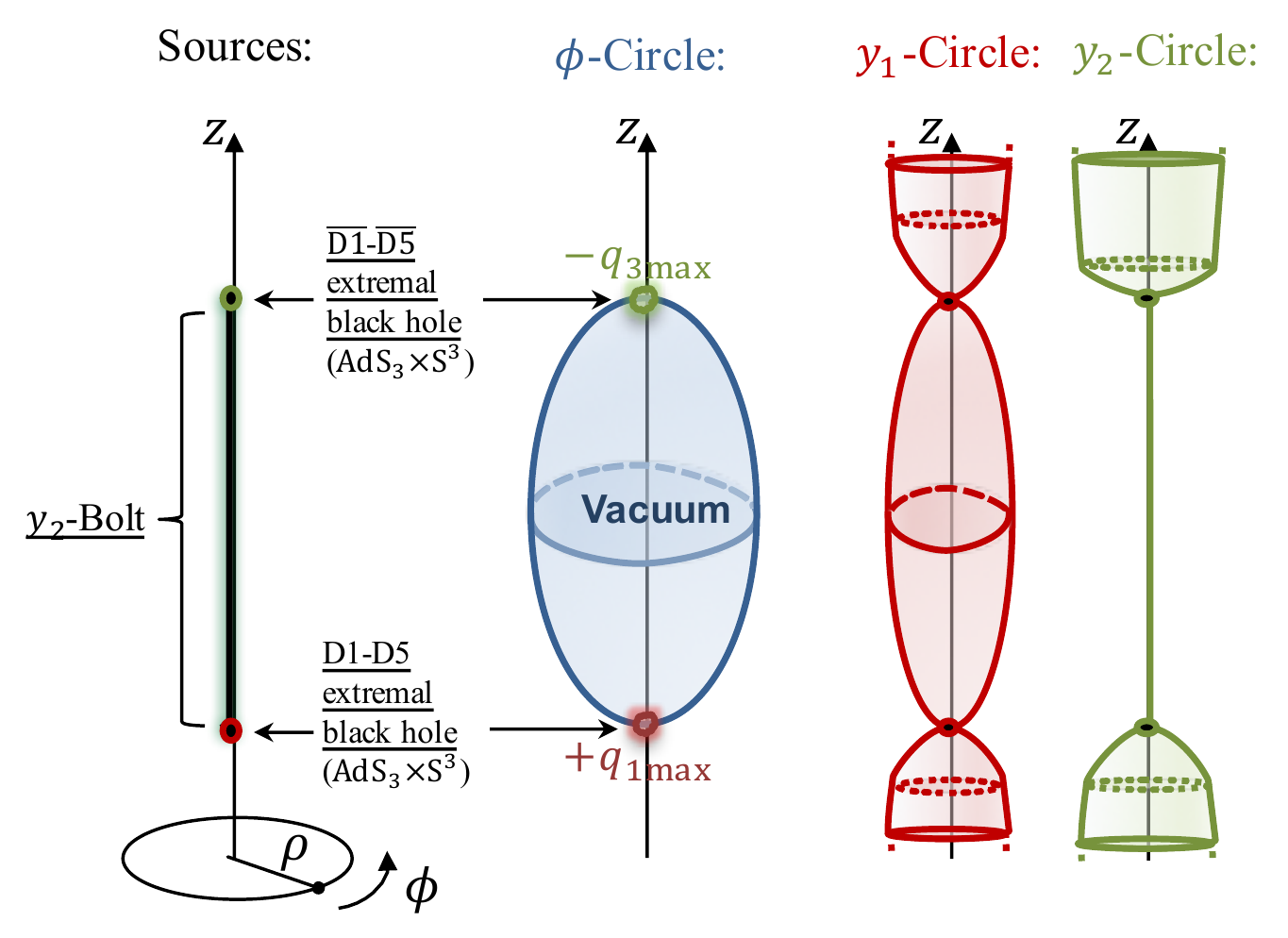}
 \caption{Bound state of BPS D1-D5 branes and anti-BPS $\overline{\text{D1}}$-$\overline{\text{D5}}$ anti-branes separated by a bolt, corresponding to two extremal D1-D5 black holes with charges of opposite sign, held apart by a vacuum bubble. We have represented the profile of the sources and topology on the $z$-axis of the T$^2$ circles, $y_1$ and $y_2$, and the circle of symmetry of the three-dimensional base, $\phi$.}
 \label{fig:B/aBboundstatesintro}
\end{figure}

The near-horizon regions correspond to warped AdS$_3 \times $S$^3$ geometries.  In the type IIB uplift, we identify the near horizons of a BPS D1-D5 black hole in the south pole and a $\overline{\text{D1}}$-$\overline{\text{D5}}$ anti-BPS black hole in the north pole of the vacuum bubble.  These are microscopic black holes with vanishing horizon area for which the microstate structure is well-established \cite{Lunin:2001fv,Taylor:2005db,Mathur:2005ai}. To our knowledge, these are the first solutions which combine them in asymptotically-neutral bound states.

We further specialize to neutral bound states where the charges of opposite signs are equal in magnitude.  This is studied in detail in section \ref{sec:neutralBaB}, and we discuss the Schwarzschild-like nature of the solutions. Excitingly, the four-dimensional ADM mass, $\cM$, can be dialed independently from the asymptotic radius of the T$^2$ circle that shrinks at the vacuum bubble, $R_{y_2}$.  Therefore, we can have macroscopic bound states with arbitrarily small extra dimensions, $\cM \gg R_{y_2}$.

First, we characterize their redshift, i.e. the norm of the timelike killing vector, as plotted in Fig.\ref{fig:B/aBRedshiftintro}. The redshift becomes very large as one approaches the vacuum bubble, and thus the solutions exhibit a Schwarzschild-like throat. It is of order $\cM/R_{y_2}\gg 1$ at the bubble locus and is infinite at its poles. The ``fangs'' in the redshift plot correspond to the infinite AdS$_3$ throats associated to the extremal black holes.  

\begin{figure}[h]\centering
  \includegraphics[width=0.65\textwidth]{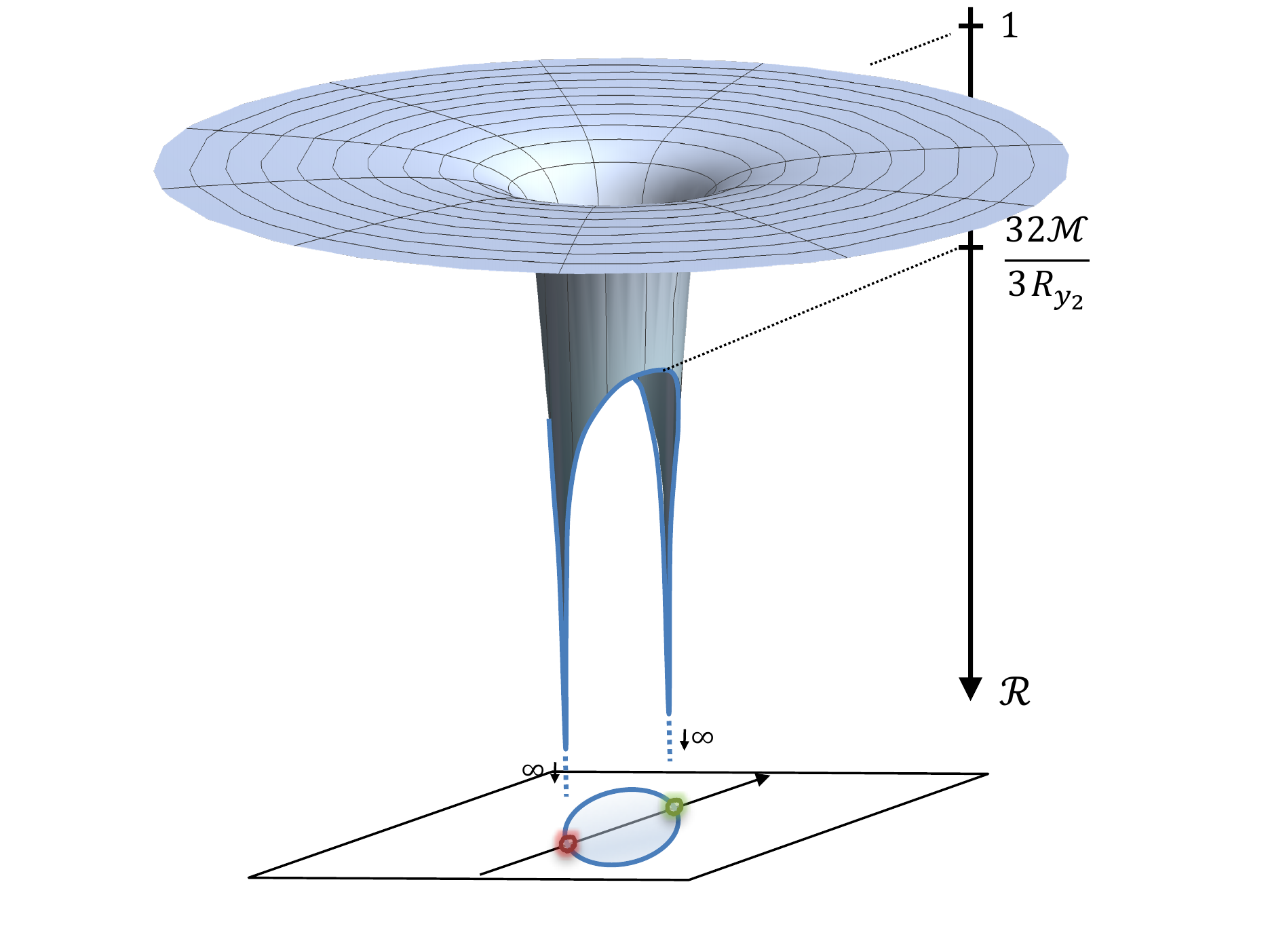}
 \caption{Redshift factor of a neutral bound state of two extremal D1-D5 and $\overline{\text{D1}}$-$\overline{\text{D5}}$ black holes  at the poles a vacuum bubble.}
 \label{fig:B/aBRedshiftintro}
\end{figure}

Second, we also show that the solutions are ultra-compact. We characterize their effective sizes by evaluating the radius of the S$^2$ as we approach the objects. Slightly away from the bubble locus, the geometries have a minimal S$^2$, at a radius 1.52 times its Schwarzschild radius.

Nevertheless, the neutral BPS/anti-BPS D1-D5 black hole bound states show significant differences from a Schwarzschild black hole. While the total charges vanish, the bound states carry small electromagnetic dipole charges of order $\sqrt{R_{y_2}/\cM}\times\cM^2 \ll \cM^2$ which could be observable asymptotically. Moreover, they have large even gravitational multipole moments and the S$^2$ becomes strongly asymmetric very close to the bubble locus.  There are other interesting properties of these solutions, which the impatient reader can jump to section \ref{sec:neutralBaB} to learn about.

\subsubsection*{Neutral bubbling solution} 

In section \ref{sec:arbitraryB}, we present a class of non-BPS bubbling solutions with arbitrary charges. They arise from the same three-rod ansatz as before but with the outer rods now of finite size, away from the BPS bound (see Fig.\ref{fig:BothConfintro}). We force them to describe regular charged bolts where $y_1$, the other T$^2$ direction, degenerates with a conical defect of order $k_1\in\mathbb{N}$. They induce two non-BPS bubbles carrying arbitrary electric and magnetic charges, or D1-D5 brane charges from a type IIB perspective.  The middle rod still induces a vacuum bubble where $y_2$ shrinks.  Our construction leads to smooth horizonless three-bubble geometries, depicted in  Fig.\ref{fig:BothConfintro}, for which the two outer bubbles can carry arbitrary charges. In section \ref{sec:neutralB}, we specialize to the neutral solutions where the charges cancel out asymptotically, thereby defining a class of smooth neutral bubbling geometries with internal flux.

\begin{figure}[h]\centering
   \includegraphics[width=0.55\textwidth]{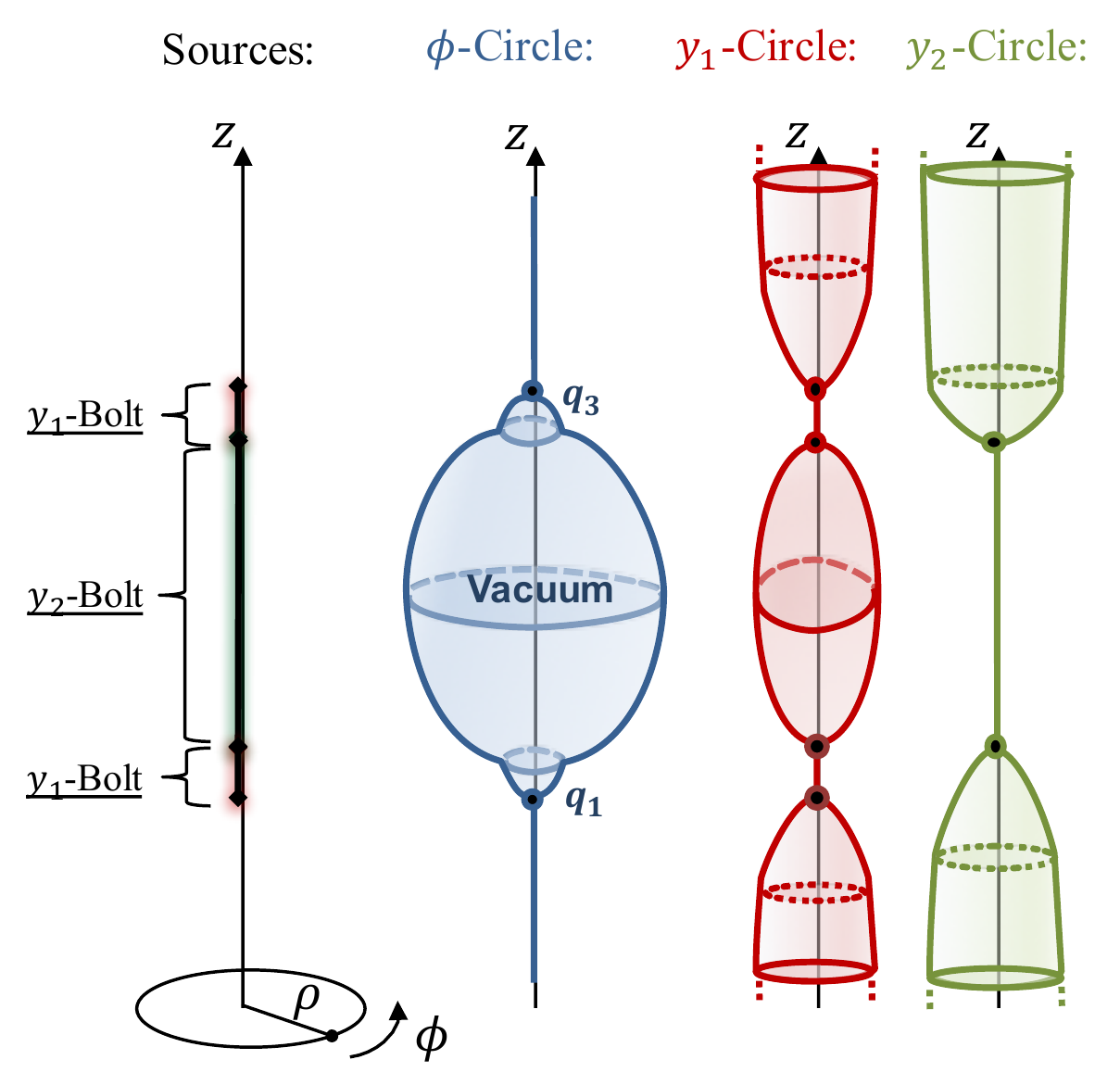}
 \caption{Schematic description of the smooth three-bubble configuration and the behavior of the $\phi$, $y_1$ and $y_2$ circles on the $z$-axis.}
 \label{fig:BothConfintro}
\end{figure}

The solution space is rich with many interesting corners to explore.  The phase space of solutions can be characterized in terms of three asymptotic parameters: the ADM mass $\mathcal{M}$, and the asymptotic radii of the two extra dimensions, $R_{y_1}$ and $R_{y_2}$. An important point is that the ADM mass is constrained to be at the same scale as the radii of the extra dimensions, unless a conical defect is introduced.  Physically, this is a manifestation of two facts.  First, a bubbling solution corresponds to a specific state of the quantum theory.  We expect dimensionful parameters to be fixed in terms of the only scales in the system, i.e the radii $R_{y_a}$.  We can, however, obtain large macroscopic bubbling solution by allowing for a conical deficit of order $k_1$ on the the outer bubbles.  Such an orbifold can be resolved by replacing the fixed points with $(k_1-1)$  Gibbons-Hawking bubbles \cite{Bah:2020ogh,Bah:2020pdz}. In this regime we have families of neutral three-bubble solutions labeled by $k_1$ with $\cM \gg R_{y_2},R_{y_1} $.

We show that as $k_1$ gets larger and larger, $k_1 R_{y_1} \gg \cM$, the outer bubbles get smaller and smaller. The bubbling geometries scale towards the bound states of extremal two-charge black holes of the previous section but resolve both horizons into small non-BPS D1-D5 bubbles with opposite charges. They share therefore the same properties by being Schwarzschild-like geometries, but they are entirely-smooth topological solitons. They have a very large and finite redshift at the bubble loci as depicted in Fig.\ref{fig:BuRedshiftIntro}, and have a minimal S$^2$ of radius 1.52 times bigger their Schwarzschild radius. We study several properties of the three-bubble configurations in section \ref{sec:arbitraryB}.  The interested reader can jump to this section, which is written to be almost self-contained.  

\begin{figure}[h]\centering
   \includegraphics[width=0.65\textwidth]{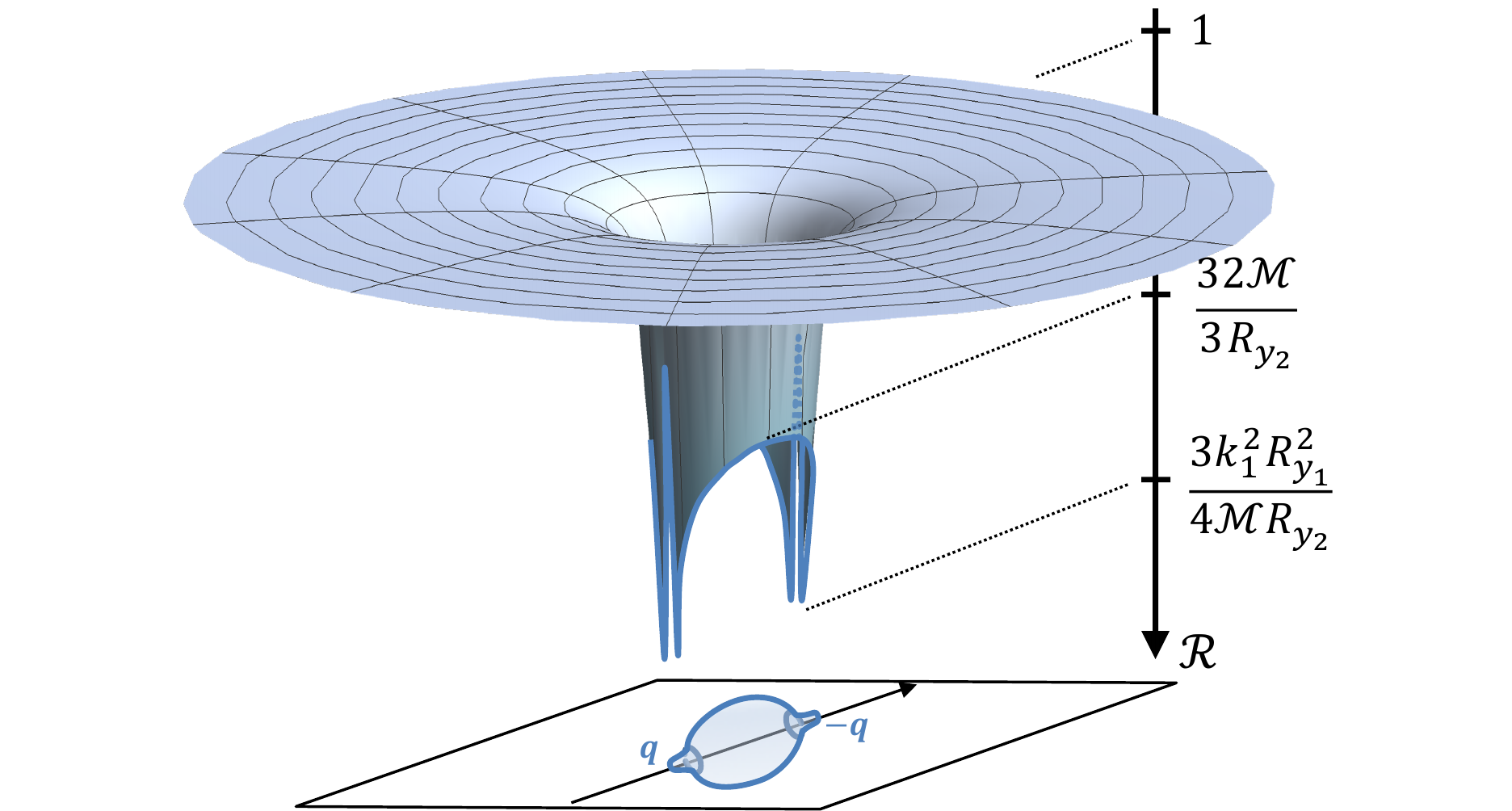}
 \caption{Redshift factor of the neutral three-bubble solutions with $\cM \gg R_{y_1}, R_{y_2}$ and $k_1 R_{y_1} \gg \cM$. The redshift is almost indistinguishable from the BPS/anti-BPS black hole bound states but the divergences are resolved by the two non-BPS bubbles replacing the extremal horizons.}
 \label{fig:BuRedshiftIntro}
\end{figure}

Our solutions are the first examples of states of a quantum gravity theory that are coherent enough to admit classical descriptions as smooth horizonless geometries with internal flux but with the same conserved charges as a Schwarzschild black hole. Other solutions of this type exist without internal flux, such as the bubble of nothing, but we believe that flux is needed for having potentially meta-stable states. First, adding electromagnetic flux to a vacuum bubble removes its instability, and a charged non-BPS bubble, when isolated, has been shown to be a meta-stable vacuum of the theory \cite{Stotyn:2011tv,Bah:2021irr}. Second, even if the present solutions have a vacuum bubble in the middle of the configurations, it has been proposed in \cite{Elvang:2002br} that its pressure to expand can be counterbalanced by the attraction force between the two bodies surrounding it. However, asserting that the present neutral solitons are indeed meta-stable vacua will require more than hints, and we anticipate further studies on the subject.

\subsubsection*{Approximate singular solutions} 

Both classes of solutions introduced in the paper, the smooth bubbling geometries and the extremal black hole bound states, have another surprising scaling property. We show that, as we increase $\cM$ with respect to the extra-dimension radii, $R_{y_1}$ and $R_{y_2}$, both types of solutions can get infinitesimally close to the following singular vacuum geometries: 
\begin{align}
ds_{6}^2 = &\left(1-\frac{4\cM}{3r} \right)\left[- dt^2 + dy_1^2\right] +\  dy_2^2 \nn \\
&\hspace{-0.4cm}+\frac{r^4}{\left[\left( r-\frac{2\cM}{3}\right)^2-\left( \frac{2\cM}{3}\right)^2 \cos^2\theta\right]^2}\left(dr^2+r^2 \left(1-\frac{4\cM}{3r} \right)d\theta^2\right) + r^2 \left(1-\frac{4\cM}{3r} \right)^{-1} \sin^2\theta \,d\phi^2,\nonumber\\
 F_3 \= & 0\,,
\end{align}
Our solutions then resolve the singularity at $r=\frac{4\cM}{3}$ either by a vacuum bubble with two extremal black holes at its poles or with the three-bubble geometries.

These singular geometries are certainly not Schwarzschild geometries, even if they share similar properties. It will be interesting to probe their gravitational footprints since we have now established that they can be classically resolved into well-defined smooth and horizonless solutions. In the future we plan to study the geodesics, scalar wave emission and ultimately gravitational waves of such backgrounds in order to investigate their potential presence in the universe.


\section{Charged Weyl framework}

\label{sec:Weyl6d}

In this section, we detail the framework in which our static axisymmetric solutions will be constructed. As discussed in previous work \cite{Bah:2021owp,Bah:2021rki}, smooth non-BPS bubbling solutions can be obtained from spacetimes with at least two extra compact dimensions in addition to the four infinite dimensions. Our general setup is obtained from the following six-dimensional Einstein theory coupled to a two-form gauge field\footnote{ The norm is given by $
\left|F_3\right|^2 =  \frac{1}{3!}\,F_{3\,\mu\nu\sigma}\,F_3^{\,\mu\nu\sigma}\,.$}
\begin{equation}
 \,S_{6} \= \frac{1}{16 \pi G_6} \int d^6x \sqrt{-\det g}\,\left(R \-  \frac{1}{2} \,\left|F_3\right|^2\,\right)\,,
\label{eq:Action6d}
\end{equation}
where $F_3=dC^{(2)}$ is the field strength of the gauge field. The action can be seen as the minimal pure $\cN = (2, 0)$ six-dimensional supergravity with the extra assumption that $F_3$ is self-dual which we will assume latter on. This theory arises as a consistent
truncation of type IIB supergravity on a rigid T$^4$.

We focus on solutions that are asymptotic to $\IR^{1,3} \times \text{T}^2$. The T$^2$ will be parametrized by $(y_1,y_2)$ with $2\pi R_{y_a}$ periodicities.

\subsection{Ansatz in various dimensions}
\label{sec:EOM}

First, we describe the solution ansatz in the six-dimensional Weyl frame, its embedding in type IIB supergravity and its reduction to four dimensions.  We consider solutions with magnetic and electric charges for $F_3$ along a circle of the torus with the Weyl ansatz\footnote{In previous work \cite{Bah:2021owp,Bah:2021rki}, a KK magnetic vector along $y_2$ has also been turned on. For the purpose of this paper, the vector has been turned off as it will not be used.}
\begin{align}
ds_{6}^2 = &\frac{1}{Z} \left[- W_1\,dt^2 + \frac{ dy_1^2}{W_1} \right] +Z\left[ \frac{1}{W_2}\, dy_2^2 + W_2 \,\left(e^{2(\nu_Z+\nu_W)} \left(d\rho^2 + dz^2 \right) +\rho^2 d\phi^2\right) \right],\nonumber\\
 F_3 \= & d\left[ H \,d\phi \wedge dy_2 \+ T \,dt \wedge dy_1 \right]\,,\label{eq:WeylSol2circle}
\end{align}
where $(\rho,z,\phi)$ are the cylindrical Weyl's coordinates of an asymptotically-$\IR^3$ base.  The warp factors $(Z,W_1,W_2,\nu_Z,\nu_W)$ and  gauge potentials $(H,T)$ are functions of $\rho$ and $z$ alone. 

\noindent We restrict\footnote{This restriction is just for convenience and for a simpler type IIB description.  If lifted, it leads to a decomposition of $Z$ into two decoupled branches $Z=\sqrt{Z_H Z_T}$ where $Z_H$ and $Z_T$ are coupled to $H$ and $T$ respectively. See \cite{Heidmann:2021cms} for more details.}  $F_3$ to be self-dual, $\star F_3 =F_3$, which implies
\begin{equation}
dH \=\rho \,Z^2\,\star_2 dT\qquad \Leftrightarrow \qquad F_3 \=dT \wedge dt \wedge dy_1+\rho \,Z^2\,(\star_2 dT)\wedge d\phi \wedge dy_2,
\label{eq:EMdual}
\end{equation}
where $\star_2$ is the Hodge dual in the flat $(\rho,z)$-subspace. 

\subsubsection*{Type IIB supergravity embedding}
The system above can be embedded in type IIB supergravity such as
\begin{equation}
ds_{10}^2 = ds_6^2 + dx_1^2+dx_2^2+dx_3^2+dx_4^2 \,,\qquad F_3=dC^{(2)} \,,
\label{eq:typeIIBembedding}
\end{equation}
where the $x_a$ parametrize the directions of the T$^4$, $C^{(2)}$ corresponds to the Ramond-Ramond two-form gauge field, and all other type IIB fields are zero. From this perspective, $H$ corresponds to a D5-brane magnetic potential while $T$ is a D1-brane electric potential where the common direction of the D1 and D5 branes is $y_1$. Our six-dimensional set-up describes then \emph{axisymmetric and static D1-D5 systems in type IIB}. Moreover, the self-duality of $F_3$ implies that all D1-D5 sources of the solutions carry \emph{equal D1 and D5 brane charges}. The asymptotic type IIB charges in unit of volume are given by
\begin{equation}
\cQ_{D1} \= \frac{1}{8\pi^2 R_{y_2} \text{Vol}(T^4)} \,\int_{S_\infty^2 \times S^{y_2}\times T^4}\,\star_{10} F_3 \quad = \quad \cQ_{D5} \=  \frac{1}{8\pi^2 R_{y_2}} \,\int_{S_\infty^2 \times S^{y_2}}\,F_3\,.
\label{eq:typeIIBcharges}
\end{equation}

\subsubsection*{Kaluza-Klein reduction and conserved charges}

We give the profile of the solution in four dimensions after reduction on the T$^2$. The truncation of the six-dimensional theory \eqref{eq:Action6d} is detailed in Appendix \ref{App:Profile4d}. In four dimensions, solutions given by \eqref{eq:WeylSol2circle} correspond to geometries with two scalars arising from the metric components along the T$^2$ and two electromagnetically dual one-form gauge fields:
\begin{align}
ds_4^2 &=  -\sqrt{\frac{W_1}{W_2}}\,\frac{dt^2}{Z}  \+Z\,\sqrt{\frac{W_2}{W_1}}\,\left[e^{2(\nu_Z+\nu_W)} \left(d\rho^2 + dz^2 \right) +\rho^2 d\phi^2 \right] , \label{eq:4dFrameworkcharged}\\
e^{\sqrt{3}\,\Phi_1} &= Z\,\sqrt{W_1^3 W_2}, \qquad e^{\sqrt{\frac{3}{2}}\Phi_2}  = \frac{W_2}{Z}, \qquad F^{(m)} \= dH \wedge d\phi\,,\qquad  F^{(e)} = dT \wedge dt\,. \nn
\end{align}

To compute the conserved charges, we expand the solution at large distance $r\to \infty$, where $r$ is the asymptotic spherical coordinates, $ 
\rho \equi r \, \sin \theta \,,\,\, z \equi r \, \cos \theta$. With the conventions of \cite{Myers:1986un}, the conserved quantities can be read from the expansion as follows
\begin{align}
\frac{1}{Z}\sqrt{\frac{W_1}{W_2}} &\,\sim\, 1 - \frac{2G_4 \cM}{r}\,,\quad F^{(m)} \sim -\cQ_m \,\sin\theta  \,d\theta \wedge d\phi \,,\quad  F^{(e)} \sim  \frac{\cQ_e}{r^2} \,dt\wedge dr \,,
 \label{eq:AsymptoticExpGen}
\end{align}
where $\cM$ is the four-dimensional ADM mass, $\cQ_e$ is the electric charge, and $\cQ_m$ is the magnetic charge. They are given in unit of the four-dimensional electromagnetic coupling $e= \sqrt{16\pi G_4}$, where $G_4 \equiv \frac{G_6}{(2\pi)^2 R_{y_1}R_{y_2}} $. Because of self-duality, the solutions have equal electric and magnetic charges $\cQ_e=\cQ_m$. Moreover, they can be directly related to the asymptotic D1 and D5 brane charges in unit of volume \eqref{eq:typeIIBcharges} such that $\cQ_m=\cQ_{D5}$ and $\cQ_e=\cQ_{D1}$.

\subsubsection*{Equations of motion}

The six-dimensional Einstein-Maxwell equations can be decomposed into the following sectors:
\begin{align}
&\text{\underline{Vacuum sector:}} \quad \Delta \log W_I  \= 0\,, \qquad \partial_z \nu_W \= \frac{\rho}{2}\,  \sum_{I=1}^2\partial_\rho \log W_I\,\partial_z \log W_I,  \nonumber\\
& \hspace{2.90cm} \partial_\rho \nu_W \= \frac{\rho}{4}\,  \sum_{I=1}^2 \left( (\partial_\rho \log W_I)^2- (\partial_z \log W_I )^2\right) \,, \nn \\
&\text{\underline{Maxwell sector:}} \quad  \Delta \log Z \= \- \rho\, Z^2\,\left[ (\partial_\rho T)^2 + (\partial_z T)^2 \right] \,,\label{eq:EOMWeyl}\\
&\hspace{2.90cm} \partial_\rho \left( \rho\, Z^2\,\partial_\rho T \right)\+\partial_z \left( \rho\, Z^2\,\partial_z T\right)  \=0\,,\nonumber\\
&\hspace{2.90cm}  \partial_z \nu_Z \=\rho \,\partial_\rho \log Z\,\partial_z \log Z -\rho\, Z^2\,\partial_\rho T \partial_z T\,, \nn\\
&\hspace{2.90cm}  \partial_\rho \nu_Z \= \frac{\rho}{2} \left( ( \partial_\rho \log Z)^2- (\partial_z \log Z )^2\right)- \frac{\rho\, Z^2}{2} \left( (\partial_\rho T)^2-( \partial_z T )^2\right)\,, \nonumber
\end{align}
where $\Delta$ is the cylindrical Laplacian for a flat three-dimensional base
\begin{equation}
\Delta \equi \frac{1}{\rho} \,\partial_\rho \left( \rho \,\partial_\rho \right) + \partial_z^2 \,.
\end{equation}

\subsection{Solution scheme}
\label{sec:SolScheme}

In this paper, we are interested in the construction and study of solutions to the previous system of equations which are regular, neutral, and have internal flux. They will be induced by three sources, two of which may carry electromagnetic charges, or D1-D5 brane charges from a type IIB perspective. The details of our approach to solving the equations are given in Appendix \ref{App:tworodansatz}. In this section, we simply discuss the solution scheme before explicitly presenting solutions.

\begin{figure}
\centering
\begin{tabular}{ccc}
\addlinespace[1ex]
\subf{\includegraphics[height=75mm]{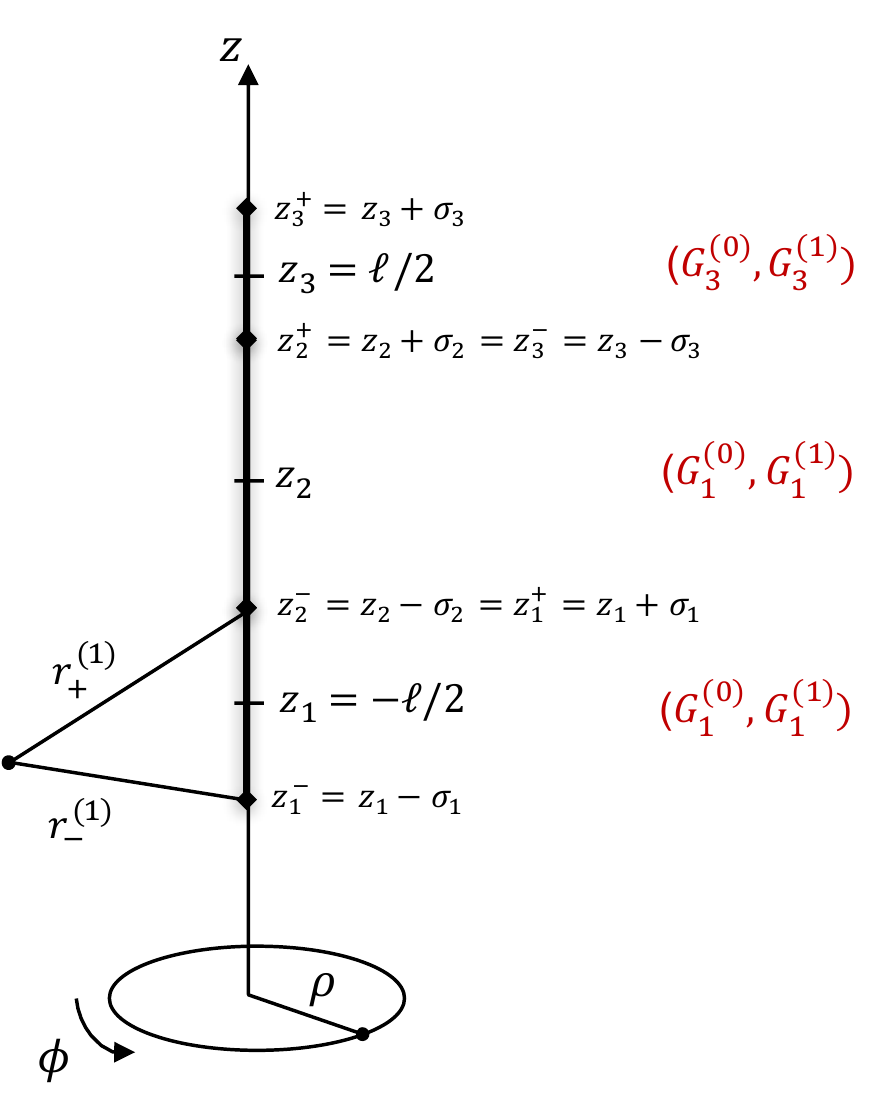}}
     {\\
     (a) The three connected rods for the \\ vacuum sector $(W_1,W_2,\nu_W)$.}
&
\subf{\includegraphics[height=75mm]{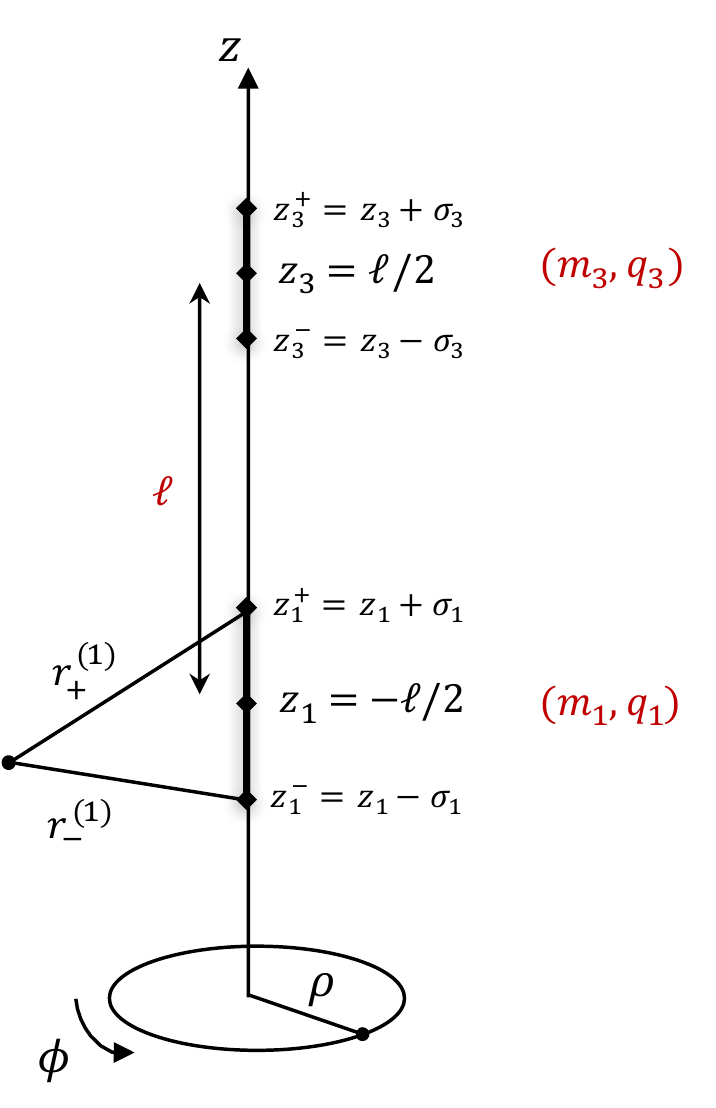}}
     {\\
     (b) The two dyonic rod sources for the \\ Maxwell sector $(Z,T,H,\nu_Z)$.}
\\
\addlinespace[1ex]
\end{tabular}
\noindent\hfil\rule{0.7\textwidth}{.6pt}\hfil
\caption{Schematic description of the rod sources used to generate the three-body solutions in the paper.}
\label{fig:rodsources}
\end{figure}

\subsubsection{Solving the vacuum sector}

It is important to note that the $W_I$ are pure ``vacuum'' warp factors, in that they are not related to the gauge potentials, whereas $Z$ is generated and induced by $T$. The vacuum warp factors and their corresponding base warp factor $\nu_W$ satisfy harmonic equations of motion, $\Delta \log W_I =0$, associated to vacuum Weyl solutions \cite{Weyl:book,Emparan:2001wk}. Generic solutions are induced by an arbitrary number $n$ of ``rods'', that is segments on the $z$-axis of size $2\sigma_i$ centered around $z=z_i$. We consider solutions induced by a maximum of $n=3$ connected rod sources (see Fig.\ref{fig:rodsources}(a)). The distance to the rod endpoints, $r_\pm^{(i)}$, and the spherical coordinates centered around each rod, $(r_i,\theta_i)$, are given by
\begin{equation}\label{eq:r_defs}
\begin{split}
r_\pm^{(i)} &\equi \sqrt{\rho^2+(z-(z_i\pm\sigma_i))^2}\,,\qquad r_i \equi \frac{r_+^{(i)}+r_-^{(i)}}{2}\,,\qquad \cos \theta_i \equi \frac{r_-^{(i)}-r_+^{(i)}}{2\sigma_i}\,.
\end{split}
\end{equation}
The functions $\log \frac{r_i + \sigma_i}{r_i - \sigma_i} $ satisfy $\Delta\left(\log \frac{r_i + \sigma_i}{r_i - \sigma_i}  \right)=0$, and can be used as the building blocks for $W_I$ and we find $\nu_W$ by integrating \eqref{eq:EOMWeyl}:
\begin{equation}
W_I = \prod_{i=i}^3\left(1-\frac{2\sigma_i}{r_i + \sigma_i}\right)^{- G^{(I)}_i}\,, \qquad e^{2\nu_W} \=\prod_{i,j=1}^3\, \left(  \frac{E_{+-}^{(i,j)}E_{-+}^{(i,j)}}{E_{++}^{(i,j)}E_{--}^{(i,j)}}\right)^{\frac{1}{2}\,\left(G_i^{(1)}G_j^{(1)}+G_i^{(2)}G_j^{(2)} \right)}  \,,
\label{eq:HarmFuncGen}
\end{equation}
where the constants, $G^{(I)}_i$, define the \emph{two weights} of the $i^\text{th}$ rod, and we have introduced the generating functions
\begin{equation}
E_{\pm \pm}^{(i,j)} \equi r_\pm^{(i)} r_\pm^{(j)} + \left(z-(z_i \pm \sigma_i)  \right)\left(z-(z_j \pm \sigma_j)  \right) +\rho^2\,.
\end{equation} 
At the $i^\text{th}$ rod, $r_i-\sigma_i\to 0$ ($\rho=0$, $z_i-\sigma_i \leq  z \leq z_i+\sigma_i$), the warp factors are singular $W_I \sim (r_i-\sigma_i)^{- G^{(I)}_i}$, $e^{2\nu_W}\sim(r_i-\sigma_i)^{{G_i^{(1)}}^2+{G_i^{(2)}}^2}$. For specific choices of weights, these singularities can become regular coordinate singularities where the $y_1$ or $y_2$ fibers degenerate, inducing \emph{bolts}, i.e. smooth ``bubble'' loci \cite{Bah:2021owp,Bah:2021rki}.

\subsubsection{Solving the Maxwell sector}

The Maxwell sector is a non-trivial set of coupled nonlinear differential equations. In \cite{Bah:2020pdz,Bah:2021owp,Bah:2021rki,Heidmann:2021cms}, a procedure was found to extract linear solutions in closed form without resorting to BPS solutions. This allowed the generation of large families of non-BPS bubbling solutions with an arbitrary number of smooth bolts on the $z$-axis in six dimensions \cite{Bah:2021owp,Bah:2021rki} or in string theory backgrounds \cite{Heidmann:2021cms}. However, the price to pay for linearity was that all sources had the same mass-to-charge ratio. While it was already remarkable that one could construct bound states of bubbles with less charge than mass, the total asymptomatic charge in these solutions was always non-vanishing. To obtain neutral solitons, it was necessary that all the internal bubbles were vacuum bubbles, known to be unstable \cite{Witten:1981gj}. 

We will get around this problem by leaving linearity aside and using known solutions to the Maxwell sector of \eqref{eq:EOMWeyl}, which correspond to sources that can have different mass-to-charge ratios across different rods. More precisely, we will adapt known results about superposition of two arbitrary Reissner-Nordstrom black holes in four-dimensional GR \cite{Alekseev:2007re,Alekseev:2007gt,Manko:2007hi,Manko:2008gb,PhysRevD.51.4192}. These solutions satisfy similar equations as our Maxwell sector for $(Z,H,T,\nu_Z)$. They correspond to two rod sources of size $2\sigma_i$, that we label by $i=1$ and $i=3$ (see Fig.\ref{fig:rodsources}(b)), with a distance $\ell$ between their centers such that
\begin{equation}
\begin{split}
\sigma_1 \equi \sqrt{m_1^2-q_1^2+2 q_1 \gamma }\,,\qquad  \sigma_3 \equi \sqrt{m_3^2-q_3^2-2 q_3 \gamma }\,,\qquad \gamma \equi \frac{m_3 q_1 \- m_1 q_3}{\ell+m_1+m_3} \,,
\end{split} \label{eq:parameters}
\end{equation}
where $m_i\geq 0$ are mass parameters while $q_i$ are associated to the equal magnetic and electric charges at the sources, or equivalently, the equal D1 and D5 charges carried by the sources. In order to have well-separated sources one needs $\ell\geq \sigma_1+\sigma_3$, and we further restrict\footnote{This restriction is motivated by having a branch of solutions where rods carry opposite charges with the same magnitude.}
\begin{equation}
\ell \geq m_1 +m_3\,.
\end{equation}
The expressions for $(Z,H,T,\nu_Z)$ are detailed in the Appendix \ref{App:tworodansatz}, \eqref{eq:two-rod_sols}  and \eqref{eq:tworodsMagneticPot}. We have therefore five-parameter solutions for $(Z,H,T,\nu_Z)$ that describe two generically non-BPS D1-D5 sources in type IIB. 

The reality conditions for the $\sigma_i$ \eqref{eq:parameters} lead to a BPS bound for the charged sources.  These BPS bounds are related to a supersymmetry conservation in type IIB. In Fig.\ref{fig:qregime}, we illustrate the allowed values for the D1-D5 charges $q_i$ for given $m_i$ and $\ell$ consistent with \eqref{eq:parameters}. They are contained in a ``diamond'' defined by four apexes given by:
\begin{equation}
(q_1,q_3) = \pm (m_1,m_3) \,,\qquad (q_1,q_3)= \pm (-q_{1\text{max}},q_{3\text{max}}),
\end{equation}
where $q_{i\text{max}}$ are the maximum values of the charges:
\begin{equation}
q_{i\text{max}} \equi m_i \,\sqrt{\frac{(\ell+m_j)^2-m_i^2}{(\ell-m_j)^2-m_i^2}}\,,\qquad j\neq i.
\label{eq:qmax}
\end{equation}

\begin{figure}[h]\centering
   \includegraphics[width=0.5\textwidth]{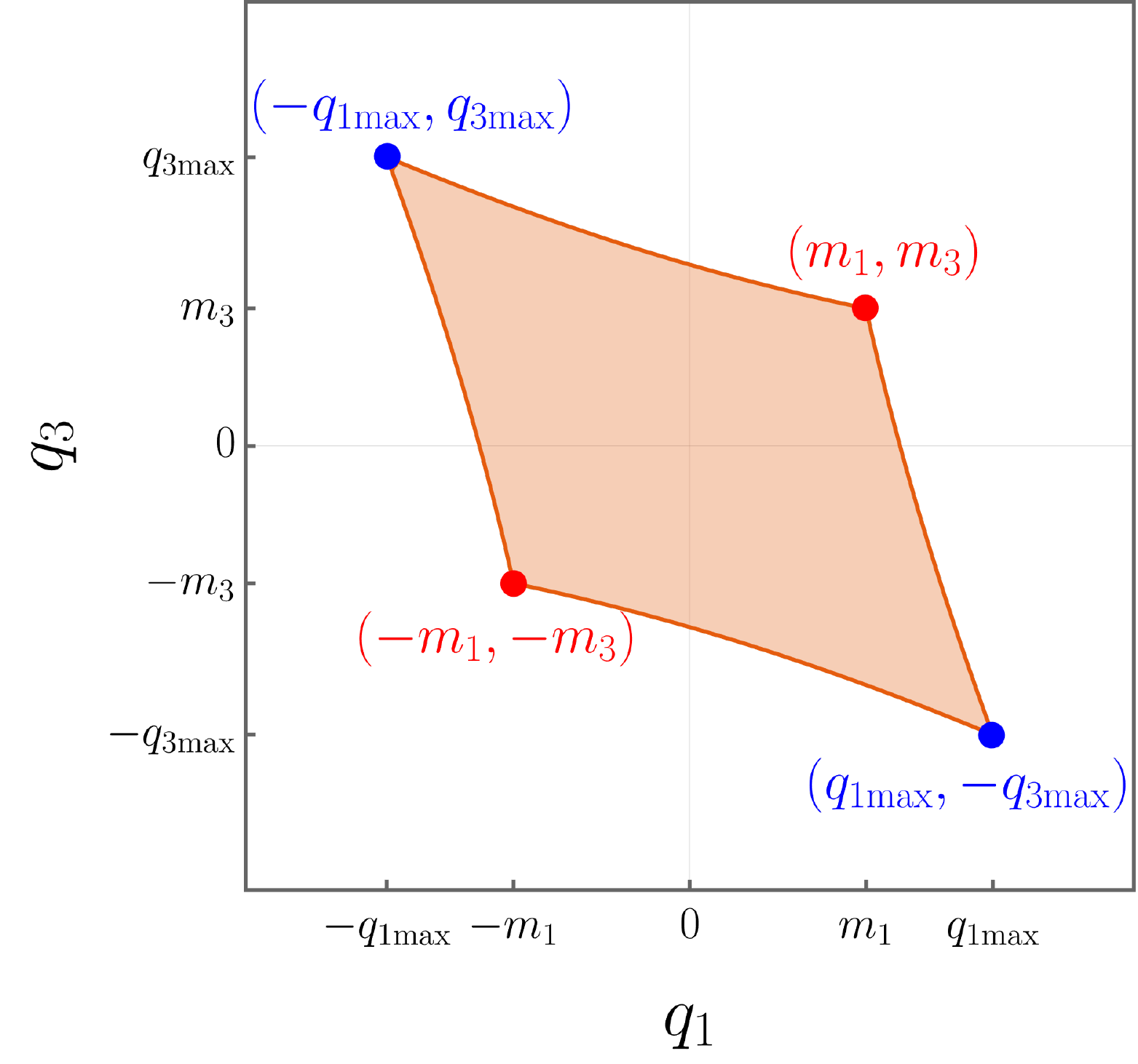}
 \caption{Parameter space of the charges $q_i$ for given $(\ell,m_1,m_3)$ and their BPS bounds.}
 \label{fig:qregime}
\end{figure}

We refer the reader to the Appendix \ref{App:qRegimes} for a detailed analysis of the regimes of charges, which we summarize here.

Inside the diamond, both $\sigma_i$ are nonzero and the charges do not saturate any BPS bounds. Therefore, the warp factors and gauge potentials $(Z,H,T,\nu_Z)$ correspond to a dyonic configuration of two non-BPS objects. In type IIB, they define two non-supersymmetric D1-D5 sources.  At this level, these sources have no physical description. This will require a specific choice for the warp factors $(W_1,W_2,\nu_W)$ \eqref{eq:HarmFuncGen} in six \eqref{eq:WeylSol2circle} or ten \eqref{eq:typeIIBembedding} dimensions in order to have objects that are regular non-BPS physical sources.  We will successfully make these objects correspond to \emph{smooth non-BPS D1-D5 bubbles in type IIB} in section \ref{sec:arbitraryB}.

At each edge of the diamond, one of the $\sigma_i$ is strictly equal to zero, and its corresponding charge saturates a BPS bound. The corresponding rod shrinks to zero size and become a point source. More precisely, one of the pair of sources becomes BPS while the other one remains non-BPS. 

At the four apexes, both $\sigma_i$ vanish and $(Z,H,T,\nu_Z)$ are sourced by two BPS point particles separated by a distance $\ell$. From a type IIB perspective, each source corresponds to a BPS D1-D5 brane or $\overline{\text{D1}}$-$\overline{\text{D5}}$ anti-brane point source, depending on the sign of the charges. The four apexes in the parameter space are the four possible BPS brane/antibrane configurations. More precisely, for $(q_1,q_3)=\pm (m_1,m_3) $, the $(Z,H,T,\nu_Z)$ are induced by two supersymmetric D1-D5 point sources (for the ``$+$'' solutions) or $\overline{\text{D1}}$-$\overline{\text{D5}}$ point sources (for the ``$-$'' solutions), while for $(q_1,q_3)= \pm (-q_{1\text{max}},q_{3\text{max}})$ they are induced by a bound state of a BPS D1-D5 point source with a anti-BPS $\overline{\text{D1}}$-$\overline{\text{D5}}$  point source. Note that, for the second scenario, $q_i$ is greater than $m_i$, and the BPS sources are carrying more charge than mass. This is due to the fact that the sources are not isolated and part of their irreducible masses become binding energy of the bound state. Once again, these BPS / anti-BPS sources have no physical descriptions in type IIB yet since one needs to decorate the solutions with appropriate $(W_1,W_2,\nu_W)$ \eqref{eq:HarmFuncGen} to have regular solutions. In the next section, we will successfully construct \emph{BPS D1-D5/ anti-BPS $\overline{\text{D1}}$-$\overline{\text{D5}}$ black hole bound states separated by a vacuum bubble}.


\section{Bound states of extremal D1-D5 and $\overline{\text{D1}}$-$\overline{\text{D5}}$ black holes on a bubble}
\label{sec:BaB}

In this section, we study the solutions obtained from the methods outlined in section \ref{sec:SolScheme} that correspond to bound states of BPS D1-D5 branes and $\overline{\text{D1}}$-$\overline{\text{D5}}$ anti-branes sitting at the poles of a vacuum bubble, as depicted in figure \ref{fig:B/aBboundstates}. The sources of the branes and the anti-branes  define the loci of extremal two-charge black holes. The details of the construction can be found in the appendix \ref{App:BaBConstruction}. We start by describing the generic bound states and their geometry before focusing specifically on configurations with vanishing net charge. The latter neutral solutions are compared with Schwarzschild geometries.

\subsection{BPS $-$ anti-BPS D1-D5 branes on a vacuum bubble}
We source the potentials $(Z,H,T,\nu_Z)$ in \eqref{eq:WeylSol2circle} with two BPS point particles with mass parameters $(m_1,m_3)$ and charges $(q_1,q_3)=  (q_{1\text{max}},-q_{3\text{max}})$ \eqref{eq:qmax}.  To support such configurations, we also source the potentials $(W_2,\nu_W)$ \eqref{eq:HarmFuncGen} with a rod between the particles as depicted in figure \ref{fig:B/aBboundstates}.  The latter object will induce a smooth bubble along the $z$-axis where the $y_2$ circle degenerates.  The potential $W_1$ is not sourced and is fixed as $W_1=1$.  

\begin{figure}[h]\centering
   \includegraphics[width=0.65\textwidth]{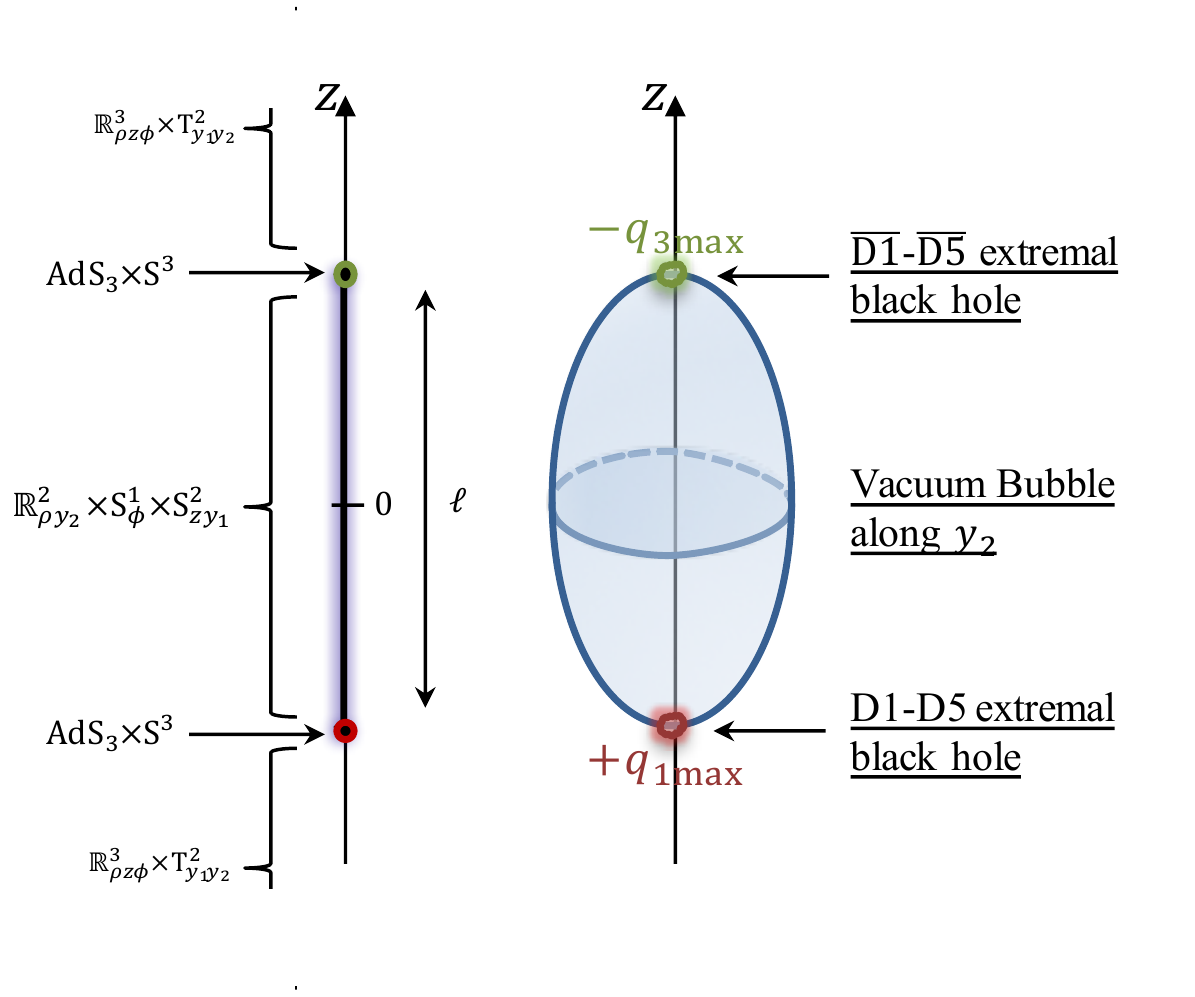}
 \caption{Profile of the sources and topology on the $z$-axis for a bound state of D1-D5 branes and $\overline{\text{D1}}$-$\overline{\text{D5}}$ anti-branes separated by a bolt. Such solutions correspond to two extremal D1-D5 black holes of opposite sign of charges hold apart by a vacuum bubble.}
 \label{fig:B/aBboundstates}
\end{figure}

The six-dimensional metric and flux \eqref{eq:WeylSol2circle} are given by
\begin{align}
ds_{6}^2 = &\frac{1}{Z} \left[- dt^2 + dy_1^2\right] + Z\left[ \frac{1}{W_2}\, dy_2^2 + W_2 \,\left(e^{2(\nu_Z+\nu_W)} \left(d\rho^2 + dz^2 \right) +\rho^2 d\phi^2\right) \right],\nonumber\\
 F_3 \= & d\left[ H \,d\phi \wedge dy_2 \+ T \,dt \wedge dy_1 \right]\,,\label{eq:BaBgeneric}
\end{align}
with 
\begin{align}
Z &\= 1+ \frac{\left(m_1^2-m_3^2 \right)(m_3 r_1-m_1 r_3)+\ell^2 (m_3 r_1+m_1 r_3+2 m_1 m_3)}{(m_3 r_1-m_1 r_3)(m_1 r_1-m_3 r_3)+\ell^2 (r_1 r_3-m_1 m_3)} \,,\hspace{1cm} \nn\\
W_2 & \= \left(1- \frac{2 \ell}{r_1+r_3+\ell} \right)^{-1} \, , \nn
\end{align}
\begin{align}
 T &= \frac{\sqrt{\left(\ell^2-(m_1+m_3)^2 \right)\left(\ell^2-(m_1-m_3)^2 \right)}\,\left( m_1 r_3-m_3 r_1 \right)}{(m_3 r_1-m_1 r_3)(m_1 r_1-m_3 r_3)+\ell^2 (r_1 r_3-m_1 m_3)}\,\frac{1}{Z}\,,\label{eq:WarpFactorBaB}\\
  H &\= \,\frac{\sqrt{\left(\ell^2-(m_1+m_3)^2 \right)\left(\ell^2-(m_1-m_3)^2 \right)}}{2} \, \nn\\
  &\hspace{0.5cm}\times \frac{(r_1 \cos \theta_1 +r_3 \cos \theta_3) (m_3 r_1-m_1 r_3)-\ell (m_3 r_1+m_1 r_3+2 r_1 r_3)}{(m_3 r_1-m_1 r_3)(m_1 r_1-m_3 r_3)+\ell^2 (r_1 r_3-m_1 m_3)}\,, \nn\\
 e^{2(\nu_Z+\nu_W)} &\=\frac{(r_1+r_3)^2-\ell^2}{4r_1r_3} \,\left( \frac{(m_3 r_1-m_1 r_3)(m_1 r_1-m_3 r_3)+\ell^2 (r_1 r_3-m_1 m_3)}{\left(\ell^2-(m_1-m_3)^2 \right)\,r_1\,r_3} \right)^2\,, \nn
\end{align}
where $(r_1,\theta_1)$ and $(r_3,\theta_3)$ are the spherical coordinates centered around the two point sources\footnote{They are identical to the coordinates given by \eqref{eq:r_defs} by taking the extremal limit $\sigma_1,\sigma_3 \to 0$.} given in terms of $ z^\ell_\pm = z \pm \frac{\ell}{2}$ as
\begin{equation}
r_1=\sqrt{\rho^2+{z^\ell_+}^2}, \qquad \cos\theta_1 \= \frac{z^\ell_+}{r_1}, \qquad r_3=\sqrt{\rho^2+{z^\ell_-}^2}, \qquad \cos\theta_3 \= \frac{z^\ell_-}{r_3}.  
\label{eq:distanceBaB}
\end{equation}
The solutions are asymptotically $\IR^{1,3}\times$T$^2$, and the ADM mass \eqref{eq:AsymptoticExpGen} and D1 and D5 brane charges \eqref{eq:typeIIBcharges} are given by
\begin{equation}
\begin{split}
\cM &= \frac{\ell+2 (m_1+m_3)}{4 G_4}, \\ \cQ_{D1} = \cQ_{D5} &= q_{1\text{max}}-q_{3\text{max}} = \left( m_1-m_3\right) \sqrt{\frac{\ell^2-(m_1+m_3)^2}{\ell^2-(m_1-m_3)^2}}. \end{split}
\label{eq:ConservedChargesBaB}
\end{equation}

One can check that the solutions are regular out of the $z$-axis since $Z,W_2,e^{2(\nu_W+\nu_Z)}>0$ for $\rho\neq 0$. However, the warp factors are singular at the sources, $\rho=0$ and $-\ell/2\leq z\leq \ell/2$, while the $\phi$-circle degenerates elsewhere on the $z$-axis, requiring a careful analysis of the topology and regularity on this axis.

\subsubsection{Topology and regularity on the $z$-axis}
\label{sec:TopBaB}

We split the analysis according to the different loci depicted in Fig.\ref{fig:B/aBboundstates}.

\begin{itemize}
\item[•] \underline{Geometry above and below the sources:}

In the region $\rho=0$ and $|z|>\ell/2$, both $Z$ and $W_2$ are non-zero. Therefore, the $\phi$-circle degenerates as the usual cylindrical coordinate degeneracy on the $z$-axis. The regularity then reduces to the study of the three-dimensional base, $ds_3^2= e^{2(\nu_Z+\nu_W)} \left(d\rho^2 + dz^2 \right) +\rho^2 d\phi^2$, where a conical singularity can appear if $e^{2(\nu_Z+\nu_W)}\neq 1$. However, one can check that $e^{2(\nu_Z+\nu_W)}=1$ on these segments, guaranteeing that $\phi$ degenerates smoothly and the semi-infinite segments have the topology of $\IR^3\times$T$^2$.

\item[•] \underline{Geometry at the vacuum bubble:}

To best describe the topology on the segment $\rho=0$ and $ |z| <\ell/2$, we adopt the following local spherical coordinates:
\begin{equation}
\rho=\rho_2\,\sqrt{\rho_2^2+\ell}\, \sin \theta_2, \qquad  z=\left(\rho_2^2+\frac{\ell}{2}\right) \cos \theta_2,
\end{equation}
and consider $\rho_2\to 0$, $0<\theta_2< \pi$.
The time slices of the metric and three-form flux \eqref{eq:BaBgeneric} give
\begin{align}
ds_6^2 \bigl|_{dt=0}&\,\propto\, d\rho_2^2 + \frac{\rho_2^2}{4\ell^2 C_2^2}\,dy_2^2  + \frac{\ell}{4} \left(d\theta_2^2 + \frac{\sin^2\theta_2}{C_2^2} \left( d\phi^2 + f_2(\theta_2)\sin^2\theta_2\, dy_1^2\right)\right),\nn\\
F_3 &\= f_1(\theta_2) \,\sin \theta_2 \,d\theta_2 \wedge dt \wedge dy_1\+ \cO(\rho_2)\,,
\end{align}
where $f_a(\theta_2)$ are well-behaved functions of $\theta_2$, and $C_2$ is a constant, $$C_2 \equi \frac{\ell^2-(m_1+m_3)^2}{\ell^2-(m_1-m_3)^2}.$$ The local geometry therefore corresponds to a bolt where the $(\theta_2,\phi,y_1)$ subspace defines a S$^2\times$S$^1$ bubble. The $(\rho_2,y_2)$ subspace gives a smooth origin of $\IR^2$ if the parameters are fixed in terms of the $2\pi R_{y_2}$ periodicity of the $y_2$-circle as
\begin{equation}
R_{y_2} \=\frac{2\ell(\ell^2-(m_1+m_3)^2)}{\ell^2-(m_1-m_3)^2}\,.
\label{eq:RegBaB}
\end{equation}
The flux is not sourced at the bolt and is regular in this region. To conclude, the segment corresponds to \emph{a vacuum  S$^1\times$S$^2$ bubble}. The poles of the  S$^2$, $\theta_2=0,\pi$, highlights special loci since the S$^1$ also shrinks here. 

\item[•] \underline{Geometry at the extremal black holes:}

We first investigate the geometry at the North pole, $\rho=0$, $z=\ell/2$, in the following local coordinates:
\begin{equation}
\rho \= \Lambda \, \rho_3^2 \,\sin(2\tau_3)\,,\quad z \=\frac{\ell}{2}+\Lambda \, \rho_3^2 \,\cos(2\tau_3)\,, \quad \Lambda \equiv  \frac{4\ell m_3^2 (\ell+m_1+m_3)^2}{(\ell-m_1+m_3)^2} 
\label{eq:localCoorBaB}
\end{equation}
and consider $\rho_3 \to 0$, $0\leq \tau_3 \leq \frac{\pi}{2}$. The metric and three-form flux \eqref{eq:BaBgeneric} give\footnote{We have used the regularity condition \eqref{eq:RegBaB} to simplify the form of the metric and flux.}
\begin{align}
ds_6^2 &\,\sim\, \frac{2 m_3 R_{y_2}\,\Sigma}{(\ell-m_1)^2-m_3^2}\,\Biggl[\frac{d\rho_3^2}{\rho_3^2}+\rho_3^2 \left(-dt^2+dy_1^2\right) \label{eq:BaBlocalGeo} \\
&\hspace{3.5cm}+ d\tau_3^2+ \frac{(\ell^2-(m_1-m_3)^2)^2}{\Sigma^2}\,\left(\sin^2 \tau_3 \,d\phi^2 +C_2^2 \cos^2 \tau_3 \,\frac{dy_2^2}{R_{y_2}^2}\right) \Biggr]\,, \nn \\
F_3 &\,=\, - 2R_{y_2}q_{3\text{max}} \Biggl[ \frac{(\ell^2-(m_1+m_3)^2)(\ell^2-(m_1-m_3)^2)}{\Sigma^2}\,\sin(2\tau_3)\,d\tau_3 \wedge d\phi\wedge \frac{dy_2}{R_{y_2}}\nn\\
&\hspace{2.8cm}-2 \rho_3 \,d\rho_3 \wedge dt \wedge dy_1 \+ \cO(\rho_3^2)\Biggr]\,,\nn
\end{align}
where $q_{3\text{max}}$ is defined in \eqref{eq:qmax} and we have introduced
\begin{equation}
\Sigma \equi \ell^2-(m_1+m_3)^2 + 4 m_1 m_3 \cos^2 \tau_3\,.
\end{equation}
We recognize a warped AdS$_3\times$S$^3$ geometry. Moreover, the flux is regular and carries an electric and magnetic charge since $F_3$ is self-dual and the integral of $F_3$ on the S$^3$ gives a finite value. Therefore the pole of the bubble corresponds to \emph{the near-horizon geometry of a two-charge extremal black hole}. From a type IIB perspective \eqref{eq:typeIIBembedding}, the charges correspond to D1 and D5 brane charges given in unit of volume by
\begin{equation}
q_{3\text{D1}} \= \frac{1}{4\pi^2 \text{Vol}(T^4)} \int_{S^3\times T^4}\,\star_{10} F_3 \quad = \quad q_{3\text{D5}}  \=  \frac{1}{4\pi^2} \int_{S^3}\,F_3 \= -2R_{y_2}\,q_{3\text{max}}\,. \nn \label{eq:BPSbranecharges}
\end{equation}
Our convention is such that negative charges are identified with the extremal black hole that corresponds to a $\overline{\text{D1}}$-$\overline{\text{D5}}$ anti-BPS system in type IIB. Moreover, even if the AdS$_3$ and S$^3$ are warped, one can check that the central charge is still given by\footnote{The central charge is derived from the generic expression $c=\frac{3 \int_{S^3} e^{4\Phi} \sqrt{g_{S^3}}}{2 G_6}$, where $e^{4\Phi}$ is obtained from the six-dimensional metric $ds_6^2 = e^{2\Phi}\left[ ds(\text{AdS}_3)^2+ds(\text{S}^3)^2 \right]$.}
\begin{equation}
c_3 \= 6 N_{\overline{\text{D1}}} N_{\overline{\text{D5}}}\,,
\end{equation}
where $N_{\overline{\text{D1}}}$ and $N_{\overline{\text{D5}}}$ are the quantized numbers of anti-D1 and anti-D5 branes, which can be expressed in terms of the string coupling, $g_s$, and the string length, $l_s$, as: $$N_{\overline{\text{D1}}} \= \frac{\text{Vol}(T^4)}{(2\pi)^4 g_s l_s^6} \,(-q_{3\text{D1}})\,,\qquad N_{\overline{\text{D5}}} \= \frac{1}{g_s l_s^2} \,(-q_{3\text{D5}} ).$$
The entropy of the extremal black hole is therefore proportional to $\sqrt{q_{3\text{D1}}q_{3\text{D5}}} = 2 R_{y_2} q_{3\text{max}}$.

At the South pole, $\rho=0$, $z=-\ell/2$, we find the same warped AdS$_3\times$S$^3$ geometry as in \eqref{eq:BaBlocalGeo}, interchanging the $i=3$ and $i=1$ indexes and using similar local coordinates to \eqref{eq:localCoorBaB}. However, the three-form flux will be given by $F_3 = +2R_{y_2}q_{1\text{max}} \left[ \ldots\right]$, and both charges will be positive. Therefore, the South pole of the bubble corresponds to an \emph{extremal D1-D5 black hole} with equal D1 and D5 brane charges given by $q_{1\text{D1}} = q_{1\text{D5}} = 2 R_{y_2} q_{1\text{max}}$. Similarly, its central charge will be given by:
\begin{equation}
c_1 \= 6 N_{\text{D1}} N_{\text{D5}}\,, \qquad N_{\text{D1}} \= \frac{\text{Vol}(T^4)\,q_{1\text{D1}}}{(2\pi)^4 g_s l_s^6} \,,\qquad N_{\text{D5}} \= \frac{q_{1\text{D5}}}{g_s l_s^2} .
\end{equation}
Note that the sum of the D1 and D5 brane charges of both black holes does not seem equal to the asymptotic values given in \eqref{eq:ConservedChargesBaB}. This is simply due to the expression of the supergravity charges in unit of volume in \eqref{eq:typeIIBcharges} compared to \eqref{eq:BPSbranecharges}.  Indeed, the volume of integration has a S$^2\times$S$^1$ topology asymptotically while it is a S$^3$ at the black holes, which produces the difference of $2R_{y_2}$. 
\end{itemize}

To conclude, the solutions given by \eqref{eq:BaBgeneric} with \eqref{eq:WarpFactorBaB} and \eqref{eq:RegBaB} correspond to bound states of two extremal two-charge black holes of mass and charges $(m_1,q_{1\text{max}})$ and $(m_3,-q_{3\text{max}})$ separated by a smooth vacuum bubble of size $\ell$. From a type IIB perspective, these black holes correspond to BPS D1-D5 and anti-BPS $\overline{\text{D1}}$-$\overline{\text{D5}}$ black holes with warped AdS$_3\times$S$^3$ near-horizon geometry. In the philosophy of \cite{Elvang:2002br,Bah:2021owp}, the vacuum bubble gives the necessary pressure to prevent the two strongly-attracting black holes from collapsing, while they in turn prevent the bubbles from expanding and eating up the whole space.

\subsection{Neutral configurations}
\label{sec:neutralBaB}

In this section, we consider the specific limit of the BPS/anti-BPS system of \eqref{eq:BaBgeneric}-\eqref{eq:WarpFactorBaB} where the net charges of \eqref{eq:ConservedChargesBaB}, $(\mathcal{Q}_{D1},\mathcal{Q}_{D5})$, vanish with the condition:
\begin{equation}
m_1 =m_3 =m\,.
\end{equation} The charges also vanish if we consider $\ell = m_1+m_3$ in \eqref{eq:ConservedChargesBaB}, however that is the strict vacuum limit where $F_3$ vanishes.  When $m_1=m_3$, the six-dimensional metric and flux are still given by \eqref{eq:BaBgeneric} with now
\begin{align}
Z &\= 1+ \frac{\ell^2 m \,( r_1+ r_3+2 m)}{m^2(r_1-r_3)^2+\ell^2 (r_1 r_3-m^2)} \,,\qquad W_2  \= \left(1- \frac{2 \ell}{r_1+r_3+\ell} \right)^{-1} \, , \nn\\
 T &= \frac{\ell m \sqrt{\ell^2-4 m^2}\,(r_3-r_1)}{m^2(r_1-r_3)^2+\ell^2 (r_1 r_3-m^2)}\,\frac{1}{Z}\,,\label{eq:WarpFactorBaBneutral}\\
  H &\= \frac{\ell\sqrt{\ell^2-4 m^2}\left[m(r_1 \cos \theta_1 +r_3 \cos \theta_3) (r_1-r_3)-\ell (m ( r_1+ r_3)+2 r_1 r_3)\right]}{2(m^2(r_1-r_3)^2+\ell^2 (r_1 r_3-m^2))}\,, \nn\\
 e^{2(\nu_Z+\nu_W)} &\=\frac{\left((r_1+r_3)^2-\ell^2\right)\left( m^2(r_1-r_3)^2+\ell^2 (r_1 r_3-m^2)\right)^2}{4\ell^4\,r_1^3r_3^3}\,, \nn
\end{align}
where $(r_1,\theta_1)$ and $(r_3,\theta_3)$ are the spherical coordinates centered around the two extremal black holes \eqref{eq:distanceBaB}.

\begin{figure}[h]\centering
   \includegraphics[width=0.65\textwidth]{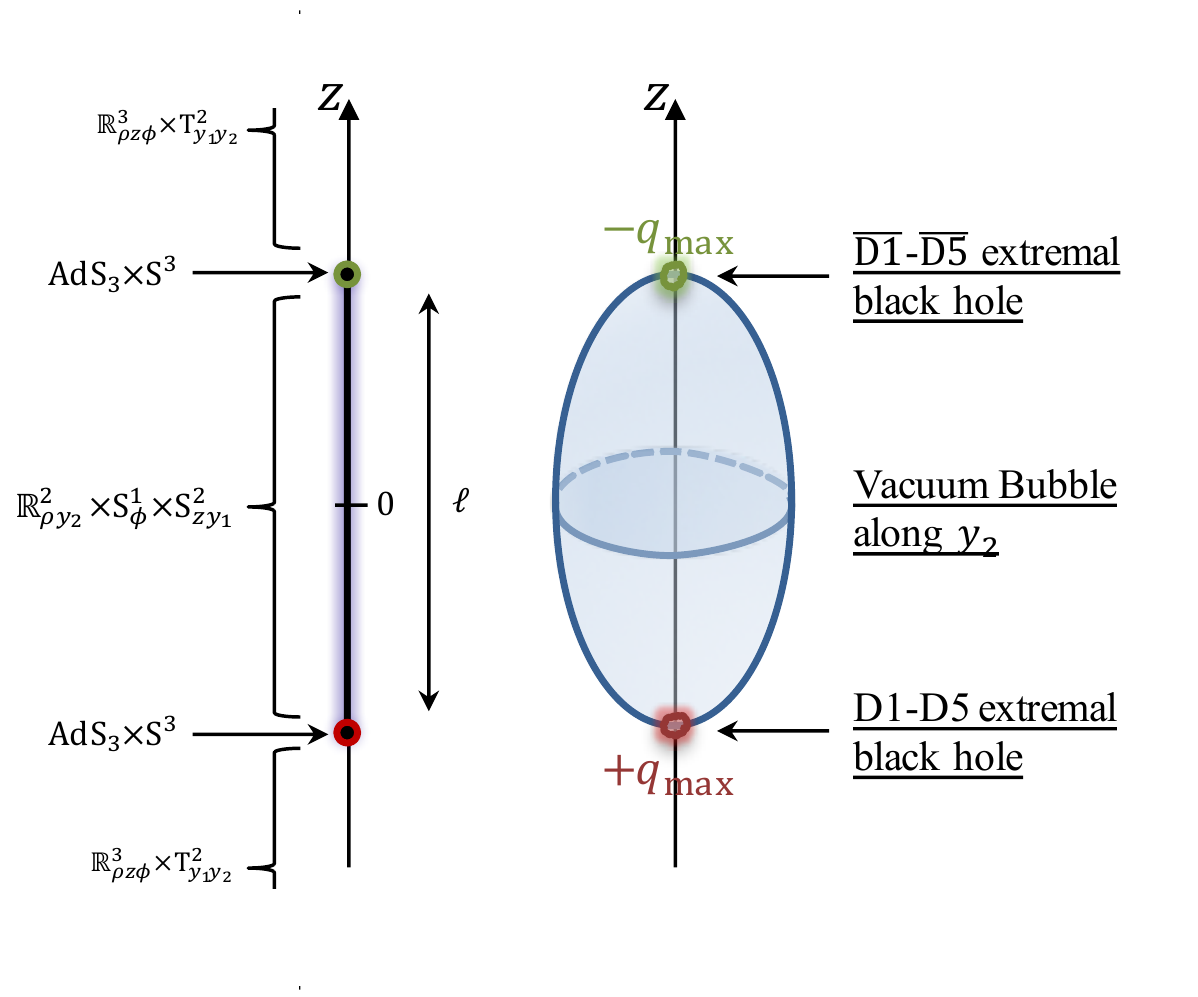}
 \caption{Profile of the sources and topology on the $z$-axis for a bound state of BPS D1-D5 branes and anti-BPS $\overline{\text{D1}}$-$\overline{\text{D5}}$ anti-branes separated by a bolt corresponding to two extremal D1-D5 black holes of opposite charges hold apart by a vacuum bubble.}
 \label{fig:B/aBboundstatesneutral}
\end{figure}

\noindent The solutions are asymptotically $\IR^{1,3}\times$T$^2$, and are neutral but massive geometries with ADM mass \eqref{eq:ConservedChargesBaB} 
\begin{equation}
\cM = \frac{\ell+4m}{4 G_4}\,,
\label{eq:MassBaBneutral}
\end{equation}
 The topology on the $z$-axis is identical to the previous solutions and is depicted in Fig.\ref{fig:B/aBboundstatesneutral}. Internally, the geometries correspond to two extremal D1-D5 and $\overline{\text{D1}}$-$\overline{\text{D5}}$ black holes at the poles of a S$^2\times$S$^1$ vacuum bubble for which the charges exactly balance each other:
\begin{equation}
q_{1\text{D1}} = q_{1\text{D5}} = 2 R_{y_2} q_\text{max} = -q_{3\text{D1}} = -q_{3\text{D5}} \,,
\end{equation}
where $q_\text{max}$ is now given by
\begin{equation}
q_\text{max}\equi  m \sqrt{\frac{\ell+2m}{\ell-2m}} \,.
\label{eq:qmaxneutral}
\end{equation}
One can therefore consider the neutral bound states as ``D1-D5 extremal black hole dipoles''. The D1 and D5 dipole charges can be read from the asymptotic expansions of the gauge potentials:
\begin{equation}
T  \underset{r\to\infty}{\sim} \frac{-\cJ_{\text{D1}} \,\cos \theta}{r^2}\,,\qquad H \underset{r\to\infty}{\sim} \frac{-\cJ_{\text{D5}}\,\sin^2\theta}{r}\,.
\label{eq:DipoleGen}
\end{equation}
where $(r,\theta)$ are the asymptotic spherical coordinates $(\rho,z)= r (\cos \theta,\sin \theta)$. We find
\begin{equation}
\cJ_{\text{D1}} \= \cJ_{\text{D5}} \= q_\text{max}\, (\ell-2m) \= m \sqrt{\ell^2-4 m^2}\,.
\end{equation}
Interestingly, the dipole charges can approach zero when $\ell \to 2m$ that does not correspond to the regime where both extremal black holes merge since $\ell$ is the distance between them. Unlike usual dipoles of charged particles, one can then have two charged objects that have finite separation but vanishing electromagnetic dipoles, that is more Schwarzchild-like. However, $\ell-2m$ is constrained by the regularity condition at the vacuum bubble \eqref{eq:RegBaB} which gives
\begin{equation}
R_{y_2} \= \frac{2(\ell^2-4m^2)}{\ell}\,,
\label{eq:RegBaBneutral}
\end{equation}
and $\ell-2m$ can only approach zero when $R_{y_2}$ is small. To better describe this limit, we can invert the perspective and consider the internal $(\ell,m)$ parameters as fixed in terms of the asymptotic quantities $\cM$ \eqref{eq:MassBaBneutral} and $R_{y_2}$.

\subsubsection{Phase space and approximate geometries}

 For simplicity, we consider $G_4=1$, and we find from \eqref{eq:MassBaBneutral} and \eqref{eq:RegBaBneutral}:
\begin{equation}
\ell= \frac{4 \cM}{3} \left(2\sqrt{(1-\epsilon_2)^2+\epsilon_2} -1+2\epsilon_2\right)\,,\quad m= \frac{2 \cM}{3} \left(2-\epsilon_2-\sqrt{(1-\epsilon_2)^2+\epsilon_2} \right)\,,
\end{equation}
where we have defined
\begin{equation}
\epsilon_2 \equi \frac{R_{y_2}}{8\cM}\,.
\label{eq:epsilon2}
\end{equation}
First, the validity of the solution, $m\geq 0$, requires $0<\epsilon_2\leq 1$, and the ADM mass is bounded below by the extra-dimension radius:
\begin{equation}
0<\epsilon_2\leq 1\qquad \Leftrightarrow \qquad \cM \geq \frac{R_{y_2}}{8}\,.
\end{equation}

At $\epsilon_2=1$, $m=0$, the two extremal black holes disappear and the solutions correspond to a vacuum S$^1\times$S$^2$ bubble in six dimensions with nothing special at its poles. For $\epsilon_2$ finite and not small, the two extremal black holes are nucleated but the whole solutions have a microscopic size of order the extra dimensions, $\cM=\cO(R_{y_2})$. 

Finally, if we consider solutions with large macroscopic size $\cM \gg R_{y_2}$,  the parameters are approximated as\footnote{We could have considered macroscopic solutions, $\cM \gg R_{y_2}$, with $\epsilon_2$ finite by allowing a conical defect of order $k_2\in \mathbb{N}$ at the vacuum bubble. This would essentially change $R_{y_2} \to k_2 R_{y_2}$ in \eqref{eq:RegBaBneutral}, and thus $\epsilon_2=\frac{k_2 R_{y_2}}{8\cM}$. With a large conical defect, one could have macroscopic solutions with finite $\epsilon_2$.  Such a conical defect will be used to generate smooth bubbling geometries in section \ref{sec:arbitraryB}, but we preferred to restrict ourselves to the minimum here.}
\begin{equation}
\ell = \frac{4 \cM}{3} \left(1 +\epsilon_2\right)\,,\quad m = \frac{2 \cM}{3} \left(1-\frac{\epsilon_2}{2} \right)\,, \label{eq:macroexpansion0}
\end{equation}
and the main quantities behave as
\begin{equation}
\begin{split}
q_\text{max} &= \frac{4 \cM}{3\sqrt{3}\,\sqrt{\epsilon_2}}\,, \quad \cJ_{\text{D1}} = \cJ_{\text{D5}} = \frac{8 \cM^2 \sqrt{\epsilon_2}}{3\sqrt{3}}, \\
\quad q_{1\text{D1}} &= q_{1\text{D5}} = -q_{3\text{D1}} = -q_{3\text{D5}}= \frac{64 \cM^2 \sqrt{\epsilon_2}}{3\sqrt{3}}.
\label{eq:macroexpansion}
\end{split}
\end{equation}
Therefore, the flux is vanishing in this regime and our bound states of extremal black holes approach a vacuum geometry. This geometry can be obtained by considering the leading order in the $\epsilon_2 \ll 1$ limit in \eqref{eq:BaBgeneric} and \eqref{eq:WarpFactorBaBneutral}: 
\begin{align}
ds_{6}^2 = &\left(1-\frac{4\cM}{3r} \right)\left[- dt^2 + dy_1^2\right] +\  dy_2^2 \nn \\
&\hspace{-0.4cm}+\frac{r^4}{\left[\left( r-\frac{2\cM}{3}\right)^2-\left( \frac{2\cM}{3}\right)^2 \cos^2\theta\right]^2}\left(dr^2+r^2 \left(1-\frac{4\cM}{3r} \right)d\theta^2\right) + r^2 \left(1-\frac{4\cM}{3r} \right)^{-1} \sin^2\theta \,d\phi^2,\nonumber\\
 F_3 \= & 0\,,\label{eq:neutralSing}
\end{align}
where $(r,\theta)$ are spherical coordinates centered around the initial vacuum bubble:
\begin{equation}
r\equi \frac{r_1+r_3+\frac{4\cM}{3}}{2}\,,\qquad \cos \theta = \frac{3}{4\cM}\left(r_1-r_3\right)\,, \qquad r\geq \frac{4\cM}{3}\,,\quad 0\leq \theta\leq \pi\,.
\label{eq:SphericalCoordinatesBaBsing}
\end{equation}
The approximate geometry is singular at the sources, $r=\frac{4\cM}{3}$, since the $\phi$-circle blows there while the $t$, $y_1$ and $\theta$ fibers shrink singularly. Therefore, the neutral bound states of extremal black holes resemble very closely this geometry up to very close to its singularity and ``resolve'' it as a vacuum bubble with two extremal D1-D5 and $\overline{\text{D1}}$-$\overline{\text{D5}}$ black holes at its poles. 
In four dimensions \eqref{eq:4dFrameworkcharged}, after compactification along the T$^2$, the approximate geometries correspond to singular neutral solutions with a dilaton:
\begin{align}
ds_{4}^2 = &-\left(1-\frac{4\cM}{3r} \right)^\frac{3}{2}\,dt^2 +\frac{r^4\,\left(1-\frac{4\cM}{3r} \right)^\frac{1}{2}}{\left[\left( r-\frac{2\cM}{3}\right)^2-\left( \frac{2\cM}{3}\right)^2 \cos^2\theta\right]^2}\left(dr^2+r^2 \left(1-\frac{4\cM}{3r} \right)d\theta^2\right) \nn\\
&+ r^2 \left(1-\frac{4\cM}{3r} \right)^{-\frac{1}{2}} \sin^2\theta \,d\phi^2,\qquad e^{\sqrt{3}\Phi_1}\=  \left(1-\frac{4\cM}{3r}\right)^{-\frac{3}{2}} \,.\label{eq:neutralSing4d}
\end{align}
It is interesting to investigate the properties of these singular geometries from a large-distance perspective and to compare them to the most famous neutral non-spinning geometries: the Schwarzschild black hole.

\subsection{Properties and comparison to Schwarzschild}
\label{sec:PropBaB}

In this section, we study the properties of the neutral solutions of two extremal D1-D5 and $\overline{\text{D1}}$-$\overline{\text{D5}}$ black holes at the poles of a vacuum bubble, given by \eqref{eq:BaBgeneric} and \eqref{eq:WarpFactorBaBneutral}, when they have a macroscopic size $\cM\gg R_{y_2}$. As long as we are focusing on regions that are not too close to the sources, one can consider the approximate geometries \eqref{eq:neutralSing} or \eqref{eq:neutralSing4d}. However, close to the sources one needs to use the generic solutions \eqref{eq:WarpFactorBaBneutral}}. Moreover, we will compare the geometries with the six-dimensional embedding of a four-dimensional Schwarzschild with the same mass, that is
\begin{equation}
ds_6^2 \= - \left( 1-\frac{2\cM}{r} \right) \,dt^2 +dy_1^2 +dy_2^2+\frac{dr^2}{1-\frac{2\cM}{r}} + r^2 \left( d\theta^2 + \sin^2\theta \,d\phi^2 \right)\,.
\label{eq:Schwarzschild}
\end{equation}
The first difference with Schwarzschild is that the bound states of extremal black holes have electromagnetic dipoles given by \eqref{eq:macroexpansion} which are suppressed by $\sqrt{\epsilon}_2$. Therefore, any charged particle moving at the vicinity of the sources should be affected by these dipoles. Ultimately, one would like to study probe geodesics or scalar waves to compare the geometries, but we leave such analysis for future projects. It will also be important to study the phenomenology of these solutions more quantitatively to understand their reasonable physical bounds.  In the present paper, we will only derive quantities which can be directly obtained from the solutions.

\begin{itemize}
\item[•] \underline{Multipole moments and redshift:}

Two key properties of the Schwarzschild geometry are that it is spherically symmetric and has no gravitational multipole moments. One can derive the multipole structure of the macroscopic bound states of extremal black holes by considering the approximate four-dimensional metric \eqref{eq:neutralSing4d}.\footnote{The multipole moments of the bound states of extremal black holes will be equal to the multipoles of the approximate solutions plus corrections of order the resolutions scale, that is $\epsilon_2$.} The metric as given is clearly axially symmetric, but is not written in ACMC coordinates\footnote{Asymptotically-Cartesian and Mass-Centered.} (see \cite{Bena:2020uup,Bah:2021jno} for more details). The ACMC coordinates $(r_S,\theta_S)$ are given by
$$
r_S \cos\theta_S \= \left(r-\frac{2\cM}{3} \right)\cos \theta\,,\qquad \sqrt{r_S^2-\frac{4\cM^2}{3}}\,\sin \theta_S \= \sqrt{r\left(r-\frac{4\cM}{3}\right)}\,\sin \theta.
$$
Then, one can read off the mass multipoles, $M_n$, from the following expansion: $$g_{t t}^{(4d)}=-1+\frac{2 M_0}{r_S}+\sum_{n \geq 1}^{\infty} \frac{2}{r_S^{n+1}}\left(M_n \,P_{n}(\cos\theta_S)+\sum_{n^{\prime}<n} c_{n n^{\prime}}^{(t t)} P_{n^{\prime}}(\cos\theta_S)\right),$$
where $P_n$ is the Legendre polynomial of degree $n$. We find that \emph{all odd mass multipoles of the solutions are zero\footnote{Plus corrections suppressed by $\epsilon_2$.} while the even moments are finite:}
\begin{equation}
M_{2n}= c_{n}\, \cM^{2n+1}\,, \quad c=\left(1,-\frac{8}{27},\frac{64}{2835},\frac{512}{56133},-\frac{118784}{42220035},-\frac{32768}{247946751},\ldots\right). \label{eq:multipolemoments}
\end{equation}
In addition to the multipole moments, one can describe the gravitational potential, or redshift factor, defined as the norm of the timelike Killing vector $\cR= -g_{tt}^{-1}$, where $g_{tt}$ is
the time component of the metric in six dimensions \eqref{eq:BaBgeneric}. To this end, we can adopt spherical coordinates centered around the vacuum bubble \eqref{eq:SphericalCoordinatesBaBsing} by replacing $4\cM/3$ with $\ell$,
\begin{equation}
r\equi \frac{r_1+r_3+\ell}{2}\,,\qquad \cos \theta = \frac{r_1-r_3}{\ell}\,, \qquad r\geq \ell\,,\quad 0\leq \theta\leq \pi\,
\label{eq:SphericalCoordinatesBaB}
\end{equation}
and the surface of the bubble is at $r=\ell\sim \frac{4\cM}{3}$. We find
\begin{equation}
\cR \= 1+ \frac{4m \left( 2(r+m)-\ell\right)}{(2r-\ell)^2-4m^2-(\ell^2-4 m^2) \cos^2\theta}\,.
\end{equation}
In Fig.\ref{fig:B/aBRedshift}, we have plotted the redshift as a function of $(r,\theta)$. One can show that the redshift is increasing as $r$ decreases with $\theta$ held constant. Moreover, the value at the bubble surface is given by
\begin{equation}
\cR \bigl|_{r=\ell} \= 1+\frac{4 m}{\left(\ell-2m\right)\sin^2\theta}\,\sim\, \frac{32\cM}{3 R_{y_2}\sin^2\theta}\,.
\end{equation}
Therefore, the redshift is indeed infinite at the poles of the vacuum bubble $\theta=0,\pi$ where the extremal D1-D5 and $\overline{\text{D1}}$-$\overline{\text{D5}}$ black holes sit. More surprisingly, the redshift is still very large elsewhere on the vacuum bubble, of order $\frac{\cM}{R_{y_2}}$.\footnote{This is surprising since a single vacuum bubble without black holes at its poles has no redshift.} Therefore, as far as the redshift is concerned, our geometries have similar properties to a Schwarzschild black hole.

\begin{figure}[h]\centering
   \includegraphics[width=0.65\textwidth]{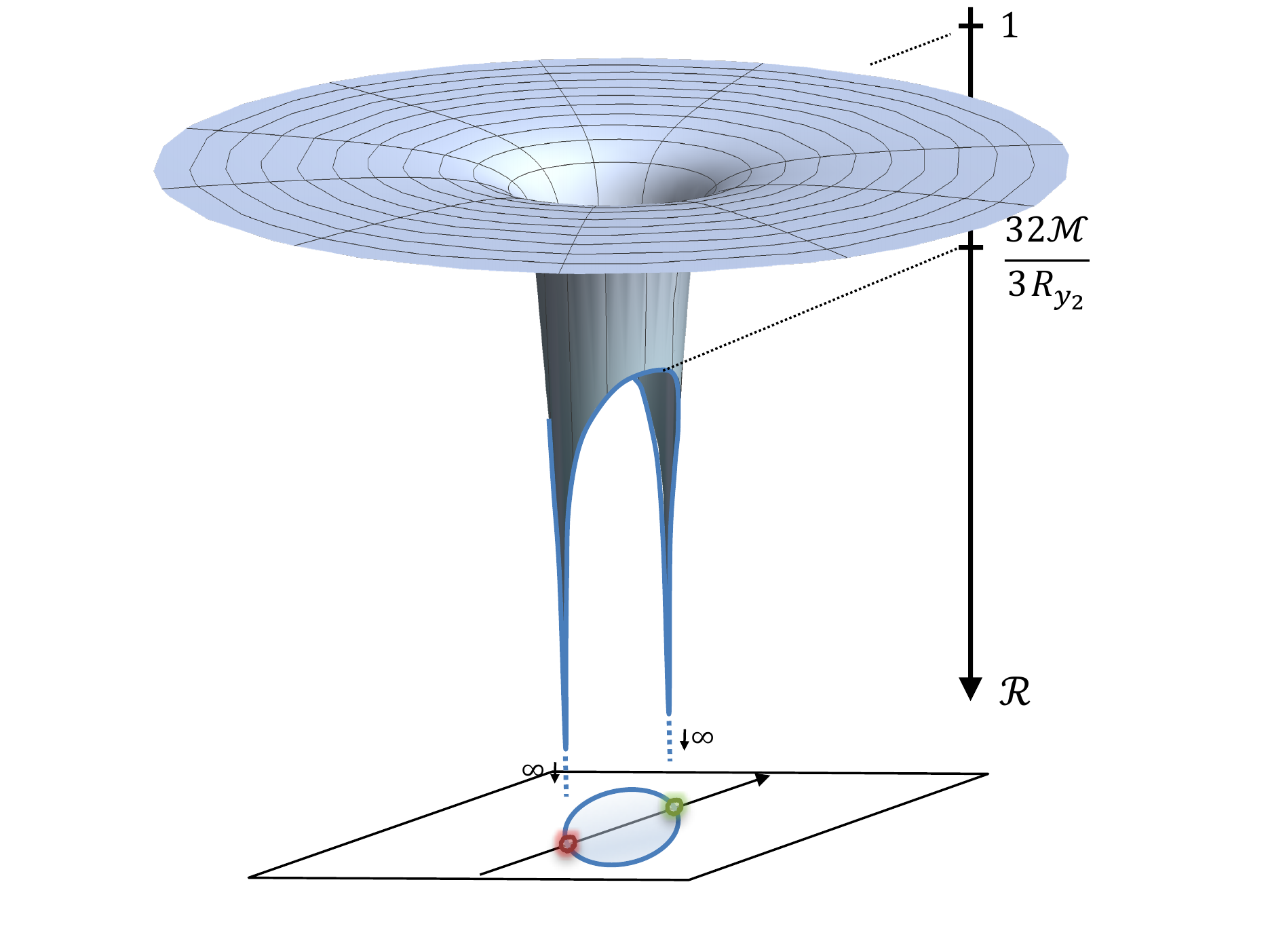}
 \caption{Redshift factor of the neutral bound states of two extremal black holes  at the poles a vacuum bubble as a function of $(r,\theta)$ \eqref{eq:SphericalCoordinatesBaB}. The surface of the bubble is at $r=\ell \sim \frac{4\cM}{3}$.}
 \label{fig:B/aBRedshift}
\end{figure}

\item[•] \underline{S$^2$ Comparison:}

Ultimately, one would like to compare the sizes of both solutions and evaluate the compactness of the neutral extremal black hole bound states. For a Schwarzschild black hole \eqref{eq:Schwarzschild}, the radial coordinate $r$ is intrinsically connected to its size since it is directly related to the area of the S$^2$,
\begin{equation}
\text{Area}_\text{Sch}(S^2)=4\pi r^2\,,
\end{equation}
and the minimal size is reached at the horizon, such that $\min(\text{Area}_\text{Sch}(S^2))=16\pi \cM^2$.
However, the radial coordinate $r$ for the bound states \eqref{eq:SphericalCoordinatesBaB} do not correspond to any physical size of the geometries, so one must compute $\text{Area}(S^2)$ and compare  its minimal value to that of Schwarzschild. From \eqref{eq:WarpFactorBaBneutral}, we find:
\begin{align}
\text{Area}(S^2) &= \int_{S^2} \sqrt{g_{\theta \theta}g_{\phi \phi }}\,d\theta d\phi = 2\pi r^2\int_{0}^\pi \frac{(2(r+m)-\ell)^2-(\ell^2-4m^2)\cos^2\theta}{(2 r-\ell)^2-\ell^2\cos^2\theta}\,\sin\theta \,d\theta\,,\nn\\
&= 4 \pi r^2\left[1-\frac{4m^2}{\ell^2}-\left(\left(\frac{4mr+\ell(\ell-2m)}{\ell^2} \right)^2-1+\frac{4m^2}{\ell^2} \right) \frac{\ell\,\log\left(1-\frac{\ell}{r}\right) }{2(2 r-\ell)}\right]\,,\nn\\
&\sim 4 \pi r^2 \,\frac{\log \left( 1-\frac{4\cM}{3 r}\right)}{\frac{4\cM}{3 r}\left( \frac{2\cM}{3 r}-1\right)}\,,\label{eq:AreaS2BaB}
\end{align}
where the last equality gives the area of the S$^2$ for the approximate metric \eqref{eq:neutralSing} which is valid when $r$ is not too close to the bubble locus $r=\ell$. In Fig.\ref{fig:B/aBAreaS2}, we have plotted the area as a function of $r$ for an illustrative value of $\epsilon_2=2\times10^{-2}$. At large distance, the area behaves as $4\pi r^2$ since the geometry is asymptotically flat. Then, the area decreases as $r$ decreases to finally blow up at the bubble locus $r=\ell$, due to the two extremal black holes at the poles of the bubble. There is therefore an inflection point in between where the area is minimum. This minimum can be obtained from the approximate metric since it is relatively far from the bubble locus. We find numerically:
\begin{equation}
\min(\text{Area}(S^2))\, \approx\, 2.32 \times 16\pi \cM^2\,\approx\, 2.32 \times \min(\text{Area}_\text{Sch}(S^2)). 
\label{eq:AreaMin}
\end{equation}
Therefore, the radius of the minimal S$^2$ of the bound states of extremal black holes is around $1.52$ times bigger than its Schwarzschild radius, which means that they are very compact geometries. 
\begin{figure}[h]\centering
   \includegraphics[width=0.65\textwidth]{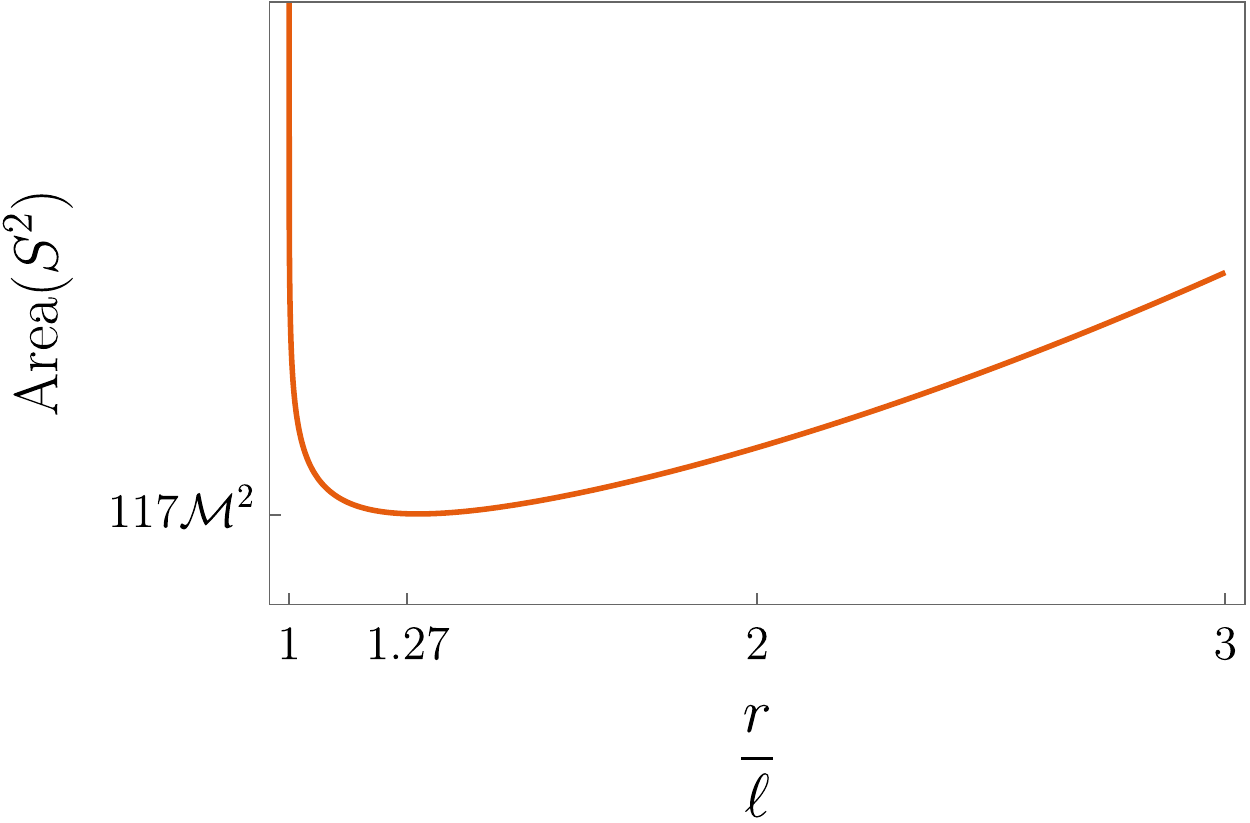}
 \caption{Area of the S$^2$ of the bound states of extremal black holes \eqref{eq:AreaS2BaB} for $\epsilon_2=2\times 10^{-2}$.}
 \label{fig:B/aBAreaS2}
\end{figure}

Another notable property of the bound state geometries is that they are asymmetric, in the sense that there is a non-trivial ratio between the North-to-South physical length and the equatorial length of the S$^2$ at a given $r$. We define the ``asymmetry factor'' of the S$^2$ as
\begin{equation}
\cA \equi \frac{\int_{\theta=0}^\pi \sqrt{g_{\theta \theta}}\,d\theta}{\int_{\phi=0}^\pi \sqrt{g_{\phi \phi}} \bigl|_{\theta=\frac{\pi}{2}}\,d\phi}\,,
\label{eq:AsymmetryBaB}
\end{equation}
which is defined to be equal to 1 for a round sphere. In Fig.\ref{fig:B/aBAssym}, we have plotted the asymmetry factor as a function of $r$. We found that the S$^2$ becomes increasingly flattened as we approach the bubble. The minimum of $\cA$ is of the order of $\frac{R_{y_2}}{\cM}$, which means that the S$^2$ is highly squashed. This is due to the strong attraction between the two extremal black holes of opposite charge. Moreover, since we have two infinite AdS$_3$ throats exactly at the poles of the bubble, $r=\ell$, the physical length between the North and South poles diverges exactly at this radius. This produces the strong inversion of the asymmetry factor very close to the bubble locus. Finally, note that the asymmetry factor at the minimum S$^2$ is about 0.77, which means that the S$^2$ here is mostly round, and it is likely that an observer would only see this almost-round ellipsoid from a distance.

\begin{figure}[h]\centering
   \includegraphics[width=0.65\textwidth]{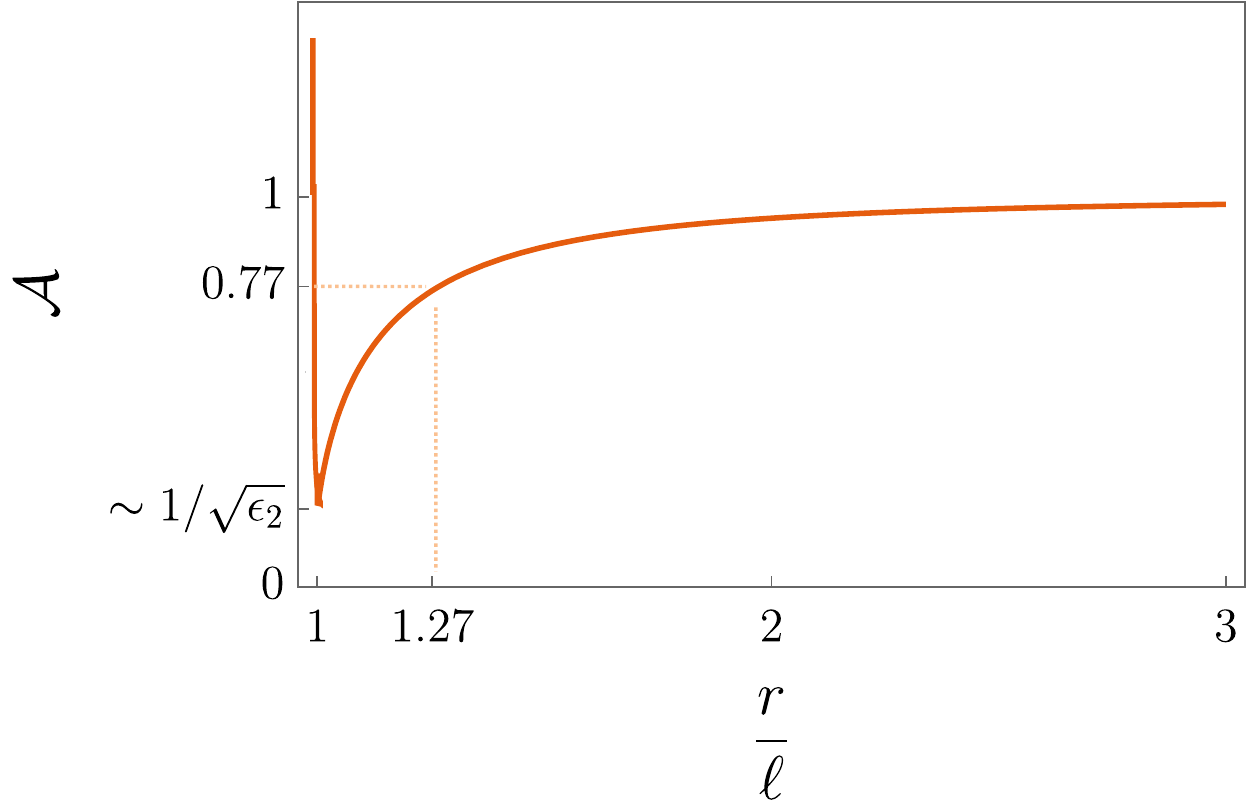}
 \caption{Asymmetry factor of the S$^2$ for the bound states of extremal black holes \eqref{eq:AsymmetryBaB} with $\epsilon_2=2 \times 10^{-2}$.}
 \label{fig:B/aBAssym}
\end{figure}

\end{itemize}

\subsection{Intermezzo}

In this section, we have constructed geometries in type IIB supergravity that correspond to bound states of two BPS D1-D5 brane and $\overline{\text{D1}}$-$\overline{\text{D5}}$ anti-brane sources and that are asymptotic to $\IR^{1,3}\times$T$^2\times$T$^4$. The two sources correspond to supersymmetric two-charge black holes that do not preserve the same supersymmetries and have charges of opposite sign. We have held them apart in a smooth manner by nucleating a vacuum bubble in between. More precisely, the extremal black holes are at the poles of a bolt where one of the T$^2$ direction smoothly degenerates.

We have focused on the solutions where the charges of the extremal black holes balance each other exactly. The solutions correspond to massive four-dimensional neutral geometries that have electromagnetic dipole moments. We have studied the phase space of solutions in terms of the ADM mass, $\cM$, and the asymptotic radius of the extra dimension that is shrinking at the bolt, $R_{y_2}$. 

We have shown that for macroscopic configurations, $\cM\gg R_{y_2}$, the geometries share similar features to a four-dimensional Schwarzschild solution of the same mass. First, they have a large redshift at the bubble locus, and this redshift diverges at the poles where infinite AdS$_3$ throats open. Moreover, these objects are ultra-compact, since the radius of the minimal S$^2$ is 1.52 times the Schwarzschild radius. However, the geometries have many differences. First, the behavior of the S$^2$ is characteristic of ``a bag spacetime'', meaning that it suddenly opens up after reaching a minimal area close to the bubble locus (see Fig.\ref{fig:B/aBAreaS2}). Moreover, it is highly squashed near the bubble locus, as it gets larger and larger towards its equator (see Fig.\ref{fig:B/aBAreaS2}). Despite this, the minimal S$^2$, which should be the visible sphere for an asymptotic observer, remains mostly round with an asymmetry factor of approximately 0.77. Finally, the solutions have non-zero even multipole moments, which should also be observable for an asymptotic observer via gravitational wave emissions.

One would ultimately like to construct neutral geometries in type IIB that are regular but also free from black hole sources and horizons. This requires that we resolve the horizons of both extremal black holes into smooth bubbling geometries. This can be achieved following two possible methods. 

The first option, which is not explored in this paper, is to resolve the horizons using only BPS ingredients. Indeed, all smooth microstates of extremal D1-D5 black holes have been constructed and classified \cite{Lunin:2001fv,Taylor:2005db,Mathur:2005ai}. They correspond to smooth geometries, called supertubes, that closely resemble the black hole, developing an AdS$_3\times$S$^3$ throat but smoothly capping off the spacetime. Therefore, one could consider replacing both the extremal D1-D5 and $\overline{\text{D1}}$-$\overline{\text{D5}}$ black holes with smooth supertube geometries, and try to identify the full solution accounting for the resulting backreaction at the poles of the vacuum bubble. 

The second method follows the approach of \cite{Bah:2020pdz,Bah:2021owp,Bah:2021rki,Heidmann:2021cms} and consists in resolving the horizons by moving away from supersymmetry. More precisely, this involves replacing the two extremal two-charge black holes by two small charged bolts where the other direction of the T$^2$ shrinks, i.e. the $y_1$-circle in our ansatz \eqref{eq:WeylSol2circle}. The bolts will correspond to small non-BPS D1-D5 bubbles that are warped by electromagnetic flux. The whole geometry will correspond to a chain of three bubbles on a line where the $y_1$ and the $y_2$ circles alternately shrink. The middle bubble will still be a vacuum bubble but the two outer ones will correspond to non-BPS D1-D5 bubbles with opposite charges. From a technical perspective, this consists in following the solution scheme of section \ref{sec:SolScheme}, but now considering three rod sources of finite size and fixing the vacuum sector warp factors such that the rod loci correspond to regular coordinate degeneracies of the $y_1$ or $y_2$ circles. 


\section{Neutral D1-D5 bubbling solutions}
\label{sec:arbitraryB}

In this section, we construct smooth non-BPS bubbling solutions in six and ten dimensions using the solution scheme in section \ref{sec:SolScheme}, and discuss their physics. This will allow later on to construct smooth neutral configurations with internal electromagnetic flux wrapping the bubbles.

\begin{figure}[h]\centering
   \includegraphics[width=0.65\textwidth]{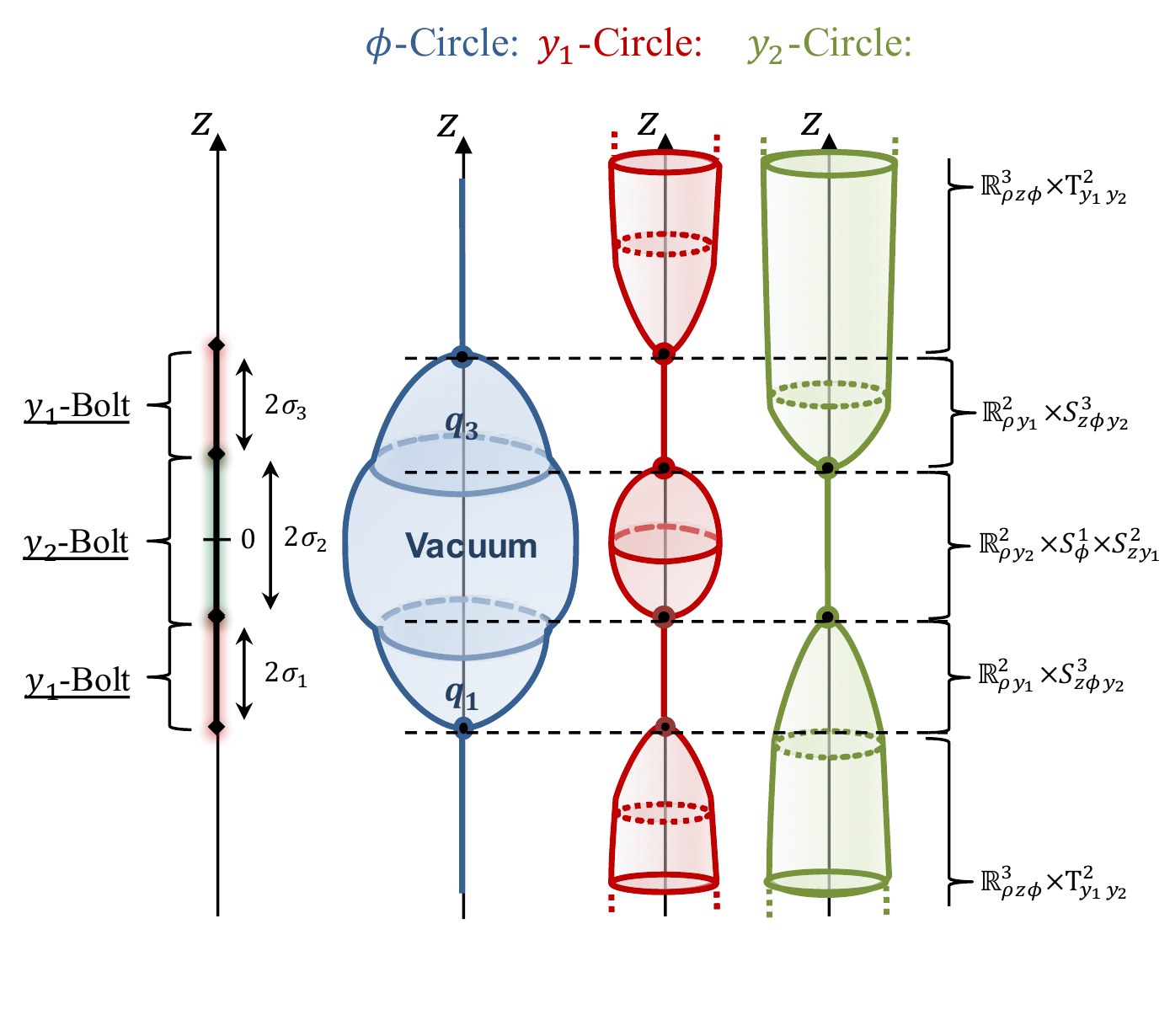}
 \caption{Schematic description of the smooth three-bubble configuration and the behavior of the $\phi$, $y_1$ and $y_2$ circles on the $z$-axis.}
 \label{fig:BothConf}
\end{figure}

\subsection{Smooth non-BPS three-bubble solutions with arbitrary charges}

We have detailed the construction of the smooth three-bubble solutions in Appendix \ref{App:BubbleConstruction}. The six-dimensional metric and gauge field are given in \eqref{eq:WeylSol2circle} and come from three finite-size rods, as shown in Fig.\ref{fig:rodsources}. The warp factors $(Z,H,T,\nu_Z)$ are sourced by the two outer rods and $(W_1,W_2,\nu_W)$ by the three connected rods. As explained in Appendix \ref{App:BubbleConstruction}, the weights $(G_i^{(1)},G_i^{(2)})$ in the $W_I$ are set to specific values so that the rods correspond to regular bolts where the $y_1$ or $y_2$ circle degenerates smoothly as shown in Fig.\ref{fig:BothConf}. To better describe the bubbling configurations, it is more convenient to introduce the functions $(\mathcal{Z},U_1,U_2)$ defined as follows 
\begin{equation}\label{eq:three_source_defs}
\mathcal{Z} = \frac{Z}{W}, \qquad U_1= \frac{1}{W_1^2}, \qquad U_2 = \frac{1}{W_2}.  
\end{equation}  The solution can then be expressed as
\begin{align}
ds_6^2& \=\frac{1}{\mathcal{Z}} \left[-dt^2+U_1 \, dy_1^2 \right]+\mathcal{Z}\, \left[ U_2\,dy_2^2+\frac{1}{U_1 U_2}\left( e^{2(\nu_Z+\nu_W)} (d\rho^2+dz^2)+\rho^2 d\phi^2\right)\right], \nn \\
F_3 &\= d\left[ H \,d\phi \wedge dy_2 \+ T \,dt \wedge dy_1 \right]\,.\label{three_source_metric}
\end{align}  
In this form, the warp factors and gauge potentials are given as 
\begin{align}
U_1\equi &\left(1-\frac{2\sigma_1}{r_1+\sigma_1}\right)\left(1-\frac{2\sigma_3}{r_3+\sigma_3}\right),  \quad U_2\equi\left(1-\frac{2\sigma_2}{r_2+\sigma_2}\right)\,,\nn\\
\mathcal{Z} \equi & \frac{(r_1+m_1)(r_3+m_3)-\left(q_1-\gamma(1+\cos \theta_3)\right)\left(q_3+\gamma(1-\cos \theta_1)\right)}{\sqrt{\left(r_1^2-\sigma_1^2+\gamma^2\sin^2 \theta_3 \right)\left(r_3^2-\sigma_3^2+\gamma^2\sin^2 \theta_1 \right)}} \,\sqrt{U_1}, \nn \\
T  \= &\frac{q_1 r_3 +q_3 r_1 +\gamma \left(m_1 \cos \theta_1 +r_1 +m_3 \cos \theta_3 -r_3 \right)}{(r_1+m_1)(r_3+m_3)-\left(q_1-\gamma(1+\cos \theta_3)\right)\left(q_3+\gamma(1-\cos \theta_1)\right)} \,,\label{eq:3rodAnsatz} \\
 H \= &\frac{r_1 \cos \theta_1 + r_3 \cos \theta_3}{2}\,T \+ \frac{\mathcal{I}}{\eta\,\cA}\,, \nonumber \\
e^{4\nu_W} \= & \frac{\left( r_2^2 -\sigma_2^2\right)\Bigl((r_1+\sigma_1)\cos^2\frac{\theta_1}{2}+(r_3+\sigma_3)\sin^2\frac{\theta_3}{2} \Bigr)\Bigl((r_1-\sigma_1)\sin^2\frac{\theta_1}{2}+(r_3-\sigma_3)\cos^2\frac{\theta_3}{2} \Bigr)}{\left(r_2^2-\sigma_2^2 \cos^2 \theta_2\right)(r_1+\sigma_1 \cos \theta_1)(r_3-\sigma_3 \cos \theta_3)}, \nn
\\
e^{4\nu_Z}  \=&  \frac{\left(r_1^2-\sigma_1^2+\gamma^2\sin^2 \theta_3 \right)\left(r_3^2-\sigma_3^2+\gamma^2\sin^2 \theta_1 \right)}{(1+2\delta)^2\,(r_1^2-\sigma_1^2 \cos^2\theta_1)(r_3^2-\sigma_3^2 \cos^2\theta_3)}  \label{eq:nu3rodsGen} \\
& \hspace{-1cm} \times \left(1+2\delta \,\frac{q_1 \left( r_1- \ell \cos \theta_1\right)+q_3 \left( r_3+ \ell \cos \theta_3\right)+\gamma \left( r_3-r_1+(l+m_3)\cos \theta_1+(l+m_1)\cos \theta_3 \right)}{q_1 r_3 +q_3 r_1 +\gamma \left(m_1 \cos \theta_1 +r_1 +m_3 \cos \theta_3 -r_3 \right)} \right)^2 . \nonumber
\end{align}
where the main constants and functions are
\begin{align}
r_i &\equi \frac{r_+^{(i)}+r_-^{(i)}}{2}\,,\qquad \cos \theta_i \equi \frac{r_-^{(i)}-r_+^{(i)}}{2\sigma_i}\,,  \qquad r_\pm^{(i)} \equi \sqrt{\rho^2+(z-(z_i\pm\sigma_i))^2}\,,\nn \\
 \sigma_1 &\equi \sqrt{m_1^2-q_1^2+2 q_1 \gamma }\,,\qquad  \sigma_3 \equi \sqrt{m_3^2-q_3^2-2 q_3 \gamma }\,, \qquad \ell \equi z_3 -z_1=\sigma_1+2 \sigma_2+\sigma_3\,,\nn\\
\gamma& \equi \frac{m_3 q_1 \- m_1 q_3}{\ell+m_1+m_3} \,,\qquad \eta \equi  \ell^2-m_1^2-m_3^2 + \left(q_1-\gamma\right)^2+ \left(q_3+\gamma\right)^2 \,,\\ \delta &\equiv \frac{m_1 m_3 -(q_1-\gamma)(q_3+\gamma)}{\eta}\,,\nn
\end{align}
and $(\mathcal{I},\cA)$ are given in \eqref{eq:I&ADeg}.

One can check that the solutions are regular out of the $z$-axis. Indeed, for $\rho>0$, we have $r_i> \sigma_i$ and therefore $U_1$, $U_2$, $\cZ$, $e^{4\nu_Z}$ and $e^{4\nu_W}$ are strictly positive. The $z$-axis corresponds to an axis of symmetry where a coordinate degenerates.  The description of the local topology and the regularity constraints requires a careful analysis which we turn to next.

\subsubsection{Regularity on the $z$-axis and topology}

On the $z$-axis, $U_1$ and $U_2$ vanish as ``Schwarzschild'' factors at the $i=1,3$ and $i=2$ rods respectively, since $r_i =\sigma_i$. From the metric \eqref{three_source_metric}, we see that this induces a change of topology by forcing one of the $y_a$ fiber to shrink while the $\phi$-circle blows up in size. The different topologies on the $z$-axis have been summarized in the right hand side of Fig.\ref{fig:BothConf}.

\begin{itemize}
\item[•] \underline{Geometry at the rods on the $z$-axis:}

We consider the local coordinates centered around the $i^\text{th}$ rod, $(\rho_i^2,\theta_i )=(r_i-\sigma_i,\theta_i )$, \be
\rho= \rho_{i}\,\sqrt{\rho_{i}^2+2\sigma_{i}} \sin \theta_i,\, \qquad z\= \left(\rho_{i}^2+\sigma_i\right) \cos \theta_{i}+z_{i}. 
\label{eq:localcoordrod}
\ee
 In the region near the rod, $\rho_i \to 0$ and $\theta_i \neq 0,\pi$, we have
\begin{equation}
r_i - \sigma_i \sim \rho_i^2\,,\quad r_i - \sigma_i \cos\theta_i>0\,,\qquad  \cos \theta_j = \text{sign}(i-j) \,, \quad r_j - \sigma_j >0 \,, \quad j \neq i\,.
\end{equation}
From \eqref{eq:3rodAnsatz} and \eqref{eq:nu3rodsGen}, we have
$$
(U_1,\rho^2) \,\propto\, \rho_i^2 \,, \quad (\cZ,U_2) \,=\,  \cO(1) \,,\quad e^{2(\nu_Z+\nu_W)} \,\propto \, \rho_i^2 \,,\quad d\rho^2 +dz^2 \propto d\rho_i^2+\frac{\sigma_i}{2} d\theta_i^2\,,
$$
for $i=1$ or $3$ and 
$$
(U_2,\rho^2) \,\propto\, \rho_i^2 \,, \quad (\cZ,U_1) \,=\,  \cO(1) \,,\quad e^{2(\nu_Z+\nu_W)} \,\propto \, \rho_i^2 \,,\quad d\rho^2 +dz^2 \propto d\rho_i^2+\frac{\sigma_i}{2} d\theta_i^2\,,
$$
for the middle rod, $i=2$. Therefore, the reduced metric on the two-dimensional subspace $(y_a,\rho_i)$, where $a=1$ for the first and third rods and $a=2$ for the middle one, describes an origin of an $\IR^2$, i.e. a \emph{bolt}, such that
\begin{equation}
ds_2^2 \sim d\rho_i^2 +\frac{\rho_i^2}{C_i^2} \,dy_a^2\,,
\end{equation}
where $C_i$ are constants that we derive in Appendix \ref{App:BubbleConstruction}. These constants must be related to the periodicity of the $y_a$-circles, $y_a \to y_a+2\pi R_{y_a}$, to have a smooth local $\IR^2$ or at least a smooth $\mathbb{Z}_{k_i}$ quotient over $\IR^2$ where $k_i$ is an integer. More precisely, it requires $C_i = k_i R_{y_a}$. These regularity conditions lead to three \emph{bubble equations}:
\begin{align}
k_1 R_{y_1} &\= \frac{ (m_1+\sigma_1)(m_3+\sigma_3+2(\sigma_1+\sigma_2))-q_1(q_3+2\gamma)}{\sqrt{\sigma_1(\sigma_1+\sigma_2)}} \,, \nn \\
k_3 R_{y_1} &\= \frac{ (m_3+\sigma_3)(m_1+\sigma_1+2(\sigma_3+\sigma_2))-q_3(q_1-2\gamma)}{\sqrt{\sigma_3(\sigma_3+\sigma_2)}} \,, \label{i13_resolution_three_bubble} \\
k_2 R_{y_2} &\=4\sqrt{\frac{\sigma_2 \left(\sigma_1+\sigma_2+\sigma_3\right)\left[(m_1+m_3)^2-(q_1+q_3)^2-(\sigma_1+\sigma_3+2\sigma_2)^2\right]}{(m_1-m_3)^2-(q_1-q_3-2\gamma)^2-(\sigma_1+\sigma_3+2\sigma_2)^2}}\,. \nn
\end{align}
The transverse spacial directions, $(\theta_i,\phi,y_2)$ for the $i=1,3$ rods and $(\theta_i,\phi,y_1)$ for the $i=2$ rod, have a finite size and define compact topological cycles or \emph{bubbles} on the $z$-axis. We have therefore three bubbles on the $z$-axis. We analyze the topology of the bubbles in the Appendix \ref{App:BubbleConstruction}. For the first and third rods, the $(\theta_i,\phi,y_2)$ has the topology of an S$^3$, while for the middle bubble, the $(\theta_i,\phi,y_1)$ has the topology of an S$^1\times$S$^2$. 

Moreover, the two outer bubbles are non-trivially wrapped by electromagnetic flux from $F_3$. Both carry two equal electric and magnetic charges.\footnote{The electric and magnetic charges are necessarily equal from self-duality of $F_3$.} From a type IIB perspective \eqref{eq:typeIIBembedding}, they correspond to D1 and D5 charges $(q_{i\text{D1}},q_{i\text{D5}})$, $i=1,3$, given in unit of volume by
\begin{equation}
q_{i\text{D1}} \= \frac{1}{4\pi^2 \text{Vol}(T^4)} \int_{S_{\theta_i \phi y_2}^3\times T^4}\,\star_{10} F_3 \quad = \quad q_{i\text{D5}}  \=  \frac{1}{4\pi^2} \int_{S_{\theta_i \phi y_2}^3}\,F_3  \label{eq:chargesBu}
\end{equation}
These yield the following charges:
\begin{equation}
q_{1\text{D1}}\= q_{1\text{D5}}\= 2 R_{y_2}\,q_1, \qquad q_{3\text{D1}}\= q_{3\text{D5}}\=  2 R_{y_2}\,q_3.
 \label{eq:rodcharges}
\end{equation}

\item[•] \underline{Geometry out of the rods on the $z$-axis:}

Above and below the rod configuration, $\rho=0$ and $z>\sigma_2+2\sigma_3$ or $z<-2\sigma_1-\sigma_2$, the main functions $(U_1,U_2,\cZ,e^{2(\nu_Z+\nu_W)})$ are finite and positive, and the $\phi$-circle degenerates as the usual cylindrical coordinate degeneracy. As for the bound states of extremal black holes in section \ref{sec:TopBaB}, $e^{2(\nu_Z+\nu_W)}=1$, so there are no conical singularities and the semi-infinite segments have a $\IR^3\times$T$^2$ topology.
\end{itemize}

\subsubsection{Conserved charges and profile in four dimensions}

In order to characterize these solutions in terms of the corresponding four-dimensional conserved charges, we use the truncation ansatz \eqref{eq:4dFrameworkcharged}, in this case
\begin{equation}\label{4d_metric_three_bubble}
\begin{split}
ds_4^2&\=-\frac{\sqrt{U_1 U_2}}{\mathcal{Z}} dt^2+\frac{ \mathcal{Z}}{\sqrt{U_1 U_2}}\left[e^{2(\nu_Z+\nu_W)} (d\rho^2 +dz^2)+\rho^2 d\phi^2\right],\\
  F^{(m)} &\= dH \wedge d\phi \,,\qquad  F^{(e)}\= dT \wedge dt,\qquad e^{\sqrt{3}\,\Phi_1} = \sqrt{\frac{ \cZ^2}{U_1^3 U_2}}, \qquad e^{\sqrt{\frac{3}{2}}\Phi_2}  = \frac{1}{\cZ U_2}. \nonumber
\end{split}
\end{equation}
Note that the solutions are singular at the rods where $U_1$ and $U_2$ vanish since we have compactified on directions which degenerate there. These singularities are resolved in six dimensions as discussed in the previous section.

Expanding in asymptotic spherical coordinates provides the ADM mass \eqref{eq:AsymptoticExpGen}, and the type IIB charges in units of volume \eqref{eq:typeIIBcharges}:
\begin{equation}
\begin{split}
\mathcal{M} &\=\frac{m_1+\sigma_2+m_3}{2G_4}\,,\qquad \cQ_\text{D1}\=\cQ_\text{D5}\= q_1+ q_3 \,.
\end{split}
\label{eq:conservedcharges3BU}
\end{equation}
Note that since $q_1$ and $q_3$ are independent real parameters, one can construct neutral solutions with arbitrary equal and opposite charges at the outer bubbles.

\subsection{Neutral bubbling solutions }\label{sec:neutralB}

\begin{figure}[h]\centering
   \includegraphics[width=0.60\textwidth]{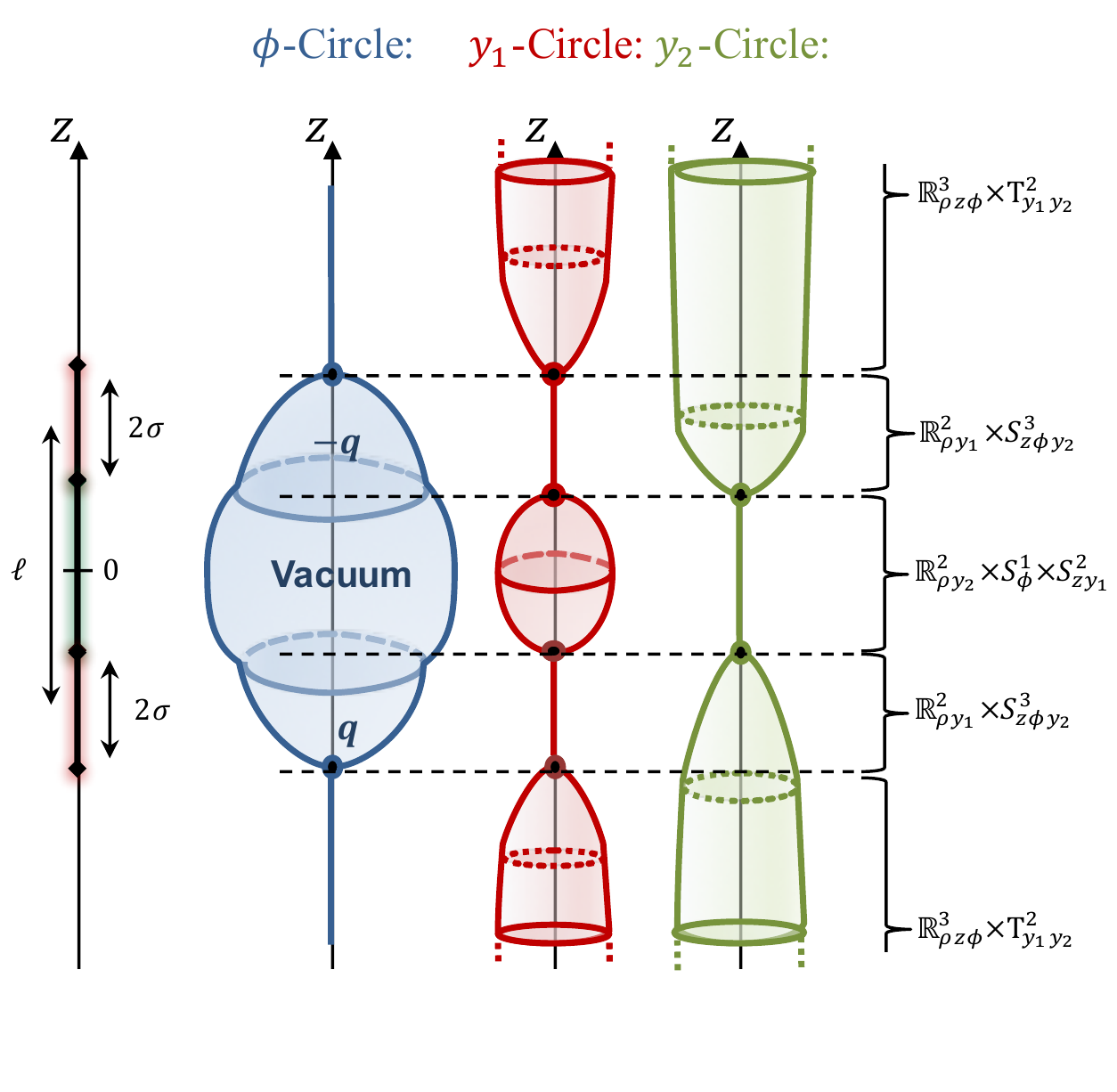}
 \caption{Schematic description of the smooth neutral three-bubble configuration and the behavior of the $\phi$, $y_1$ and $y_2$ circles on the $z$-axis.}
 \label{fig:neutralconf}
\end{figure}

The class of smooth three-bubble solutions discussed in the previous section admits a net neutral case supported by internal electromagnetic flux. Indeed, one can set the charges \eqref{eq:conservedcharges3BU} to zero by considering solutions with $q_1 = -q_3=q$. Without restriction, we can assume $q\geq 0$ by symmetry. 
Note that $F_3$ is nonzero and the bubbles are still wrapped by flux (see Fig.\ref{fig:neutralconf}).

In particular, we consider the class of solutions of the previous section with
\begin{equation}
m_1=m_3 =m \,\geq \, 0, \qquad q_1=-q_3=q \,\geq \, 0.
\label{eq:neutralassum}
\end{equation}
We have assumed for simplicity and symmetry that $m_1$ and $m_3$ are equal, so that the first and third bubbles are identical with opposite charges and $\sigma_3=\sigma_1 = \sigma$. The solutions are therefore given in terms of three parameters: the distance between the centers of the first and third rods $\ell$, also relate to the rod length of the middle rod $\ell-2\sigma$, and the mass and charge parameters, $(m,q)$, giving rise to the length of these rods. The six-dimensional metric and gauge field of the neutral bubbling geometries are still given by \eqref{three_source_metric}, which now read
\begin{align}
U_1\equi &\left(1-\frac{2\sigma}{r_1+\sigma}\right)\left(1-\frac{2\sigma}{r_3+\sigma}\right),  \quad U_2\equi\left(1-\frac{\ell-2\sigma}{r_2+\frac{\ell}{2}-\sigma}\right)\,,\nn\\
\cZ \equi & \frac{(r_1+m)(r_3+m)+\left(q-\gamma(1+\cos \theta_3)\right)\left(q-\gamma(1-\cos \theta_1)\right)}{\sqrt{\left(r_1^2-\sigma^2+\gamma^2\sin^2 \theta_3 \right)\left(r_3^2-\sigma^2+\gamma^2\sin^2 \theta_1 \right)}} \,\sqrt{U_1},\hspace{1.8cm} \nn 
\end{align}
\begin{align}
 e^{4(\nu_Z+\nu_W)} = & \frac{\left( r_2^2 -\left(\frac{\ell}{2}-\sigma\right)^2\right)\Bigl((r_1+r_3+\sigma(\cos\theta_1-\cos\theta_3))^2-\left(\ell+2\sigma\right)^2\Bigr)\left(r_1^2-\sigma^2+\gamma^2\sin^2 \theta_3 \right)}{4(1+2\delta)^2\,\left(r_2^2-\left(\frac{\ell}{2}-\sigma\right)^2 \cos^2 \theta_2\right)(r_1+\sigma \cos \theta_1)(r_3-\sigma \cos \theta_3)(r_1^2-\sigma^2 \cos^2\theta_1)}\nn\\
& \hspace{-1cm} \times \frac{\left(r_3^2-\sigma^2+\gamma^2\sin^2 \theta_1 \right)}{(r_3^2-\sigma^2 \cos^2\theta_3)}  \left(1+2\delta \,\frac{(q-\gamma) \left( r_3-r_1\right)+\left(\ell(q-\gamma) -\gamma m\right)\left( \cos \theta_1+ \cos \theta_3\right)}{(q-\gamma) \left( r_1-r_3\right)-\gamma m \left( \cos \theta_1+ \cos \theta_3\right)} \right)^2 ,\nonumber \\
T  \=& \frac{(q-\gamma)( r_3 -r_1 )+\gamma\, m (\cos \theta_1 -\cos \theta_3)}{(r_1+m)(r_3+m)+\left(q-\gamma(1+\cos \theta_3)\right)\left(q-\gamma(1-\cos \theta_1)\right)} \,,\label{eq:3rodneutralAnsatz} \\
 H \=&\frac{r_1 \cos \theta_1 + r_3 \cos \theta_3}{2}\,T_1 \+ \frac{\mathcal{I}}{\eta\,\cA}\,. \nonumber 
\end{align}
The main constants and functions also simplify to
\begin{equation}
\begin{split}
r_i &\equi \frac{r_+^{(i)}+r_-^{(i)}}{2}\,,\qquad \cos \theta_i \equi \frac{r_-^{(i)}-r_+^{(i)}}{2\sigma_i}\,,  \qquad r_\pm^{(i)} \equi \sqrt{\rho^2+(z-(z_i\pm\sigma_i))^2}\,, \\
 \sigma_1 &\= \sigma_3  \= \sigma \= \sqrt{m^2-q^2\frac{\ell-2m}{\ell+2m} }\,,\quad  \sigma_2\= \frac{\ell}{2}-\sigma \,,\quad z_2=0\,,\quad z_3=-z_1= \frac{\ell}{2}\,,\\
\gamma& \= \frac{2 q m}{\ell+2 m} \,,\qquad \eta \=  \ell^2-2m^2+ 2\left(q-\gamma\right)^2\,,\qquad  \delta \= \frac{m^2+(q-\gamma)^2}{\eta}\,.
\end{split}
\label{eq:maindataneutral}
\end{equation}

The truncation to four dimensions takes the same basic form as in \eqref{4d_metric_three_bubble}. The resulting ADM mass is given by \eqref{eq:conservedcharges3BU} while the asymptotic D1 and D5 charges are zero
\begin{equation}
\mathcal{M}=\frac{4m+\ell-2\sigma}{4G_4}, \qquad \mathcal{Q}_\text{D1}=\mathcal{Q}_\text{D5}=0\,.
\label{eq:ADMmassneutral}
\end{equation}
Because the solutions have internal charges that cancel out asymptotically, $F_3$ is sourced as an electromagnetic or D1-D5 dipole on the $z$-axis. The dipole charges can be measured from the expansions \eqref{eq:DipoleGen}, and are given by
\begin{equation}
\cJ\equi \cJ_\text{D1} \= \cJ_\text{D5} \= q \,\left( \ell- 2m\right)\,.
\label{eq:DipoleBu}
\end{equation}
We find the same dipole moments as two electromagnetic monopoles of charges $q$ and $-q$ separated by a distance $\ell-2m$. Note however that our geometries depicted in Fig.\ref{fig:neutralconf} are slightly different from this simple picture. First, there are rod sources and second, their separation is $\ell-2\sigma$. Since $\sigma \leq m$, the dipole charges can be made very small while the two bubbles are kept at finite separation with non-zero charges.\footnote{For instance, if we have $\ell-2m\propto \epsilon \to 0$ and $q\propto \epsilon^{-1/2}$ we have $\cJ \propto \epsilon^{1/2}$ while $\ell-2\sigma=\cO(1)$ and $q$ is large. As we will see, such solutions with almost zero dipole charges but with large $q$ and finite separation will exist in the phase space.}

We have therefore described smooth spacetimes that correspond to massive and neutral asymptotically $\IR^{1,3}\times$T$^2$ solutions or $\IR^{1,3}\times$T$^2\times$T$^4$ from a type IIB perspective. The interior is made of three specific loci where one of the T$^2$ directions smoothly degenerates, defining regular ends to spacetime (see Fig.\ref{fig:neutralconf}). The region $r_1=\sigma_1$ ($\rho=0$, $\frac{-\ell}{2}-\sigma \leq z \leq \frac{-\ell}{2}+\sigma$) defines a bolt where the $y_1$-circle degenerates and a non-BPS S$^3$ bubble sits. This bubble has equal electric and magnetic charges corresponding to D1 and D5 charges in type IIB given  in unit of volume by \eqref{eq:rodcharges},
\begin{equation}
q_{1\text{D1}}=q_{1\text{D5}}=2R_{y_2}q\,.\label{eq:charge1stBU}
\end{equation} 
At $r_2=\sigma_2$ ($\rho=0$, $\frac{-\ell}{2}+\sigma \leq z \leq \frac{\ell}{2}-\sigma$), we have a bolt where $y_2$ degenerates, defining a neutral S$^1\times$S$^2$ bubble. Finally, at $r_3=\sigma_3$ ($\rho=0$, $\frac{\ell}{2}-\sigma \leq z \leq \frac{\ell}{2}+\sigma$), we again have a bolt where the $y_1$ degenerates, defining a non-BPS  S$^3$ bubble. This bubble has D1 and D5 charges given in units of volume by
\begin{equation}
q_{3\text{D1}}=q_{3\text{D5}}=-2R_{y_2}q\,.\label{eq:charge3rdBU}
\end{equation}

The $\IR^2$ at the bolts corresponds to smooth $\mathbb{Z}_{k_i}$ quotients over $\IR^2$ if the constraints \eqref{i13_resolution_three_bubble} are satisfied, that is\footnote{Note that from  \eqref{i13_resolution_three_bubble} one should have three constraints. However, since we assume $m_1=m_3$ \eqref{eq:neutralassum}, the bubble equations for the first and third bubbles are identical and give the same equation if we assume the conical defect at the third bubble to be the same as the conical defect on the first one $k_3=k_1$.}
\begin{equation}
k_1 \, R_{y_1}\=\frac{\sqrt{2}\,(\ell+2m)(m+\sigma)}{\sqrt{\ell \sigma}}\,, \qquad k_2 \, R_{y_2} \=\frac{2(\ell^2-4m^2)}{\ell}\,,\qquad k_i \in \mathbb{N}\,.
\end{equation}
Moreover, we will assume that the vacuum bubble in the middle has no conical defect $k_2=1$. This will restrict the class of bubbling geometries but will drastically simplify the analysis of the phase space and the comparison with the neutral bound states of extremal black holes constructed in section \ref{sec:neutralBaB}.

We have two equations for three real and one integer variables, so we end up with a family of solutions with one real and one integer parameters. It is convenient to consider $$\alpha\equi \frac{\sigma}{m} \= \sqrt{1-\frac{q^2}{m^2}\frac{\ell-2m}{\ell+2m}},$$
as the independent real parameter such that $0\leq \alpha \leq 1$. There is only one real solution with positive $m$:
\begin{align}
m &= \frac{\alpha k_1^2 R_{y_1}^2}{(1+\alpha)\sqrt{16 \alpha k_1^2 R_{y_1}^2+(1+\alpha)^2 R_{y_2}^2}}, \qquad q^2=\frac{\alpha^2(1-\alpha) R_{y_1}^4}{(1+\alpha)^2 R_{y_2} \sqrt{16 \alpha k_1^2 R_{y_1}^2+(1+\alpha)^2 R_{y_2}^2}} 
\nonumber \\
 \ell &=\frac{R_{y_2}}{4}+\frac{8\alpha k_1^2 R_{y_1}^2+(1+\alpha)^2 R_{y_2}^2}{4(1+\alpha)\sqrt{16 \alpha k_1^2 R_{y_1}^2+(1+\alpha)^2 R_{y_2}^2}}, \qquad \sigma=m \, \alpha,
\label{eq:alphasol}
\end{align}
We have therefore defined a two-parameter family of smooth non-BPS neutral three-bubble solutions: $k_1$ is integer and corresponds to the orbifold parameter at the outer D1-D5 bubbles, while $\alpha$ is a continuous real parameter $0\leq \alpha\leq 1$.

\subsubsection{Phase space at fixed mass}

The phase space of solutions is characterized by choices for the extra dimensions, $R_{y_1}$ and $R_{y_2}$, and the ADM mass $\cM$ \eqref{eq:ADMmassneutral}. This requires to solve the expression of the ADM mass to get $\alpha$ as a function of $(\cM,R_{y_1},R_{y_2})$. This leads to a polynomial of degree four in $\alpha$ with several interesting branches. In this section, we consider the family of solutions that are macroscopic, i.e.  solutions where $\cM \gg R_{y_1}, R_{y_2}$.  For that purpose, we introduce $\epsilon_2$ as in \eqref{eq:epsilon2} and $\epsilon_1$ such as
\begin{equation}
\epsilon_1 \equi \frac{k_1R_{y_1}}{8\cM}\,,\qquad \epsilon_2 \equi \frac{R_{y_2}}{8\cM}\,.
\label{eq:epsilon}
\end{equation} We will allow the orbifold parameter $k_1$ to be large, such that $\epsilon_1$ is not necessarily small. Therefore, we expand in $\epsilon_2\ll 1$ up to linear order.\footnote{More precisely, this means that all quantities that follow should be accompanied by a $\cO(\epsilon_2^2)$.}  In this limit, the phase space of solutions from the inversion of the ADM mass formula \eqref{eq:ADMmassneutral} can be characterized by the equation
\begin{equation}
\epsilon_1 = \frac{1+\alpha}{\sqrt{\alpha}(3-\alpha)}\left(1-\frac{\epsilon_2}{2}\right), \qquad \mbox{with} \qquad 0\leq \alpha \leq 1.  \label{eq:ep1alpha}
\end{equation}  Interestingly, inverting this relation for $\alpha$ leads to branches for fixed $\epsilon_1$.  It is useful to express all parameters in terms of $\alpha$ subject to the above relation for $\epsilon_1$ \eqref{eq:ep1alpha}.  Up to linear order of $\epsilon_2$ we find
\begin{equation*}
\frac{\ell}{\mathcal{M}} =\frac{4}{3-\alpha}\left(1+(2-\alpha)\frac{ \epsilon_2}{2}\right), \quad \frac{m}{\mathcal{M} } = \frac{2}{3-\alpha}\left(1+\frac{\epsilon_2}{2}\right), \quad \frac{q^2}{\mathcal{M}^2} = \frac{16(1-\alpha^2)}{\epsilon_2(3-\alpha)^3 }\left(1-\frac{3\epsilon_2}{2}\right).  
\end{equation*}

The first important observation is that there is a minimum value for $\epsilon_1$ in \eqref{eq:ep1alpha} occurring at $\alpha = 2\sqrt{3} -3$.  This minimum value for $\epsilon_1$ implies that all solutions satisfy a bound on $\mathcal{M}$ given as
\begin{equation}
8\mathcal{M} \, \epsilon_1^0\,  \leq \, k_1 R_{y_1}, \qquad \mbox{with} \qquad \epsilon_1^0 \= \frac{1}{3} \sqrt{3 + 2 \sqrt{3} } \,\sim\, 0.85. 
\end{equation} The main conclusion is that the orbifold parameter $k_1$ at the outer bubbles needs to be large for $\cM \gg R_{y_1}$. This requirement has already been observed in related situations: one needs a large conical defect at a non-BPS bubble in order to have a configuration larger than the extra-dimension radius \cite{Bah:2020ogh}.  As mentioned in previous work, the conical defect can be resolved with $(k_1-1)$ Gibbons-Hawking bubbles that are in a phase where their characteristic size is much smaller than $R_{y_1}$.  Physically, the implication is that macroscopic states require a large number of elementary degrees of freedom, in this case bubbles.  \\

Next we consider the boundaries of the $\alpha$ interval and the solutions in those regimes. We sketch the bubbles at different values of $\alpha$ along the phase space, and their corresponding $\epsilon_1$, in Fig.\ref{fig:BuPhaseSpace}.

\begin{itemize}
\item[•] \underline{$\alpha \to 1$: the vacuum bubble limit.}

First, we consider the $\alpha \to 1$ limit, which corresponds to $\epsilon_1 \to 1$.  At leading order, the parameters $(\ell, m, \sigma)$ go to finite values that are large and fixed by the ADM mass, while the charges and the rod length of the middle bubble, $\ell-2\sigma$, vanish.  
If we take $\alpha \to 1 + \mathcal{O}(\epsilon_2^2)$, the phase space equation \eqref{eq:ep1alpha} fixes $\epsilon_1=1-\epsilon_2/2+\mathcal{O}(\epsilon_2^2)$.
This implies that the vanishing quantities behave as\footnote{We have substituted back $\epsilon_2 = \frac{R_{y_2}}{8M}$. } 
\begin{equation}
\ell-2\sigma \sim R_{y_2}, 
\qquad q \sim  \sqrt{R_{y_2} \mathcal{M} } , \qquad \mathcal{J} \sim\sqrt{R_{y_2}^3 \cM} \,.
\end{equation}

Thus in the $\alpha \to 1$ limit the dipole charges and the D1-D5 charges are heavily suppressed, as $\sqrt{R_{y_2}^3 \cM}$, while the rod length of the middle vacuum bubble is very small compared to the rod lengths of the outer bubbles.  One can show that the metric and flux \eqref{eq:3rodneutralAnsatz} approaches a solution induced by a single vacuum bubble, of radius $8\mathcal{M}$, corresponding to a bolt along the $y_1$ direction with a conical defect $k_1=8\cM/R_{y_1}$. Here, the top bubble approaches a north hemisphere and the bottom bubble a south hemisphere as depicted in Fig.\ref{fig:BuPhaseSpace}.

\begin{figure}[h]\centering
   \includegraphics[width=1\textwidth]{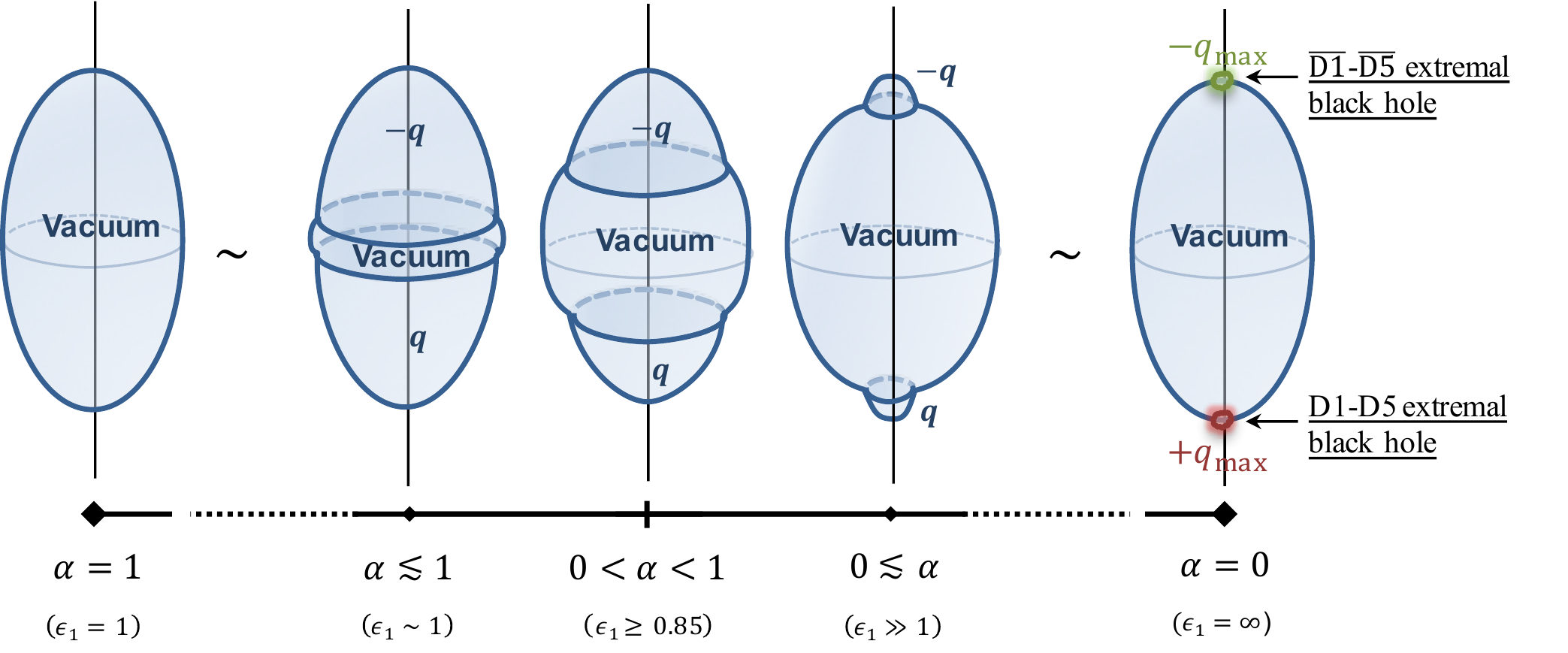}
 \caption{Schematic description of the bubbles in the phase space.  When $\alpha \to 1 $, the solutions approach a vacuum bubble.  When $\alpha \to 0$, the solutions scale towards the BPS/anti-BPS D1-D5 black hole bound states of section \ref{sec:neutralBaB}.}
 \label{fig:BuPhaseSpace}
\end{figure}

\item[•] \underline{$\alpha \to 0$: the BPS/anti-BPS black hole bound state limit.}

We now consider the limit where $\alpha \to 0$, in which $\epsilon_1 \to \infty$.  This corresponds to taking 
 \begin{equation}
 k_1 R_{y_1} \gg 8\mathcal{M} .  
 \end{equation}  We can either interpret this as taking the large $k_1$ limit or the decompactification limit for $y_1$.  We will adopt the former perspective.  From \eqref{eq:ep1alpha}, this also induces $\alpha = (3\epsilon_1)^{-2}$, that is
\begin{equation*}
\ell = \frac{4\cM}{3}\left(1 +\epsilon_2  \right)\,,\quad m = \frac{2\cM}{3}\left(1 -\frac{\epsilon_2}{2} \right)\,,\quad q = \frac{4\cM}{3\sqrt{3\epsilon_2}}\,,\quad \sigma = \frac{2 \cM}{27 \epsilon_1^2}\,, \quad \cJ= \frac{8 \cM^2 \sqrt{\epsilon_2}}{3\sqrt{3}}\,.
\end{equation*}
We see that $\sigma$ gets smaller and smaller, while $q$ is large which means that the two outer bubbles approach their BPS limit where the rods become point particles. The configurations correspond to a large vacuum bubble with two small non-BPS D1-D5 bubbles with opposite charges at its poles. 

At the hypothetical point $\epsilon_1 =\infty$, we retrieve the values for the solutions corresponding to a pair of BPS and anti-BPS D1-D5 black holes on a vacuum bubble, \eqref{eq:macroexpansion0} and \eqref{eq:macroexpansion}. More precisely, one can show that the metric and flux of the bubbling solutions \eqref{eq:3rodneutralAnsatz} are indeed identical to \eqref{eq:WarpFactorBaBneutral} up to small corrections, for all regions outside the immediate vicinity of the first and third bubbles, $r_1,r_3 \gg \epsilon_1^{-2} \cM$. 

Since $\epsilon_1$ can be made arbitrarily large without affecting the conserved charges by increasing the conical defect $k_1$, the solutions can be made arbitrarily close to bound states of extremal black holes. However, they resolve the horizons by replacing them with two small, smooth non-BPS D1-D5 bubbles. To conclude, we have a \emph{scaling family of neutral non-BPS bubbling geometries} that resemble infinitely closely the bound states of BPS D1-D5 and $\overline{\text{D1}}$-$\overline{\text{D5}}$ black holes on a vacuum bubble studied in section \ref{sec:neutralBaB} (see Fig.\ref{fig:BuPhaseSpace}).

\end{itemize}

\subsection{Properties and comparison to Schwarzschild}
\label{sec:BuPhaseSpace}

We will now describe the properties of the macroscopic neutral three-bubble geometries in the regime where $\cM \gg R_{y_1}, R_{y_2}$, with $\epsilon_1 \gg 1$ and $\epsilon_2 \ll 1$. The bubbling geometries are indistinguishable from the bound states of BPS black hole up to a small scale around the black hole horizons, and by transitivity, also indistinguishable from the singular vacuum solution \eqref{eq:neutralSing}. Therefore, most properties of the solutions will resemble those described in section \ref{sec:PropBaB} for the black hole bound states, except in the environment near the black holes.

First, they will have multipole moments in four dimensions as in \eqref{eq:multipolemoments},
with corrections of order the resolution scales, i.e. $\cO(\epsilon_1^{-1},\epsilon_2)$.

\begin{figure}[h]\centering
   \includegraphics[width=0.65\textwidth]{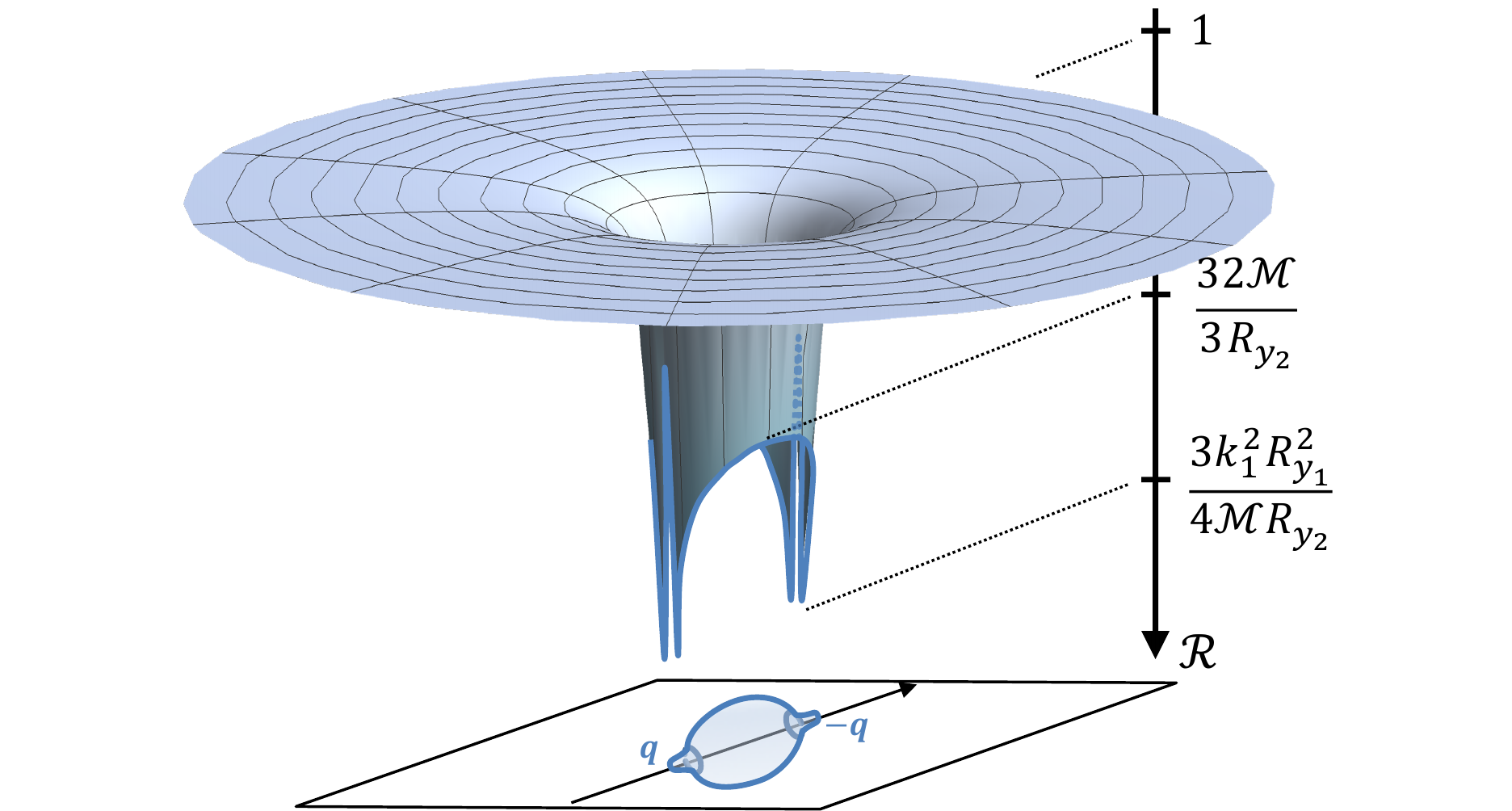}
 \caption{Redshift factor of the neutral three-bubble solutions in the spherical coordinates $(r,\theta)$ of \eqref{eq:SphericalCoordinatesBu}. The surface of the bubble is at $r=\ell +2\sigma$.}
 \label{fig:BuRedshift}
\end{figure}

Second, the redshift factor, given by the norm of the timelike Killing vector $\cR=-g_{tt}^{-1}$ in six dimensions, will get very large as we approach the bubbles. In figure \ref{fig:BuRedshift}, We have plotted the redshift for a configuration with $\epsilon_1=10$ and $\epsilon_2 = 2\times 10^{-2}$ as a function of the spherical coordinates centered around the bubble configuration:
\begin{equation}
\rho \=\sqrt{r(r-(\ell+2\sigma))}\,\sin \theta\,,\qquad z\= \left( r-\left( \frac{\ell}{2}+\sigma\right)\right)\, \cos \theta\,.
\label{eq:SphericalCoordinatesBu}
\end{equation}
The redshift at the bubble locus, $r=\ell+2\sigma$, is similar to the redshift at the vacuum bubble for the black hole bound states, and is of order $\epsilon_2^{-1}\sim \frac{\cM}{R_{y_2}}$. However, the bubble configurations resolve the AdS$_3$ throat at the poles of the vacuum bubble and cap off the redshift factor at order $\frac{\epsilon_1^2}{\epsilon_2} \sim \frac{k_1^2 R_{y_1^2}}{\cM R_{y_2}}$. It is interesting to note that the redshift is relatively small at the outer bubbles compared to the redshift along the vacuum bubble. This gives a kind of ``Isengard tower'' shape to the redshift profile (see Fig.\ref{fig:BuRedshift}). The take-away message is that the redshift is finite and regular everywhere since the geometries are everywhere smooth and is very large at the bubble loci.

Similar arguments can be made for the S$^2$ geometry described by the $(\theta,\phi)$ subspace. They bear the same properties as the black hole bound states up to a scale very close to the bubble locus. First, the area reaches a minimum at $r\approx 1.27 (\ell+2\sigma)$ which is $2.32$ bigger than the horizon area of a Schwarzschild black hole with the same mass as given in \eqref{eq:AreaMin}. From there, the S$^2$ opens up in a similar manner as depicted in Fig.\ref{fig:B/aBAreaS2} but, unlike Fig.\ref{fig:B/aBAreaS2}, ends at $r=\ell+2\sigma$ with a finite area. The asymmetry factor of the S$^2$ \eqref{eq:AsymmetryBaB} is also similar to Fig.\ref{fig:B/aBAssym}, in that the S$^2$ gets very asymmetric after it reaches its minimal area.  Therefore, the S$^2$ of the scaling neutral bubbling geometries has a ``bag'' feature: it suddenly opens up into a very asymmetric space after reaching a radius of order $1.52$ times the Schwarzshild radius and caps off smoothly as a chain of two small D1-D5 bubbles of opposite charges on a vacuum bubble.

To conclude, the family of scaling neutral bubbling geometries correspond to smooth ultra-compact geometries that are wrapped by electromagnetic flux but asymptotically neutral. They share properties with a Schwarzschild black hole: a very large redshift and an ultra-compact minimal S$^2$. However, they differ on several points, since they resemble a vacuum solution which is not Schwarzschild but instead described by a singular metric, \eqref{eq:neutralSing}.  The S$^2$ has a nontrivial asymmetry and they have large even multipole moments, scaling as $M_{2n} \sim \cM^{2n+1}$. Nevertheless, these are the first examples of ultra-compact bubbling geometries wrapped by flux for which the UV origin in type IIB string theory is well established, and which at the same time bear properties reminiscent of astrophysical black holes. 

We would like to take these investigations further by studying the gravitational footprints of these solutions in more detail. This could be done by deriving null geodesics in these backgrounds, as well as scalar modes and ultimately gravitational wave emissions. Furthermore, it will be crucial to study the phenomenology of these solutions and to understand their physical bounds to be stable states in the theory. For this last point, we hope that they correspond to meta-stable solutions following from the work in \cite{Bah:2021irr,Stotyn:2011tv,Elvang:2002br}. Indeed, these geometries consist of three basic ingredients: two charged bubbles on a vacuum bubble. First, a charged bubble is itself a meta-stable soliton of the theory \cite{Bah:2021irr,Stotyn:2011tv}. Second, even if a vacuum bubble has a quantum instability that forces it to expand when isolated \cite{Witten:1981gj},  the surrounding charged bubbles compress it and could stabilize it under their attraction.  A similar mechanism was proposed in \cite{Elvang:2002br} for two small vacuum black holes on a vacuum bubble. Nevertheless, a thorough analysis of the classical and quantum instability of these solutions would be needed to validate such a claim.


\section{Outlook}

The main achievement of this paper has been to successfully generate the first neutral solitons with internal flux in supergravity theories. This was made possible by the fact that the Einstein-Maxwell equations could be decomposed into decoupled sectors of four-dimensional Ricci-flat equations when axial symmetry, staticity, and an appropriate choice of gauge potentials are assumed \cite{Bah:2021owp,Bah:2021rki,Heidmann:2021cms}. Using and simplifying known solutions to this system of equations \cite{Alekseev:2007re,Alekseev:2007gt,Manko:2007hi,Manko:2008gb,PhysRevD.51.4192}, we have constructed and studied classes of solutions induced by three regular sources in six dimensions and in type IIB. The first class corresponds to bound states of a BPS D1-D5 black hole and a anti-BPS $\overline{\text{D1}}$-$\overline{\text{D5}}$ black hole on a vacuum bubble for which the net charges are zero. The second are smooth neutral three-bubble geometries where the two outer bubbles carry opposite D1-D5 charges and are separated by a vacuum bubble. These preliminary constructions open up a vast field of further projects that we aim to investigate in the future.

The first direction we would like to follow is to generalize our construction of BPS / anti-BPS black hole bound states. In this paper, we have focused on D1-D5 black holes that have equal D1 and D5 charges for the sake of simplicity. However, from \cite{Heidmann:2021cms}, it is \textit{a priori} also possible to consider D1-D5-P black holes with different charges, and generate bound states of a BPS D1-D5-P black hole and an anti-BPS $\overline{\text{D1}}$-$\overline{\text{D5}}$-$\overline{\text{P}}$ black hole on a vacuum bubble. This would be particularly interesting because unlike their D1-D5 partners, BPS D1-D5-P black holes have a large macroscopic area. Moreover, a large part of their microstates, called superstrata \cite{Bena:2015bea,Bena:2016ypk,Bena:2016agb,Bena:2017xbt,Ceplak:2018pws,Heidmann:2019zws,Heidmann:2019xrd}, correspond to smooth geometries which resolve their horizons into a smooth global AdS$_3\times$S$^3$ cap in their near-horizon region. One can therefore hope to construct, at least numerically, bound states of a superstratum with an anti-BPS superstratum separated by a smooth vacuum bubble for which the net charges can be tuned to zero. 

\noindent A crucial question to address will be that of classical and quantum stability, which we would like to study numerically or analytically in a later project. We would also like to analyze the gravitational footprints of the new bound state solutions by studying probe geodesics, scalar waves, and quasi-normal modes in these backgrounds. This would be of particular interest with the advent of gravitational wave detection and black hole imaging.

The second direction concerns smooth topological solitons without horizon. In this paper we have focused on the minimum to generate Schwarzschild-like states in gravity, and many corners remain to be explored. First, the phase space of the  three-bubble solutions we have constructed is larger if we do not restrict ourselves to macroscopic solutions, $\cM \gg R_{y_1}, R_{y_2}$, and also if we allow conical defects at the vacuum bubble. Generic solutions can range from the particle scale, of order the size of the extra-dimensions, to the astrophysical scale. Some may have small or large electromagnetic dipoles, small or large internal charges, small or large bubble size, etc. A wide variety of solutions can be constructed, not only those resembling the BPS / anti-BPS black hole boundary states studied in the section \ref{sec:BuPhaseSpace}. Second, many other families of neutral topological solitons can be generated from other supergravity frameworks, such as in \cite{Heidmann:2021cms}, where the equations of motion also decompose into sectors of four-dimensional Ricci flat equations. They will have their own features that will be interesting to explore. Moreover, we have used here a specific two-rod ansatz with arbitrary charges which was originally developed to describe superposition of two Reissner-Nordstr\"om black holes in four dimensions \cite{Alekseev:2007re,Alekseev:2007gt,Manko:2007hi,Manko:2008gb,PhysRevD.51.4192}. We can use its generalization to an arbitrary number of sources with arbitrary charges \cite{PhysRevD.51.4192} to construct neutral smooth bubbling geometries with an arbitrary number of bubbles. 

\noindent Moreover, it will be crucial to study the classical and quantum stability of these neutral bubbling geometries in the manner of \cite{Bah:2021irr}. We would also like to have a better understanding of the geometries by studying null geodesics, quasi-normal modes and other features characterizing their gravitational footprints, following the roadmap of \cite{Mayerson:2020tpn}.


\vspace{-0.3cm}
\section*{Acknowledgments}
The work of IB, PH, and PW is supported in part by NSF grant PHY-2112699. The work of IB is also supported in part by the Simons Collaboration on Global Categorical Symmetries. We are grateful to Iosif Bena, Anindya Dey, Daniel Mayerson and Nick Warner for interesting and stimulating discussions. We thank F. Bacchini for pointing out a typo in a solution.

\vspace{1cm}

\appendix
\leftline{\LARGE \bf Appendices}
\section{Reduction to four dimensions}
\label{App:Profile4d}

We describe the truncation of the six-dimensional theory in \eqref{eq:Action6d} to five and four dimensions that will provide lower-dimensional descriptions of our constructions. We will truncate all fields that are not turned on in the ansatz \eqref{eq:WeylSol2circle}.

\subsubsection*{Reduction to five dimensions}

The KK reduction of the action \eqref{eq:Action6d} on $y_2$ for an ansatz of the form \eqref{eq:WeylSol2circle} gives the following five-dimensional theory 
\begin{align}
S_{5}^\text{KK} &= \frac{1}{16 \pi G_5 } \int d^5x \sqrt{-\det g}\, \left(R - \mathcal{L}_5 \right) \\
\mathcal{L}_5 &= \frac{1}{2}\,\partial_\mu \Phi_2 \,\partial^\mu \Phi_2+ \frac{e^{\frac{\sqrt{2}}{\sqrt{3}}\,\Phi_2}}{2}\,\left| F^{(m)}\right|^2  + \frac{e^{-\frac{\sqrt{2}}{\sqrt{3}}\,\Phi_2}}{2}\,\left| F^{(e1)}\right|^2 \nonumber 
\end{align}
with $G_5 \equiv \frac{G_6}{2\pi R_{y_2}}$.  The five-dimensional theory contains a dilaton, $\Phi_2$, a pair of one-form gauge fields with field strength $F^{(m)}$, and a two-form gauge field with field strength $F^{(e1)}$.  These are identified with the flux in six dimensions as
\begin{equation}
F_3 \= F^{(m)} \wedge dy_2 \+ F^{(e1)} .
\end{equation}
In this framework, the solution ansatz \eqref{eq:WeylSol2circle} reduces to 
\begin{align}
ds_{5}^2 &\= \left(\frac{Z}{W_2} \right)^{\frac{1}{3}}\left[\frac{1}{Z} \left(- W_1\,dt^2 + \frac{ dy_1^2}{W_1} \right)  + Z W_2 \,\left(e^{2(\nu_Z+\nu_W)} \left(d\rho^2 + dz^2 \right) +\rho^2 d\phi^2\right) \right] ,  \nn \\
 e^{\sqrt{\frac{3}{2}}\Phi_2}  &\= \frac{W_2}{Z}, \qquad F^{(m)} \= dH \wedge d\phi\,,\qquad  F^{(e1)} = dT \wedge dt \wedge dy_1\,. \label{eq:5dRed}
\end{align}

\subsubsection*{Reduction to four dimensions}

After compactification on $y_1$, we restrict to a truncation of the KK theory to the Einstein-Maxwell-Dilaton theory
\begin{align}
S_{4}^\text{KK} &=  \frac{1}{\left(16 \pi G_4\right) } \int d^4x \sqrt{-\det g}\,\left( R - \mathcal{L}_4 \right), \\
\mathcal{L}_4 &= \frac{1}{2}\,\partial_\mu \Phi_2 \,\partial^\mu \Phi_2 + \frac{1}{2}\,\partial_\mu \Phi_1 \,\partial^\mu \Phi_1 + \frac{e^{\frac{\sqrt{2}}{\sqrt{3}}\,\Phi_2-\frac{1}{\sqrt{3}}\,\Phi_1}}{2}\,\left| F^{(m)}\right|^2 + \frac{e^{-\frac{\sqrt{2}}{\sqrt{3}}\,\Phi_2+\frac{1}{\sqrt{3}}\,\Phi_1}}{2}\,\left| F^{(e)}\right|^2. \nonumber 
\end{align}
with $G_4 \equiv \frac{G_6}{(2\pi)^2 R_{y_1}R_{y_2}} $. The gauge fields are identified as 
\begin{equation}
 F_3 \= F^{(m)} \wedge dy_2 \+ F^{(e)}\wedge dy_1.
\end{equation} Note that we have turned off a KK vector associated to $y_1$ and the components of $F_3$ orthogonal to $y_1$ and $y_2$ in the truncation.  The solution ansatz \eqref{eq:WeylSol2circle} is 
\begin{align}
ds_4^2 &=  -\sqrt{\frac{W_1}{W_2}}\,\frac{dt^2}{Z}  \+Z\,\sqrt{\frac{W_2}{W_1}}\,\left[e^{2(\nu_Z+\nu_W)} \left(d\rho^2 + dz^2 \right) +\rho^2 d\phi^2 \right] , \label{eq:4dFrameworkchargedApp}\\
e^{\sqrt{3}\,\Phi_1} &= Z\,\sqrt{W_1^3 W_2}, \qquad e^{\sqrt{\frac{3}{2}}\Phi_2}  = \frac{W_2}{Z}, \qquad F^{(m)} \= dH \wedge d\phi\,,\qquad  F^{(e)} = dT \wedge dt\,. \nn
\end{align}


\section{Maxwell sector ansatz with arbitrary charges}
\label{App:tworodansatz}

In this appendix, we present the ansatz we use to solve the Maxwell sector of \eqref{eq:EOMWeyl}. The equations \eqref{eq:EOMWeyl} admit the same structure as four-dimensional Einstein-Maxwell equations, allowing us to make use of known solutions describing bound states of four-dimensional Reissner-Nordstr\"om black holes \cite{Alekseev:2007re,Alekseev:2007gt,Manko:2007hi,Manko:2008gb,PhysRevD.51.4192}.
While the linear ansatz of \cite{Bah:2020pdz,Bah:2021owp,Bah:2021rki,Heidmann:2021cms} require fixed charge to mass ratios for all rod sources, the Reissner-Nordstr\"om bound state  ansatz employed here allows sources of arbitrary charges.

\subsection{The ansatz}

The solutions constructed and studied in \cite{Alekseev:2007re,Alekseev:2007gt,Manko:2008gb} can be expressed in terms of a warp factor $Z$, two electric and magnetic gauge potentials $T$ and $H$, and a base warp factor $\nu_Z$ satisfying the same equations given in the Maxwell sector of \eqref{eq:EOMWeyl}. They describe a pair of dyonic Reissner-Nordstr\"om black holes of arbitrary charge. We use identical warp factors and gauge potentials in our six-dimensional constructions.

In either the four-dimensional or six-dimensional context, these Maxwell sector solutions are sourced by a pair of charged rods of size $2\sigma_i$ aligned on the $z$-axis and centered at $z=z_i$ (see Fig.\ref{fig:rodsources}(b)). Overall, they are characterized by five parameters: two mass parameters $m_i\geq 0$, an inter-center distance $\ell$, and two charge parameters $q_i$ sourcing the electric gauge field and its magnetic dual. The rod widths are fixed in terms of these parameters as
\begin{equation}
\sigma_1 \equi \sqrt{m_1^2-q_1^2+2 q_1 \gamma }\,,\qquad  \sigma_3 \equi \sqrt{m_3^2-q_3^2-2 q_3 \gamma }, \qquad
\gamma \equi \frac{m_3 q_1 \- m_1 q_3}{\ell+m_1+m_3} .
\end{equation}
In anticipation of the insertion of a bubble between these rod sources sourced in the vacuum sector we have adopted the labels $i=1$ and $i=3$, reserving $i=2$ for this intermediate source.  Note that for the rod sources to be non-overlapping, $\ell$ is bounded by below as
\begin{equation}
\ell \geq \sigma_1 +\sigma_3.
\end{equation}
Consistency with neutral solutions, $q_1=-q_3$, further requires that we restrict
\begin{equation}
\ell \geq m_1 +m_3,
\end{equation} 
to ensure real $\sigma_i$. For a detailed account of the range of possible $q_i$ values, we refer the reader to the next subsection.

In \cite{Alekseev:2007re}, the solutions are written in terms of the electric gauge potential, $T$. Here we first summarize and simplify the electric two-rod solutions of \cite{Alekseev:2007re}, and then use \cite{Manko:2008gb} to derive an expression for the magnetic gauge potential. As in \eqref{eq:r_defs}, we denote the distances to the rod endpoints $r_\pm^{(i)}$ and the spherical coordinates centered around each rod $(r_i,\theta_i)$, with the locus of the sources located at $r_i=\sigma_i$. With these preliminaries aside, the ansatz for the warp factors $Z$, $\nu_Z$ and the gauge potential $T$ is given by
\begin{align}\label{eq:two-rod_sols}
Z & \=  \frac{(r_1+m_1)(r_3+m_3)-\left(q_1-\gamma(1+\cos \theta_3)\right)\left(q_3+\gamma(1-\cos \theta_1)\right)}{\sqrt{\left(r_1^2-\sigma_1^2+\gamma^2\sin^2 \theta_3 \right)\left(r_3^2-\sigma_3^2+\gamma^2\sin^2 \theta_1 \right)}}\,, \nonumber\\
T & \= \frac{q_1 r_3 +q_3 r_1 +\gamma \left(m_1 \cos \theta_1 +r_1 +m_3 \cos \theta_3 -r_3 \right)}{(r_1+m_1)(r_3+m_3)-\left(q_1-\gamma(1+\cos \theta_3)\right)\left(q_3+\gamma(1-\cos \theta_1)\right)} \,,\\
e^{4\nu_Z} & \=  \frac{\left(r_1^2-\sigma_1^2+\gamma^2\sin^2 \theta_3 \right)\left(r_3^2-\sigma_3^2+\gamma^2\sin^2 \theta_1 \right)}{(1+2\delta)^2\,(r_1^2-\sigma_1^2 \cos^2\theta_1)(r_3^2-\sigma_3^2 \cos^2\theta_3)}  \nonumber\\
& \hspace{-0.4cm}\times \left(1+2\delta \,\frac{q_1 \left( r_1- \ell \cos \theta_1\right)+q_3 \left( r_3+ \ell \cos \theta_3\right)+\gamma \left( r_3-r_1+(l+m_3)\cos \theta_1+(l+m_1)\cos \theta_3 \right)}{q_1 r_3 +q_3 r_1 +\gamma \left(m_1 \cos \theta_1 +r_1 +m_3 \cos \theta_3 -r_3 \right)} \right)^2 .\nonumber
\end{align}
Note that we have introduced one additional parameter, 
\begin{equation}
\delta \equi \frac{m_1 m_3 -(q_1-\gamma)(q_3+\gamma)}{\ell^2-m_1^2-m_3^2+(q_1-\gamma)^2+(q_3+\gamma)^2}\,.
 \label{eq:parametersApp}
\end{equation}
For subsequent discussions of regularity it is useful to observe the behavior of the functions $Z$, $\nu_Z$ on the $z$-axis. One finds that while $Z$ is singular at the rod sources, as $Z \sim \rho^{-1} \sim (r_i-\sigma_i)^{-1/2}$, it remains a finite, non-vanishing function of $z$ between, above, and below the sources. The warp factor $e^{2\nu_Z}$ exhibits particularly simple behavior on the $z$-axis, vanishing near the sources as $e^{2\nu_Z}\sim \rho \sim (r_i-\sigma_i)^{1/2}$, while 
\begin{align}
e^{2\nu_Z}=1 \text{ above and below the rods, }&\qquad  e^{2\nu_Z}=\frac{1-2\delta}{1+2\delta} \text{ between the rod sources.} 
\end{align}
In a given six-dimensional solution constructed using this ansatz, the weights $G_i^{(I)}$ in the vacuum sector solution \eqref{eq:HarmFuncGen} must be tuned to ensure regular, coordinate singularities at the locus of the rod sources (and on the intervening segment). Away from the $z$-axis, note that $Z$ and $e^{2\nu_Z}$ are automatically positive, finite, and non-vanishing. They satisfy the appropriate asymptotic behavior as well, with $Z$, $e^{2\nu}$ approaching unity far from the sources.

Since we are interested in solutions with electromagnetic flux, one needs to derive the magnetic dual of $T$, \eqref{eq:EMdual}. In practice, obtaining $H$ via integration of \eqref{eq:EMdual} does not appear feasible, so we instead use the known result obtained from inverse-scattering methods. The magnetic dual of $T$ is given by
\begin{equation}
H \=\frac{r_1 \cos \theta_1 + r_3 \cos \theta_3}{2}\,T \+ \frac{\mathcal{I}}{\eta\,\cA}\,,
\label{eq:tworodsMagneticPot}
\end{equation}
where we have defined
\begin{align*}
\eta \equi  &\ell^2-m_1^2-m_3^2 + \left(q_1-\gamma\right)^2+ \left(q_3+\gamma\right)^2 \,,
\end{align*}
\begin{align}
\cA \equi  & (r_1+m_1)(r_3+m_3)-\left(q_1-\gamma(1+\cos \theta_3)\right)\left(q_3+\gamma(1-\cos \theta_1)\right)\label{eq:I&ADeg} \\
&+ \delta \Bigl[-\eta +(r_1+2 m_1 )r_1 +(r_3+2 m_3 )r_3 - \left( 2 m_1 \ell+ \sigma_1^2 \cos \theta_1 -2 \gamma( q_3+\gamma)\right) \cos \theta_1 \nonumber\\ 
& \hspace{0.8cm}+  \left( 2 m_3 \ell- \sigma_3^2 \cos \theta_3 +2 \gamma( q_1-\gamma)\right) \cos \theta_3 + 2 \eta \delta \cos \theta_1 \cos \theta_3 \Bigr]. \nonumber
\end{align}
\begin{align*}
\mathcal{I} \equi   & \gamma \eta^2 (4\delta^2-1)+ \left( m_3 \ell (q_1-\gamma)-\gamma( \eta \delta+m_3^2)+\gamma (q_1-\gamma)^2\right) \left(r_3^2-\sigma_3^2 \cos^2 \theta_3\right) \nonumber\\
& - \left( m_1 \ell (q_3+\gamma)+\gamma(  \eta \delta +m_1^2)-\gamma (q_3+\gamma)^2\right) \left(r_1^2-\sigma_1^2 \cos^2 \theta_1\right)\nonumber \\
&+ \gamma \eta (1-2 \delta )\left((\eta \delta+\gamma^2)\cos \theta_1\,\cos \theta_3- r_1 r_3 \right) \nonumber \\
&+\left(m_1(q_1-\gamma)(2\gamma^2-\eta)-2m_3(q_3+\gamma)\sigma_1^2 \right) r_3 \cos \theta_1\nonumber \\
&+\left(m_3(q_3+\gamma)(2\gamma^2-\eta)-2m_1(q_1-\gamma)\sigma_3^2 \right) r_1 \cos \theta_3\nonumber \\
& + \left[ \gamma \bigl(2 \delta (2m_3-m_1)+ (2m_1-m_3) \bigr)- \frac{\ell}{2} \left(2 \delta (q_1-\gamma)- (q_3+\gamma)	 \right)\right] \eta\, r_1\nonumber \\
& + \left[ \gamma \bigl(2 \delta (2m_1-m_3)+ (2m_3-m_1) \bigr)+ \frac{\ell}{2} \left(2 \delta (q_3+\gamma)- (q_1-\gamma)	 \right)\right]\eta\, r_3\nonumber \\
& + \Bigl[ \left( 2\delta (q_1-\gamma)+ (q_3+\gamma)\right)\sigma_1^2 - \left( (q_1-\gamma) \delta \ell -(2 m_1 +m_3) \gamma \delta + \frac{m_1 \gamma }{2}\right) \ell\nonumber \\
&\hspace{0.5cm}+(q_1-q_3-2\gamma)\gamma^2(2\delta-1)\Bigr]\eta \, \cos \theta_1 \nonumber \\
& + \Bigl[ \left( 2\delta (q_3+\gamma)+ (q_1-\gamma)\right)\sigma_3^2 - \left( (q_3+\gamma) \delta \ell +(2 m_3 +m_1) \gamma \delta - \frac{m_3 \gamma}{2}\right) \ell\nonumber \\
&\hspace{0.5cm}+(q_3-q_1+2\gamma)\gamma^2(2\delta-1)\Bigr] \eta \, \cos \theta_3 ,
\end{align*}

One can check that $H \sim q_i \cos \theta_i + f(\theta_i)$ and $T \sim \text{cst} + \cO(r_i-\sigma_i)$ at the $i^\text{th}$ rod, $r_i \to \sigma_i$, where $f(\theta_i)$ is a function that does not contribute to the integral of the flux.\footnote{More precisely, $f(\pi)=f(0)$.} This means that each rod carries \emph{two equal electric and magnetic charges given by $q_i$}. Moreover, at large distance $r=\sqrt{\rho^2+z^2}\to \infty $, we have $Z^{-1}\sim 1 -\frac{m_1+m_3}{r}$. In other words, $m_i$ are the \emph{four-dimensional mass induced by both rods} \eqref{eq:AsymptoticExpGen}. Therefore, the ansatz described superpositions of \emph{massive two-charge objects on a line where the charges are equal to $q_i$ and the mass are given by $m_i$}, as claimed previously.

Unlike the linear ansatz derived in \cite{Bah:2020pdz,Bah:2021owp,Bah:2021rki,Heidmann:2021cms}, there is no constraint on the sign of the charges carried by the rod sources. Moreover, the linear ansatz can be obtained as a special case of the present solution by fixing the sources to have the same mass-to-charge ratio, $\frac{m_1}{q_1} = \frac{m_3}{q_3}$. This restriction  sets $\gamma=0$ and drastically stimplifies the form of the solutions. 

\subsection{Regimes of charges and BPS limits}
\label{App:qRegimes}

The ansatz presented above requires the size of the rods, $\sigma_i$ \eqref{eq:parametersApp}, to be well-defined. This constrains non-trivially the values of the charges carried by the objects with respect to their induced masses $m_i$ and the distances $\ell$.
In particular, $\sigma_i\in \IR$ requires that 
\begin{equation}
m_i^2 -q_i^2 \,\frac{\ell+m_i-m_j}{\ell+m_i+m_j} -q_i q_j \frac{2m_i}{\ell+m_i+m_j}\,\geq \,0 \,,\qquad i\neq j\,,
\end{equation}
with $\ell \geq m_1 +m_3$ to ensure consistency with  neutrality, $q_1=-q_3$. These relations constrain the values for the charge parameters $q_i$ to lie within the ``diamond'' depicted in Fig.\ref{fig:qregime}. 
In the remainder of this appendix, we consider the form of the arbitrary charge ansatz at the four  BPS points in the $q_i$ parameter space. As noted in the main text, these points are the four apexes of the diamond. Recall also that from the IIB perspective, these four points correspond to brane/brane, antibrane/brane, brane/antibrane, and antibrane/antibrane configurations. 

\subsubsection{Bound states of brane/brane or anti-brane/anti-brane}

We first consider the two points in the parameter space Fig.\ref{fig:qregime} with $(q_1,q_3)=\pm (m_1,m_3)$. The warp factors and gauge potentials $(Z,H,T,\nu_Z)$ given in \eqref{eq:two-rod_sols} and \eqref{eq:tworodsMagneticPot} drastically simplify,  to
\begin{equation}
Z \= 1+ \frac{m_1}{r_1}+\frac{m_3}{r_3}\,,\quad T \= \pm \left( 1-\frac{1}{Z}\right)\,,\quad H \= \mp \left(m_1 \cos \theta_1 +m_3 \cos \theta_3 \right)\,,\quad e^{2\nu_Z} =1\,.
\end{equation}
We recognize the supersymmetric solutions corresponding to two BPS dyonic point particles. In the type IIB embedding, they correspond to two supersymmetric D1-D5 point sources (for the ``$+$'' solutions) or $\overline{\text{D1}}$-$\overline{\text{D5}}$ point sources (for the ``$-$'' solutions). 
One could further develop these bound states of two D1-D5 black strings or two $\overline{\text{D1}}$-$\overline{\text{D5}}$ black strings at the poles of vacuum bubble, by turning on appropriate sources for the vacuum sector $(W_I, \nu_W)$, but since the solutions cannot be made neutral, we will instead focus on bound states consisting of one D1-D5 and one $\overline{\text{D1}}$-$\overline{\text{D5}}$ black string.

\subsubsection{Bound states of anti-brane/brane or brane/anti-brane}
We now consider the pair of points in the parameter space Fig.\ref{fig:qregime} given by $(q_1,q_3)=\pm (q_{1\text{max}},-q_{3\text{max}})$, with $q_{i\text{max}}$ as in \eqref{eq:qmax}. The warp factors and gauge potentials \eqref{eq:two-rod_sols} and \eqref{eq:tworodsMagneticPot},  give\footnote{We have removed an irrelevant constant in the magnetic gauge potential $H$ in \eqref{eq:tworodsMagneticPot} by shifting $H \to H \mp \ell \,\sqrt{\frac{\ell^2-(m_1+m_3)^2}{\ell^2-(m_1-m_3)^2}} $.}
\begin{align*}
Z &\= 1+ \frac{\left(m_1^2-m_3^2 \right)(m_3 r_1-m_1 r_3)+\ell^2 (m_3 r_1+m_1 r_3+2 m_1 m_3)}{(m_3 r_1-m_1 r_3)(m_1 r_1-m_3 r_3)+\ell^2 (r_1 r_3-m_1 m_3)} \,,\\
 e^{2\nu_Z} &\=\left( \frac{(m_3 r_1-m_1 r_3)(m_1 r_1-m_3 r_3)+\ell^2 (r_1 r_3-m_1 m_3)}{\left(\ell^2-(m_1-m_3)^2 \right)\,r_1\,r_3} \right)^2\,.
\end{align*}
\begin{equation}
\begin{split}
 T &\=\pm \frac{\sqrt{\left(\ell^2-(m_1+m_3)^2 \right)\left(\ell^2-(m_1-m_3)^2 \right)}\,\left( m_1 r_3-m_3 r_1 \right)}{(m_3 r_1-m_1 r_3)(m_1 r_1-m_3 r_3)+\ell^2 (r_1 r_3-m_1 m_3)}\,\frac{1}{Z}\,,\\
  H &\= \pm\,\frac{\sqrt{\left(\ell^2-(m_1+m_3)^2 \right)\left(\ell^2-(m_1-m_3)^2 \right)}}{2} \,\\
  &\hspace{0.5cm}\times \frac{(r_1 \cos \theta_1 +r_3 \cos \theta_3) (m_3 r_1-m_1 r_3)-\ell (m_3 r_1+m_1 r_3+2 r_1 r_3)}{(m_3 r_1-m_1 r_3)(m_1 r_1-m_3 r_3)+\ell^2 (r_1 r_3-m_1 m_3)}\,,\\
\end{split}
\label{eq:BaBMaxSec}
\end{equation}
The configurations are not BPS anymore since $Z$ is not harmonic and $e^{2\nu_Z}\neq 1$ (and thus the three-dimensional base is not flat). While they are sourced by two BPS two-charge sources, the sources are of opposite supersymmetry.  In type IIB supergravity, we have bound states of a BPS D1-D5 point source with a anti-BPS $\overline{\text{D1}}$-$\overline{\text{D5}}$  point source. To make these point sources correspond to regular bound states of extremal D1-D5 and $\overline{\text{D1}}$-$\overline{\text{D5}}$ black strings, one needs to nucleate a vacuum bubble where $y_2$ smoothly degenerates in between them. We detail this construction in the next Appendix.


\section{BPS D1-D5 brane $-$  anti-BPS $\overline{\text{D1}}$-$\overline{\text{D5}}$ anti-brane on a regular bolt}
\label{App:BaBConstruction}

In this appendix, we discuss how the BPS point source solutions to the Maxwell sector given in \eqref{eq:BaBMaxSec} can be incorporated into a full six-dimensional construction describing bound states of a BPS D1-D5 brane and a anti-BPS $\overline{\text{D1}}$-$\overline{\text{D5}}$ anti-brane at the poles of a vacuum bubble.

Our starting point is the generic Weyl ansatz \eqref{eq:WeylSol2circle}, repeated here for convenience,
\begin{align}
ds_{6}^2 = &\frac{1}{Z} \left[- W_1\,dt^2 + \frac{ dy_1^2}{W_1} \right] + \frac{Z}{W_2}\, dy_2^2 + W_2 Z\,\left[e^{2(\nu_Z+\nu_W)} \left(d\rho^2 + dz^2 \right) +\rho^2 d\phi^2\right] ,\nonumber\\
 F_3 \= & d\left[ H \,d\phi \wedge dy_2 \+ T \,dt \wedge dy_1 \right]\,,
\end{align}
with the solutions for $(Z, H,T,\nu_Z)$ supplied by \eqref{eq:BaBMaxSec}. That is, we fix the charge parameters to extremal values of opposite sign, $(q_1,q_3)=(q_{1\text{max}},-q_{3\text{max}})$, with $q_{i\text{max}}$ as in \eqref{eq:qmax}. Without loss of generality, we have chosen to label the source with positive charge parameter $i=1$, as in section \ref{sec:BaB}. Choosing $z=0$ midway between the sources, we thus have a system of two BPS point sources located at $z_1=-\frac{\ell}{2}$ and $z_3=\frac{\ell}{2}$.

\subsection*{Insertion of a regular bolt}
In order to ensure that the metric is well-behaved in the region local to the BPS point sources, we must turn on the vacuum sector warp factors $W_I$, $e^{2\nu_W}$. To see this clearly, adopt local coordinates
\begin{equation}
\rho=\Lambda \rho_3^2 \sin(2\tau_3) , \qquad z=\frac{\ell}{2}+\Lambda \rho_3^2 \cos(2\tau_3), \qquad \Lambda \equiv 4\ell m_3^2 \frac{(\ell +m_1+m_3)^2}{(\ell-m_1 +m_3)^2}
\end{equation}
as in \eqref{eq:localCoorBaB}, and consider the geometry local to the point source at $\rho=0$, $z=\ell/2$, or equivalently $\rho_3 = 0$, $0 \leq \tau_3 \pi/2$. Under this change of coordinates, the $\rho$, $z$ portion of the line element transforms as
\begin{equation}
d\rho^2+dz^2 =4\Lambda^2 \rho_3^4 \left( \frac{d\rho_3^2}{\rho_3^2}+d\tau_3^2 \right).
\end{equation}
Near the locus $\rho_3 \rightarrow 0$, $Z \sim \rho_3^{-2}$, so unless we at least turn on $W_2$, the metric coefficient $g_{y_2 y_2}$ will be singular. 
To determine the appropriate vacuum sector sources  to avoid such behavior, note that if
\begin{equation}
e^{2\nu_W} \sim \rho_3^{\alpha_0}, \qquad  W_I \sim \rho_3^{\alpha_I}, 
\end{equation}
near the point source, for some numbers $\alpha_0$, $\alpha_1$, $\alpha_2$, then our metric coefficients will scale as
\begin{align}
g_{tt} \sim \rho_3^{2+\alpha_1}, \qquad g_{y_1 y_1} \sim \rho_3^{2-\alpha_1}, \qquad  g_{y_2 y_2} \sim \rho_3^{-2-\alpha_2}, 
\nonumber
\\
g_{\phi \phi} \sim \rho_3^{2+\alpha_2}, \qquad g_{\tau_3 \tau_3} \sim \rho_3^{2+\alpha_0+\alpha_2}, \qquad g_{\rho_3 \rho_3} \sim \rho_3^{\alpha_0+\alpha_2}.
\end{align}
We have used the fact that near this locus $e^{2\nu_Z} \sim \mathcal{O}(1)$.
We see that when $\alpha_2=-2$, the $\rho_3$-dependence of the $y_2$ and $\phi$ fibers is simultaneously removed. If in addition $\alpha_0=0$, then the entire $(\tau_3, \phi, y_2)$ subspace is free from any scaling with $\rho_3$. As seen in the main text, this subspace can take the form of a warped $S^3$. Furthermore, if $\alpha_1=0$, the scaling of $g_{y_1 y_1}$, $g_{tt}$ and $g_{\rho_3 \rho_3}$ with $\rho_3$ will be appropriate to describe a local AdS$_3$ geometry.

One needs to use the vacuum sector specifically to achieve such a local AdS$_3 \times S^3$ geometry for both point sources. If we insert a $i=2$ rod source for the vacuum sector, centered at $z_2=0$ of width $\sigma_2=\ell/2$, then solutions for the warp factors $(W_I, \nu_W)$ take the form \eqref{eq:HarmFuncGen} with a single pair of weight parameters $G^{(I)}$,
\begin{equation}
W_I=\left(1-\frac{\ell}{r_2+\ell/2}\right)^{-G^{(I)}}, \qquad e^{2\nu_W}=\left(  \frac{E_{+-}^{(2,2)}E_{-+}^{(2,2)}}{E_{++}^{(2,2)}E_{--}^{(2,2)}}\right)^{\frac{1}{2}\,\left(G^{(1)^2}+G^{(2)^2}\right)}  \,.
\end{equation}
At the endpoints of this new source, say at $\rho=0$, $z=\ell/2$, these functions scale with the local coordinate $\rho_3$ as 
\begin{equation}
W_I \sim \rho_3^{-2G^{(I)}}, \qquad e^{2\nu_W} \sim \mathcal{O}(1).
\end{equation}
In other words, inserting such a source sets  $\alpha_0=0$ and $\alpha_I=-2G^{(I)}$. 
We see that a desirable scaling can be achieved by choosing the weights as
\begin{equation}\label{eq:BaBweights}
G^{(1)}=0, \qquad G^{(2)}=1.
\end{equation}
This is also the choice of weights needed for the metric to be smooth along the segment of the $z$-axis between the point sources. For example, if we would keep the vacuum sector turned off, in particular fixing $e^{2\nu_W}=1$, the line element on the three-dimensional $(\rho,z,\phi)$ base  would take the form $ds_3^2=e^{2\nu_Z} (d\rho^2 +dz^2)+\rho^2 d\phi^2$. Since $e^{2\nu_Z}=\frac{1-2\delta}{1+2\delta}$ along this portion of the $z$-axis, such a solution generically describes a strut between the sources, inducing a conical singularity for the $\phi$-circle. However, the insertion of a vacuum sector rod spanning $-\ell/2 \leq z \leq \ell/2$ with weights \eqref{eq:BaBweights} means that in this region
\begin{equation}
W_2 \sim \rho_2^{-2}, \qquad e^{2\nu_W} \sim \rho_2^2,
\end{equation}
so that $g_{y_2 y_2} \sim \rho_2^2$ while all other metric coefficients, including $g_{\phi \phi}$, remain finite and non-vanishing on the interior of the segment. As shown in the main text, this region corresponds to a bolt where the $y_2$ circle smoothly degenerates and the remaining angular directions form a vacuum bubble of topology S$^2\times$S$^1$.

In summary, we can ensure a regular metric between the BPS and anti-BPS point sources by inserting an intervening $i=2$ rod sourcing the vacuum sector. Our complete six-dimensional solution is thus obtained by taking \eqref{eq:BaBMaxSec} for $(Z,H,T,\nu_Z)$ and choosing a vacuum sector solution
\begin{equation}\label{eq:BaBVacSec}
W_2=\left(1-\frac{\ell}{r_2-\ell/2}\right)^{-1}, \qquad e^{2\nu_W}=\frac{4r_2^2-\ell^2}{4r_2^2-\ell^2 \cos^2 \theta_2}.
\end{equation}
This solution can be expressed in terms of the distances $r_1$ and $r_3$ to the BPS point sources,  as in \eqref{eq:WarpFactorBaB}. One could have also tried using KKm point charges to regularize the geometry. However, for a pair of brane and anti-brane sources, the KKm charges will resolve the divergence at the point sources into black strings but will not provide the necessary repulsion between them. Therefore, such a solution will require an intermediate strut (that is, a singular string with negative tension) to prevent the sources from collapse.


\section{Non-BPS D1-D5 bubbling solutions with arbitrary charges}
\label{App:BubbleConstruction}

This appendix details the construction of the three-bubble solutions constructed in \ref{sec:arbitraryB}.
In the generic Weyl ansatz \eqref{eq:WeylSol2circle}, namely
\begin{align}
ds_{6}^2 = &\frac{1}{Z} \left[- W_1\,dt^2 + \frac{ dy_1^2}{W_1} \right] + \frac{Z}{W_2}\, dy_2^2 + W_2 Z\,\left[e^{2(\nu_Z+\nu_W)} \left(d\rho^2 + dz^2 \right) +\rho^2 d\phi^2\right] ,\nonumber\\
 F_3 \= & d\left[ H \,d\phi \wedge dy_2 \+ T \,dt \wedge dy_1 \right]\,,
\end{align}
we take $(Z, H,T,\nu_Z)$ as in \eqref{eq:two-rod_sols} and \eqref{eq:tworodsMagneticPot}. In other words, we source these Maxwell sector warp factors and gauge potentials with a pair of rod sources of width $\sigma_1$, $\sigma_3$.
For the vacuum sector $(W_I, \nu_W)$, we use solutions of the form \eqref{eq:HarmFuncGen}, sourced by three connected rods of width $\sigma_1$, $\sigma_2$, and $\sigma_3$, so that the $i=1,3$ vacuum sector sources exactly overlap with those of the Maxwell sector. There are generically six weights, $(G_i^{(1)},G_i^{(2)})$ for $i=1$ to $3$, which must be specified to complete the solution. We first explain how to fix these weights, before considering the local behavior of the solutions near the sources.

\subsection{Fixing weights}
Our choice for the weights $(G_i^{(1)},G_i^{(2)})$ is fixed by the condition that each of the three segments $z_i -\sigma_i \leq  z \leq z_i+\sigma_i$ correspond to regular bolts on the $z$-axis, where either the $y_1$ or the $y_2$-circle smoothly degenerates while all other metric coefficients remain finite and non-vanishing. In terms of local coordinates
\begin{equation}
\rho= \rho_{i}\,\sqrt{\rho_{i}^2+2\sigma_{i}} \sin \theta_i,\, \qquad z\= \left(\rho_{i}^2+\sigma_i\right) \cos \theta_{i}+z_{i},
\label{eq:localcoordrodApp}
\end{equation}
the locus of these segments consists of $\rho_i \rightarrow 0$ with $0 \leq \theta_i \leq \pi$. To fix the weights, it will be sufficient to consider how the metric coefficients
\begin{equation}
-g_{tt}=\frac{W_1}{Z}, \qquad g_{y_1 y_1} = \frac{1}{Z \, W_1}, \qquad g_{y_2 y_2} = \frac{Z}{W_2}
\end{equation}
scale with $\rho_i$ on each segment.
The scalings of the vacuum sector warp factors with $\rho_i$ near a given source are set by the weights, as
\begin{equation}
W_I \sim \rho_i^{-2G_i^{(I)}}, \qquad e^{2\nu_W} \sim \rho_i^{2 (G_i^{(1)^{2}}+ G_i^{(2)^{2}})}.
\end{equation}
We know from Appendix \ref{App:tworodansatz} that on the $i=1,3$ segments,
\begin{equation}
Z \sim \rho_i^{-1}, \qquad e^{2\nu_Z} \sim \rho_i
\end{equation}
while $Z,\, \nu_Z \sim \mathcal{O}(1)$ on the intermediate $i=2$ segment. Let us focus for the moment on the $i=1,3$ segments where the Maxwell sector is sourced. We require the overall scaling of the metric coefficient $g_{tt}$ to be independent of $\rho_i$ near these sources, so
\begin{equation}
W_1 \sim Z \sim \rho_i^{-1} \qquad \Rightarrow \qquad G_{i=1,3}^{(1)}=\frac{1}{2}.
\end{equation}
By the same token, on the intermediate $i=2$ segment we must have
\begin{equation}
W_1 \sim Z \sim \mathcal{O}(1) \qquad \Rightarrow \qquad G_{i=2}^{(1)}=0.
\end{equation}
With all three weights $G_i^{(1)}$ now fixed, $W_1$ and thus also the $g_{y_1 y_1}$ metric coefficient are specified completely. In particular, $g_{y_1 y_1}$ shrinks as $\rho_i^2$ on the $i=1,3$ segments but scales independently from $\rho_2$ on the $i=2$ segment. We see that at the $i=1,3$ segments we have bolts where the $y_1$-circle degenerates. In order for $g_{y_2 y_2}$ to remain non-vanishing at the $i=1,3$ sources, we must have
\begin{equation}
W_2 \sim Z \sim \rho_i^{-1} \qquad \Rightarrow \qquad G_{i=1,3}^{(2)}= \frac{1}{2}
\end{equation}
here, so that $g_{y_2 y_2} \sim \mathcal{O}(1)$.  Similarly, to generate a bolt where the $y_2$ circle degenerates as $\rho_2^2$ on the intermediate segment, we must have
\begin{equation}
W_2 \sim \rho_2^{-2} \qquad \Rightarrow \qquad G_{i=2}^{(2)}=1
\end{equation}
in this region. With the definitions \eqref{eq:three_source_defs}, this choice of weights reduces the ansatz to that of \eqref{eq:3rodAnsatz} and \eqref{eq:nu3rodsGen}. As a cursory regularity check, note that fixing the weights as described above ensures that on the three rods of the $z$-axis
\begin{equation}
e^{2(\nu_Z+\nu_W)} \sim \rho_i^2 \qquad \Rightarrow \qquad g_{\rho_i \rho_i}, \, g_{\theta_i \theta_i} ,\, g_{\phi \phi} \sim \mathcal{O}(1),
\end{equation}
as required for these sources to be bubbles where the $\phi$-circle has been blown up. 

\subsection{Local behavior at the sources}
While a variety of basic regularity conditions for these three-bubble solutions have been addressed in the main text, the local behavior of the metric and gauge potentials near the rod sources merits further elaboration. Adopting local coordinates to a given source as in \eqref{eq:localcoordrodApp}, the line element on the $\rho$, $z$ subspace transforms as
\begin{equation}
d\rho^2+dz^2 = 2\sigma_i \sin^2 \theta_i \left( d\rho_i^2+\frac{\sigma_i}{2} d\theta_i^2 \right)
\end{equation}
near the locus of the source. We first consider the middle segment $-\sigma_2 \leq z \leq \sigma_2$, before discussing the $i=1,3$ segments $z_i-\sigma_i \leq z \leq z_i +\sigma_i$.

\begin{enumerate}

\item[•] \textbf{Geometry at the vacuum bubble:}
The metric in the region local to the $i=2$ rod source takes the form
\begin{align}
& ds_6^2|_{dt=0} \propto d\rho_2^2 +\frac{\rho_2^2}{C_2^2} dy_2^2  + \tilde{g}_{\phi \phi}(\theta_2) d\phi^2+\frac{\sigma_2}{2} \left[ d\theta_2^2+\tilde{g}_{y_1 y_1}(\theta_2) \sin^2 \theta_2 dy_1^2 \right]
\nonumber \\
& F_3 = t_2(\theta_2) \, \sin \theta_2 \, d\theta_2 \wedge dt \wedge dy_1+\mathcal{O}(1),
\end{align}
where $t_2(\theta_2)$ is a well-behaved function over the entire range $0 \leq \theta_2 \leq \pi$, so that component of $F_3$ along $dy_1$ vanishes at the points of intersection with the charged rods sitting above and below. Note that since $F_3$ has no leg on either $dy_2$ or $d\phi$, and $\star F_3$ has no non-vanishing leg on $d\theta_2$, the $i=2$ source does not carry D1 and D5 brane charges \eqref{eq:typeIIBcharges}. The local metric coefficients $\tilde{g}_{\phi \phi}$, $\tilde{g}_{y_1 y_1}$ can be written
\begin{align}
\tilde{g}_{\phi \phi}&=\frac{32\sigma_2}{C_2^2} \left(\sigma_1+\sigma_2 \cos^2 \frac{\theta_2}{2}\right)\left(\sigma_3+\sigma_2 \sin^2 \frac{\theta_2}{2}\right)
\nonumber
\\
\tilde{g}_{y_1 y_1}&= \frac{16\sigma_2^2 }{C_2^2} \, \frac{ \left(\sigma_1+\sigma_2 \cos^2 \frac{\theta_2}{2}\right)\left(\sigma_3+\sigma_2 \sin^2 \frac{\theta_2}{2}\right) }{\left[q_1 q_3 -\left(m_1+\sigma_1+2\sigma_2 \cos^2 \frac{\theta_2}{2}\right)\left(m_3+\sigma_3+2\sigma_2 \sin^2 \frac{\theta_2}{2}\right)
\right]^2},
\end{align}
while $C_2$ is a constant given by
\begin{align}
C_2 &= 4 \sqrt{\sigma_2(\sigma_1+\sigma_2+\sigma_3)} \sqrt{\frac{1-2\delta}{1+2\delta}} \nonumber \\
&=4\sqrt{\frac{\sigma_2 \left(\sigma_1+\sigma_2+\sigma_3\right)\left[(m_1+m_3)^2-(q_1+q_3)^2-(\sigma_1+\sigma_3+2\sigma_2)^2\right]}{(m_1-m_3)^2-(q_1-q_3-2\gamma)^2-(\sigma_1+\sigma_3+2\sigma_2)^2}}\,.
\end{align}
Fixing $k_2 R_{y_2}=C_2$ as in \eqref{i13_resolution_three_bubble} thus ensures that the time slices of the six-dimensional metric feature a regular bolt, where the $(\rho_2, y_2)$ subspace describes a smooth $\mathbb{R}^2$ origin with line element $d\rho_2^2 +\rho_2^2\frac{dy_2^2}{R_{y_2}^2}$.  The topology of the $(\phi,\theta_2, y_1)$ subspace can be deduced by noting that $\tilde{g}_{\phi \phi}(\theta_2)$,  $\tilde{g}_{y_1 y_1}(\theta_2)$ are positive over the entire range $0 \leq \theta_2 \leq \pi$. The local metric on this subspace is thus that of a warped S$^1 \times$S$^2$.

In summary, the metric near the $i=2$ rod source describes a vacuum bubble of topology S$^1 \times$S$^2$, where the $y_2$-circle smoothly degenerates.

\item[•] \textbf{Geometry at the charged bubbles:}
In this local region, the time slices of the metric take an analogous form, namely
\begin{align}\label{eq:chargedBubMetric}
&ds_6^2|_{dt=0} \propto d\rho_i^2 +\frac{\rho_i^2}{C_i^2} dy_1^2+ \frac{\sigma_i}{2} d\theta_i^2  + (1\pm \cos \theta_i) F_i(\theta_i) d\phi^2+(1\mp \cos \theta_i) G_i(\theta_i) dy_2^2 ,
\nonumber \\
& F_3=-q_i R_{y_2} \left[ (\sin \theta_i +h_i(\theta_i) ) \, d\theta_i \wedge d\phi \wedge 
\frac{dy_2}{R_{y_2}} 
+\rho_i \, t_i(\theta_i) \, d\rho_i \wedge dt \wedge dy_1 \right]+\mathcal{O}(\rho_i)
\end{align}
where the upper sign is taken for $i=1$ and the lower for $i=3$. $F_i(\theta_i)$ and $G_i(\theta_i)$ are smooth and strictly positive functions. $h_i(\theta_i)$ and $t_i(\theta_i)$ are likewise well-behaved, though not positive over the entire range $0 \leq \theta_i \leq \pi$. Instead, $\int_0^\pi h_i(\theta_i) d\theta_i=0$, allowing us to read off the D5-brane charges given in  \eqref{eq:rodcharges}. The components of $F_3$ in \eqref{eq:chargedBubMetric} are related by Hodge duality, as 
\begin{equation}
\star_6 \left[ \rho_i \, t_i(\theta_i) \, d\rho_i \wedge dt \wedge dy_1 \right] = (\sin \theta_i+h_i(\theta_i)) \, d\theta_i \wedge d\phi \wedge dy_2,
\end{equation}
allowing us to read off the D1-brane charges in \eqref{eq:rodcharges} as well.

The constants $C_i$ in \eqref{eq:chargedBubMetric} are found to be given by
\begin{align}
C_1 &\= \frac{ (m_1+\sigma_1)(m_3+\sigma_3+2(\sigma_1+\sigma_2))-q_1(q_3+2\gamma)}{\sqrt{\sigma_1(\sigma_1+\sigma_2)}} \,, \nonumber \\
C_3 &\= \frac{ (m_3+\sigma_3)(m_1+\sigma_1+2(\sigma_3+\sigma_2))-q_3(q_1-2\gamma)}{\sqrt{\sigma_3(\sigma_3+\sigma_2)}}.
\end{align}
With the appropriate constraints \eqref{i13_resolution_three_bubble} on $k_i R_{y_i}$ we again find regular bolts, where the $(\rho_i, y_1)$ subspace smoothly degenerates as a $\mathbb{Z}_{k_i}$ quotient over $\mathbb{R}^2$.
The topology of the $(\theta_i, \phi,y_2)$ subspace is controlled by the $1 \pm \cos \theta_i$ factors in the local metric. In particular, we have a S$^3$ topology, with the $y_2$-circle shrinking on the inner edges of each rod source ($\theta_1=0$ and $\theta_3=\pi$, respectively) and the $\phi$ circle shrinking at the outer edges ($\theta_1=\pi$ and $\theta_3=0$). We conclude that the $(\theta_i, \phi,y_2)$ subspace local to  the $i=1,3$ sources defines S$^3$ bubbles carrying non-trivial D1 and D5 charges.

\end{enumerate}



\bibliographystyle{utphys}      

\bibliography{microstates}       

\providecommand{\href}[2]{#2}\begingroup\raggedright\begin{thebibliography}{10}

\bibitem{Lin:2004nb}
H.~Lin, O.~Lunin, and J.~M. Maldacena, ``{Bubbling AdS space and 1/2 BPS
  geometries},'' {\em JHEP} {\bfseries 10} (2004) 025,
\href{http://arxiv.org/abs/hep-th/0409174}{{\ttfamily arXiv:hep-th/0409174}}.

\bibitem{Lunin:2001jy}
O.~Lunin and S.~D. Mathur, ``{AdS/CFT duality and the black hole information
  paradox},'' \href{http://dx.doi.org/10.1016/S0550-3213(01)00620-4}{{\em Nucl.
  Phys.} {\bfseries B623} (2002) 342--394},
\href{http://arxiv.org/abs/hep-th/0109154}{{\ttfamily arXiv:hep-th/0109154}}.

\bibitem{Bena:2022ldq}
I.~Bena, E.~J. Martinec, S.~D. Mathur, and N.~P. Warner, ``{Snowmass White
  Paper: Micro- and Macro-Structure of Black Holes},''
  \href{http://arxiv.org/abs/2203.04981}{{\ttfamily arXiv:2203.04981
  [hep-th]}}.

\bibitem{Warner:2019jll}
N.~P. Warner, ``{Lectures on Microstate Geometries},''
  \href{http://arxiv.org/abs/1912.13108}{{\ttfamily arXiv:1912.13108
  [hep-th]}}.

\bibitem{Bena:2007kg}
I.~Bena and N.~P. Warner, ``{Black holes, black rings and their microstates},''
  \href{http://dx.doi.org/10.1007/978-3-540-79523-0}{{\em Lect. Notes Phys.}
  {\bfseries 755} (2008) 1--92},
\href{http://arxiv.org/abs/hep-th/0701216}{{\ttfamily arXiv:hep-th/0701216}}.

\bibitem{Bah:2020ogh}
I.~Bah and P.~Heidmann, ``{Topological Stars and Black Holes},''
  \href{http://arxiv.org/abs/2011.08851}{{\ttfamily arXiv:2011.08851
  [hep-th]}}.

\bibitem{Bah:2020pdz}
I.~Bah and P.~Heidmann, ``{Topological Stars, Black holes and Generalized
  Charged Weyl Solutions},'' \href{http://arxiv.org/abs/2012.13407}{{\ttfamily
  arXiv:2012.13407 [hep-th]}}.

\bibitem{Bah:2021owp}
I.~Bah and P.~Heidmann, ``{Smooth Bubbling Geometries Without Supersymmetry},''
  \href{http://arxiv.org/abs/2106.05118}{{\ttfamily arXiv:2106.05118
  [hep-th]}}.

\bibitem{Bah:2021rki}
I.~Bah and P.~Heidmann, ``{Bubble Bag End: A Bubbly Resolution of Curvature
  Singularity},'' \href{http://arxiv.org/abs/2107.13551}{{\ttfamily
  arXiv:2107.13551 [hep-th]}}.

\bibitem{Heidmann:2021cms}
P.~Heidmann, ``{Non-BPS Floating Branes and Bubbling Geometries},''
  \href{http://arxiv.org/abs/2112.03279}{{\ttfamily arXiv:2112.03279
  [hep-th]}}.

\bibitem{Weyl:book}
H.~Weyl {\em Ann. Phys. (Leipzig)} {\bfseries 54} (1917) 117.

\bibitem{Papapetrou:1953zz}
A.~Papapetrou, ``{Eine rotationssymmetrische losung in der allgemeinen
  relativitatstheorie},'' {\em Annals Phys.} {\bfseries 12} (1953) 309--315.

\bibitem{Emparan:2001wk}
R.~Emparan and H.~S. Reall, ``{Generalized Weyl solutions},''
  \href{http://dx.doi.org/10.1103/PhysRevD.65.084025}{{\em Phys. Rev. D}
  {\bfseries 65} (2002) 084025},
  \href{http://arxiv.org/abs/hep-th/0110258}{{\ttfamily arXiv:hep-th/0110258}}.

\bibitem{Serini}
R.~Serini, ``{Euclideita dello spazio completamente vuto nella relativita
  general di Einstein},'' {\em Atti dela Academia dei Lincei Ser. 5 Rendiconti}
  {\bfseries 27} (1918) 235--238.

\bibitem{Einstein}
A.~Einstein, ``{Demonstration of the non-existence of gravitational fields with
  a non-vanishing total mass free of singularities},'' {\em Univ. Nac. Tucumn.
  Revista A} {\bfseries 2} (1941) 5--15.

\bibitem{Einstein:1943ixi}
A.~Einstein and W.~Pauli, ``{On the Non-Existence of Regular Stationary
  Solutions of Relativistic Field Equations},''
  \href{http://dx.doi.org/10.2307/1968759}{{\em Annals Math.} {\bfseries 44}
  no.~2, (1943) 131}.

\bibitem{Lichnerowicz}
A.~Lichnerowicz {\em Theories Relativiste de la Gravitation et de
  l’Electromagnetisme} (1955) .

\bibitem{Gibbons:2013tqa}
G.~Gibbons and N.~Warner, ``{Global structure of five-dimensional fuzzballs},''
  \href{http://dx.doi.org/10.1088/0264-9381/31/2/025016}{{\em
  Class.Quant.Grav.} {\bfseries 31} (2014) 025016},
\href{http://arxiv.org/abs/1305.0957}{{\ttfamily arXiv:1305.0957 [hep-th]}}.

\bibitem{Stotyn:2011tv}
S.~Stotyn and R.~B. Mann, ``{Magnetic charge can locally stabilize
  Kaluza\textendash{}Klein bubbles},''
  \href{http://dx.doi.org/10.1016/j.physletb.2011.10.015}{{\em Phys. Lett. B}
  {\bfseries 705} (2011) 269--272},
  \href{http://arxiv.org/abs/1105.1854}{{\ttfamily arXiv:1105.1854 [hep-th]}}.

\bibitem{Bah:2021irr}
I.~Bah, A.~Dey, and P.~Heidmann, ``{Stability of topological solitons, and
  black string to bubble transition},''
  \href{http://arxiv.org/abs/2112.11474}{{\ttfamily arXiv:2112.11474
  [hep-th]}}.

\bibitem{Witten:1981gj}
E.~Witten, ``{Instability of the Kaluza-Klein Vacuum},''
  \href{http://dx.doi.org/10.1016/0550-3213(82)90007-4}{{\em Nucl. Phys. B}
  {\bfseries 195} (1982) 481--492}.

\bibitem{Alekseev:2007re}
G.~A. Alekseev and V.~A. Belinski,
  \href{http://dx.doi.org/10.1142/9789812834300_0022}{``{Superposition of
  fields of two Reissner - Nordstrom sources},''} in {\em {11th Marcel
  Grossmann Meeting on General Relativity}}, pp.~2490--2492.
\newblock 10, 2007.
\newblock \href{http://arxiv.org/abs/0710.2515}{{\ttfamily arXiv:0710.2515
  [gr-qc]}}.

\bibitem{Alekseev:2007gt}
G.~A. Alekseev and V.~A. Belinski, ``{Equilibrium configurations of two charged
  masses in General Relativity},''
  \href{http://dx.doi.org/10.1103/PhysRevD.76.021501}{{\em Phys. Rev. D}
  {\bfseries 76} (2007) 021501},
  \href{http://arxiv.org/abs/0706.1981}{{\ttfamily arXiv:0706.1981 [gr-qc]}}.

\bibitem{Manko:2007hi}
V.~S. Manko, ``{The Double-Reissner-Nordstrom solution and the interaction
  force between two spherically symmetric charged particles},''
  \href{http://dx.doi.org/10.1103/PhysRevD.76.124032}{{\em Phys. Rev. D}
  {\bfseries 76} (2007) 124032},
  \href{http://arxiv.org/abs/0710.2158}{{\ttfamily arXiv:0710.2158 [gr-qc]}}.

\bibitem{Manko:2008gb}
V.~S. Manko, E.~Ruiz, and J.~Sanchez-Mondragon, ``{Analogs of the
  double-Reissner-Nordstrom solution in magnetostatics and dilaton gravity:
  mathematical description and some physical properties},''
  \href{http://dx.doi.org/10.1103/PhysRevD.79.084024}{{\em Phys. Rev. D}
  {\bfseries 79} (2009) 084024},
  \href{http://arxiv.org/abs/0811.2029}{{\ttfamily arXiv:0811.2029 [gr-qc]}}.

\bibitem{PhysRevD.51.4192}
E.~Ruiz, V.~S. Manko, and J.~Mart\'{\i}n, ``Extended n-soliton solution of the
  einstein-maxwell equations,''
  \href{http://dx.doi.org/10.1103/PhysRevD.51.4192}{{\em Phys. Rev. D}
  {\bfseries 51} (Apr, 1995) 4192--4197}.
  \url{https://link.aps.org/doi/10.1103/PhysRevD.51.4192}.

\bibitem{Elvang:2002br}
H.~Elvang and G.~T. Horowitz, ``{When black holes meet Kaluza-Klein bubbles},''
  \href{http://dx.doi.org/10.1103/PhysRevD.67.044015}{{\em Phys. Rev. D}
  {\bfseries 67} (2003) 044015},
  \href{http://arxiv.org/abs/hep-th/0210303}{{\ttfamily arXiv:hep-th/0210303}}.

\bibitem{Lunin:2001fv}
O.~Lunin and S.~D. Mathur, ``{Metric of the multiply wound rotating string},''
  \href{http://dx.doi.org/10.1016/S0550-3213(01)00321-2}{{\em Nucl. Phys.}
  {\bfseries B610} (2001) 49--76},
\href{http://arxiv.org/abs/hep-th/0105136}{{\ttfamily arXiv:hep-th/0105136}}.

\bibitem{Taylor:2005db}
M.~Taylor, ``{General 2 charge geometries},'' {\em JHEP} {\bfseries 03} (2006)
  009,
\href{http://arxiv.org/abs/hep-th/0507223}{{\ttfamily arXiv:hep-th/0507223}}.

\bibitem{Mathur:2005ai}
S.~D. Mathur, ``{The quantum structure of black holes},''
  \href{http://dx.doi.org/10.1088/0264-9381/23/11/R01}{{\em Class. Quant.
  Grav.} {\bfseries 23} (2006) R115},
\href{http://arxiv.org/abs/hep-th/0510180}{{\ttfamily arXiv:hep-th/0510180}}.

\bibitem{Myers:1986un}
R.~C. Myers and M.~Perry, ``{Black Holes in Higher Dimensional Space-Times},''
\href{http://dx.doi.org/10.1016/0003-4916(86)90186-7}{{\em Annals Phys.}
  {\bfseries 172} (1986) 304}.

\bibitem{Bena:2020uup}
I.~Bena and D.~R. Mayerson, ``{Black Holes Lessons from Multipole Ratios},''
  \href{http://arxiv.org/abs/2007.09152}{{\ttfamily arXiv:2007.09152
  [hep-th]}}.

\bibitem{Bah:2021jno}
I.~Bah, I.~Bena, P.~Heidmann, Y.~Li, and D.~R. Mayerson, ``{Gravitational
  Footprints of Black Holes and Their Microstate Geometries},''
  \href{http://arxiv.org/abs/2104.10686}{{\ttfamily arXiv:2104.10686
  [hep-th]}}.

\bibitem{Bena:2015bea}
I.~Bena, S.~Giusto, R.~Russo, M.~Shigemori, and N.~P. Warner, ``{Habemus
  Superstratum! A constructive proof of the existence of superstrata},''
  \href{http://dx.doi.org/10.1007/JHEP05(2015)110}{{\em JHEP} {\bfseries 05}
  (2015) 110},
\href{http://arxiv.org/abs/1503.01463}{{\ttfamily arXiv:1503.01463 [hep-th]}}.

\bibitem{Bena:2016ypk}
I.~Bena, S.~Giusto, E.~J. Martinec, R.~Russo, M.~Shigemori, D.~Turton, and
  N.~P. Warner, ``{Smooth horizonless geometries deep inside the black-hole
  regime},'' \href{http://dx.doi.org/10.1103/PhysRevLett.117.201601}{{\em Phys.
  Rev. Lett.} {\bfseries 117} no.~20, (2016) 201601},
\href{http://arxiv.org/abs/1607.03908}{{\ttfamily arXiv:1607.03908 [hep-th]}}.

\bibitem{Bena:2016agb}
I.~Bena, E.~Martinec, D.~Turton, and N.~P. Warner, ``{Momentum Fractionation on
  Superstrata},'' \href{http://dx.doi.org/10.1007/JHEP05(2016)064}{{\em JHEP}
  {\bfseries 05} (2016) 064},
\href{http://arxiv.org/abs/1601.05805}{{\ttfamily arXiv:1601.05805 [hep-th]}}.

\bibitem{Bena:2017xbt}
I.~Bena, S.~Giusto, E.~J. Martinec, R.~Russo, M.~Shigemori, D.~Turton, and
  N.~P. Warner, ``{Asymptotically-flat supergravity solutions deep inside the
  black-hole regime},'' \href{http://dx.doi.org/10.1007/JHEP02(2018)014}{{\em
  JHEP} {\bfseries 02} (2018) 014},
\href{http://arxiv.org/abs/1711.10474}{{\ttfamily arXiv:1711.10474 [hep-th]}}.

\bibitem{Ceplak:2018pws}
N.~Ceplak, R.~Russo, and M.~Shigemori, ``{Supercharging Superstrata},''
\href{http://arxiv.org/abs/1812.08761}{{\ttfamily arXiv:1812.08761 [hep-th]}}.

\bibitem{Heidmann:2019zws}
P.~Heidmann and N.~P. Warner, ``{Superstratum Symbiosis},''
\href{http://arxiv.org/abs/1903.07631}{{\ttfamily arXiv:1903.07631 [hep-th]}}.

\bibitem{Heidmann:2019xrd}
P.~Heidmann, D.~R. Mayerson, R.~Walker, and N.~P. Warner, ``{Holomorphic Waves
  of Black Hole Microstructure},''
  \href{http://dx.doi.org/10.1007/JHEP02(2020)192}{{\em JHEP} {\bfseries 02}
  (2020) 192}, \href{http://arxiv.org/abs/1910.10714}{{\ttfamily
  arXiv:1910.10714 [hep-th]}}.

\bibitem{Mayerson:2020tpn}
D.~R. Mayerson, ``{Fuzzballs and Observations},''
  \href{http://dx.doi.org/10.1007/s10714-020-02769-w}{{\em Gen. Rel. Grav.}
  {\bfseries 52} no.~12, (2020) 115},
  \href{http://arxiv.org/abs/2010.09736}{{\ttfamily arXiv:2010.09736
  [hep-th]}}.

\end{thebibliography}\endgroup



\end{document}